\documentclass{article}

\usepackage{arxiv}

\usepackage[utf8]{inputenc} 
\usepackage[T1]{fontenc}    
\usepackage{hyperref}       
\usepackage{url}            
\usepackage{booktabs}       
\usepackage{amsfonts}       
\usepackage{nicefrac}       
\usepackage{microtype}      
\usepackage{lipsum}
\usepackage{graphicx}
\usepackage{lipsum}
\usepackage{graphicx}
\usepackage{subfigure}
\usepackage{natbib} 
\usepackage{amsthm,amsmath,amsfonts,amssymb}
\newtheorem{theorem}{Theorem}
\graphicspath{ {./images/} }
\usepackage{algpseudocode}
\usepackage[linesnumbered,ruled,vlined]{algorithm2e}
\SetKwInput{KwInput}{Input}        
\SetKwInput{KwOutput}{Output}  
\usepackage{appendix}

\title{Adaptive Grid Designs for Classifying Monotonic Binary Deterministic Computer Simulations}

\author{
Tian Bai\thanks{Joint first authors.}\\
School of Mathematics and Statistics\\
Beijing Institute of Technology\\
\texttt{3120215737@bit.edu.cn}
\And
Dianpeng Wang\footnotemark[1]\\
School of Mathematics and Statistics\\
Beijing Institute of Technology\\
\texttt{wdp@bit.edu.cn}
\And
Kuangqi Chen\\
School of Mechanical Engineering\\
Beijing Institute of Technology\\
\texttt{3220225081@bit.edu.cn}
\And
Xu He\thanks{Corresponding author.}\\
State Key Laboratory of Mathematical Science (SKLMS)\\ 
Academy of Mathematics and Systems Science\\
Chinese Academy of Sciences\\
\texttt{hexu@amss.ac.cn}
}

\begin{document}
\maketitle
\begin{abstract}
This research is motivated by the need for effective classification in ice-breaking dynamic simulations, aimed at determining the conditions under which an underwater vehicle will break through the ice. This simulation is extremely time-consuming and yields deterministic, binary, and monotonic outcomes. Detecting the critical edge between the negative-outcome and positive-outcome regions with minimal simulation runs necessitates an efficient experimental design for selecting input values.
In this paper, we derive lower bounds on the number of functional evaluations needed to ensure a certain level of classification accuracy for arbitrary static and adaptive designs. 
We also propose a new class of adaptive designs called adaptive grid designs, which are sequences of grids with increasing resolution such that lower resolution grids are proper subsets of higher resolution grids. 
By prioritizing simulation runs at lower resolution points and skipping redundant runs, adaptive grid designs require the same order of magnitude of runs as the best possible adaptive design, which is an order of magnitude fewer than the best possible static design. 
Numerical results across test functions, the road crash simulation and the ice-breaking simulation validate the superiority of adaptive grid designs. 
\end{abstract}


\section{Introduction}
Simulation is a widely employed technique to investigate complex processes and systems across various fields~\citep{Santner2003}. 
Many simulations are deterministic in that there is no random error for the outcome. 
When the simulation yields deterministic, binary, and monotonic outcomes for all input variables, practitioners often confront the challenge of detecting the edge that separates the negative-outcome and positive-outcome regions. 

\begin{figure}
	\centering
	\includegraphics[width=1\linewidth]{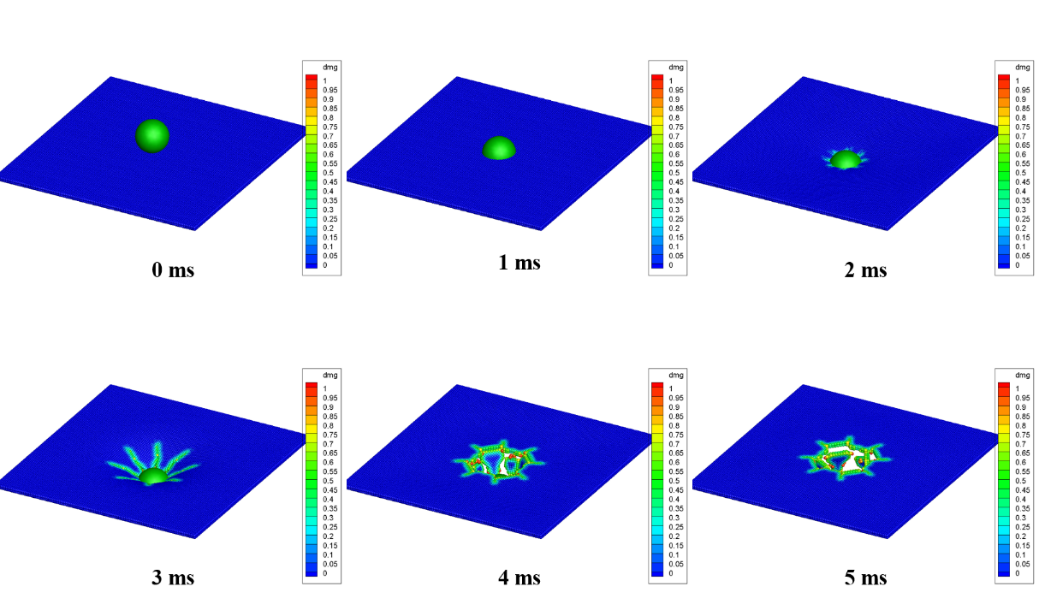}
	\caption{Simulation on the local damage of the ice sheet at different times. \label{fig:icesheet}}
\end{figure}

One compelling example is the ice-breaking simulation, which motivates this research. 
In order to guarantee the success of high-speed underwater ice-breaking vehicles, engineers are eager to understand the successful conditions under which the underwater vehicle will break through the ice~\citep{DONG2024119659}.
In particular, the interest centers around a few key controllable factors that significantly affect the status of the launch, such as the initial velocity of the vessel, the ice thickness, and the elastic modulus of the ice, which measures how much the ice deforms under a certain stress~\citep{CHEN2021108811}.
The resulting response is binary: either success or failure in breaking through the ice.  
The ice-breaking process is governed by a nonlinear, monotonic, and threshold-driven failure mechanism, where small changes in input parameters near the transition zone lead to abrupt changes in system response. 
Physically, this corresponds to the transition from subcritical impact, where the ice undergoes elastic deformation and recovery, to supercritical impact, where crack initiation, propagation, and fracture lead to catastrophic failure. 
The simulation is deterministic since it consistently produces the same output given the same input.
Moreover, the response is monotonic, as higher initial velocity, lower ice thickness, and smaller ice elastic modulus increase the likelihood of ice breakage. 
Regrettably, simulating ice-breaking dynamics is exceedingly time-consuming, as a single simulation run requires over 7 days of computation using 120 CPU cores. 
To streamline computation, we adopt a simplified ice-breaking dynamic by using a sphere to replace the actual vehicle. 
Even with the simplified dynamic, a single simulation run requires nearly 10 hours of computation, parallelized across 120 CPU cores, to complete. 
For visualization, Fig.~\ref{fig:icesheet} illustrates the progression of the local damage on the ice sheet at various time points throughout one simulation run. 
Other examples of deterministic, binary, and monotonic simulations include road crash simulations~\citep{imberg2022active}, flow regime transition simulation~\citep{ZHANG2023}, and onset of cavitation simulation~\citep{HE2024}, among others.

To accurately classify the outcome using as few simulation runs as possible, it is crucial to use an adequate model for outcome prediction along with an efficient experimental design to determine the input values of simulation runs.
Because the simulations are deterministic, ideally the model should correctly predict all outcomes of completed experiments. 
For physical experiments with binary outputs, logistic linear regression models are prevalent~\citep{Stufken2012}.
However, as parametric models, logistic linear regression models are incapable to predict complex edges that separate the negative-outcome and positive-outcome regions. 
Furthermore, even when a hyperplane or a quadratic curve can correctly partition the obtained outcomes, there is neither unique maximum likelihood estimator nor unique separating edge for logistic linear regression models~\citep{Silvapulle1981}. 
This is because logistic regression assumes that the underlying probability of a positive outcome lies between zero and one, whereas in the problems we consider throughout this paper, the underlying probabilities are either exactly zero or exactly one.
For deterministic simulations, Gaussian processes models are commonly adopted~\citep{Santner2003}. Gaussian process models that incorporate monotonicity information have been proposed in \cite{Golchi2015},  \cite{Wang2016}, and \cite{Shape-Constrained}.  
However, ordinary Gaussian process models and their variants are not optimal for our problems since they assume each output follows a Gaussian distribution, while our problem involves binary outputs. 
The generalized Gaussian process model~\citep{Sung2020}, proposed for computer experiments with binary outcomes, is also unsuitable because it assumes each outcome follows a Bernoulli distribution with a probability that is between zero and one. 
Models that handle deterministic binary outcome while leveraging monotonicity include monotonic random forest \citep{Francisco2015}, XGBoost with monotonic constraints \citep{XGBoost2016}, 
as well as those described in \cite{cano2019monotonic} and the references therein. 
In this work, we use the support vector classification method~\citep{Corinna1995} due to its high efficiency with small sample sizes and ease of implementation~\citep{CERVANTES2020189}.

The main theoretical contribution of this study lies in developing and evaluating experimental design methods for efficiently classifying monotonic binary deterministic computer simulations. 
Space-filling designs such as Latin hypercube designs \citep{Mckay1979} and maximin-distance Latin hypercube designs~\citep{Shan2015} that featured at scattering points uniformly in the design space are prevailing for general computer simulations~\citep{Santner2003}. 
For detecting the edge, however, high-resolution grid points are more popular. 
Consider two design points $\mathbf{x}$ and $\mathbf{y}$, where $\mathbf{x}$ is no higher than $\mathbf{y}$ in all dimensions.
According to the monotonic and deterministic properties, if the outcome of $\mathbf{y}$ is negative, we can infer without running any experiment that the outcome of $\mathbf{x}$ is also negative. 
Likewise, if the outcome of $\mathbf{x}$ is positive, we can skip running the run at $\mathbf{y}$ because the outcome must also be positive. Therefore, for monotonic binary deterministic simulations, \emph{adaptive designs}, which select input values based on obtained outcomes, outperform \emph{static designs}.

Despite obvious advantages, few adaptive design methods have been proposed for monotonic binary deterministic computer simulations. 
Adaptive designs for logistic models~\citep{Probal1993,yang2012optimal} or ordinary Gaussian process models~\citep{ Jones1998, Lee2023} do not work well for our problem because the model assumptions are violated. 
There are no adaptive design methods proposed for models that handle deterministic binary outcomes while leveraging monotonicity.
\cite{DEROCQUIGNY2009363} summarized adaptive design methods for monotonic simulations, but among them, only the adaptive Monte Carlo method is applicable for binary simulations. 
We may use minimum energy designs \citep{Roshan2015}, which sequentially sample points from a known distribution, by setting the distribution to be the uniform distribution that is supported on the uncertain area. 
Another approach is to utilize active learning techniques, such as uncertainty sampling \citep{Lewis1994} and contrastive active learning \citep{Margatina2021}, along with other methods suited for binary classification, as summarized in \cite{Kumar2020}.
However, neither minimum energy designs nor active learning methods exploit the monotonic property. 
Additionally, to the best of our knowledge, no theoretical results have been provided on how many functional evaluations is needed to ensure a certain level of classification accuracy. 

In this paper, 
we propose a new class of adaptive designs called adaptive grid designs, which are sequences of grids with increasing resolution such that lower resolution grids are proper subsets of higher resolution grids. 
By prioritizing simulation runs at lower resolution points and skipping redundant runs, adaptive grid designs are significantly superior to static grid designs. 
Furthermore, we derive lower bounds on the number of functional evaluations needed to ensure a certain level of classification accuracy for arbitrary static and adaptive designs. 
We show that adaptive grid designs require the same order of magnitude of runs as the best possible adaptive design, which is an order of magnitude fewer than the best possible static design. 
Finally, we validate the advantage of adaptive grid designs using numerical results on test functions,  the road crash simulation, and the ice-breaking simulation. 

The rest of this article is organized as follows.  
In Section~\ref{sec:design}, we propose the construction of adaptive grid designs after summarizing several other types of designs and derive their theoretical properties. 
In Section~\ref{sec:comp}, we provide numerical results to compare different design approaches. 
In Section~\ref{sec:road} and~\ref{sec:ice}, we apply the proposed adaptive grid designs to the road crash simulation and the ice-breaking simulation.  
Some final remarks are given in Section~\ref{sec:conc}. 
Additional design construction methods as well as theoretical properties and proofs of theorems are provided in the Appendix.


\section{Design Methodology}\label{sec:design}

In this section, we provide the construction and properties of several types of designs.

\subsection{An Illustration}

Throughout this paper, let $f({\mathbf{x}}) \in \{-1,1\}$ denote the binary outcome corresponding to design point ${\mathbf{x}}$. 
Without loss of generality, throughout this section we assume the design space is $[0,1]^p$ and the simulation is monotonic non-decreasing, i.e., $f(x_1,\ldots,x_p) \leq f(y_1,\ldots,y_p)$ if $x_k\leq y_k$ for any $1\leq k\leq p$. 
Let $\mathbf{\Omega}$ denote the set of such functions. 
Nevertheless, we shall discuss our remedy when the input space is not $[0,1]^p$ or the function is monotonic non-increasing for some input variables in Sections~\ref{sec:road}--\ref{sec:ice}. 

Let $\mathbf{A} = \{ \mathbf{z} : \mathbf{z} \in [0,1]^p , f(\mathbf{z}) = -1 \}$ and $\mathbf{B} = \{ \mathbf{z} : \mathbf{z} \in [0,1]^p , f(\mathbf{z}) = 1 \}$ denote the negative-outcome and positive-outcome regions, respectively. 
Due to the monotonic property, after obtaining the outcomes corresponding to a design $\mathbf{D} \subset [0,1]^p$, we will know for sure that $f(\mathbf{z})=-1$ for $\mathbf{z} \in \cup_{\mathbf{x} \in \mathbf{D}, f(\mathbf{x})=-1} \prod_{k=1}^p [0,x_k]$ and $f(\mathbf{z})=1$ for $\mathbf{z} \in \cup_{\mathbf{x} \in \mathbf{D}, f(\mathbf{x})=1} \prod_{k=1}^p [x_k,1]$. 
Clearly, $\cup_{\mathbf{x} \in \mathbf{D}, f(\mathbf{x})=-1} \prod_{k=1}^p [0,x_k] \subset \mathbf{A}$, $\cup_{\mathbf{x} \in \mathbf{D}, f(\mathbf{x})=1} \prod_{k=1}^p [x_k,\allowbreak 1] \subset \mathbf{B}$, $\mathbf{B} = [0,1]^p \setminus \mathbf{A}$, and we are unsure of the outcome only at the uncertain area 
\[ \mathbf{U} = \left\{ \mathbf{A} \setminus \left( \cup_{\mathbf{x} \in \mathbf{D}, f(\mathbf{x})=-1} \prod_{k=1}^p \left[0,x_k\right] \right) \right\}
\cup \left\{ \mathbf{B} \setminus \left( \cup_{\mathbf{x} \in \mathbf{D}, f(\mathbf{x})=1} \prod_{k=1}^p \left[x_k,1\right] \right) \right\}. \]
Let $V(\mathbf{Z})$ denote the volume of a region $\mathbf{Z} \subset [0,1]^p$. 
Clearly, the accuracy of a given classifier is closely related to the volume of uncertainty area, $V(\mathbf{U})$, which is determined by $\mathbf{D}$ and $f(\cdot)$. 
Since we have no prior knowledge on $f(\cdot)$, an ideal design should lead to low $V\{\mathbf{U}(\mathbf{D},f)\}$
regardless of $f(\cdot)$. 
It is thus reasonable to use $\text{sup}_{f \in \mathbf{\Omega}} V\{\mathbf{U}(\mathbf{D},f)\}$ to measure the efficiency of the design $\mathbf{D}$. 

\begin{figure}
	\centering
	\subfigure[$\mathbf{D}_{\text{SG},2,16}$, $v=0.188$.]{
		\begin{minipage}[t]{0.3\linewidth}
			\centering
			\includegraphics[width=\linewidth]{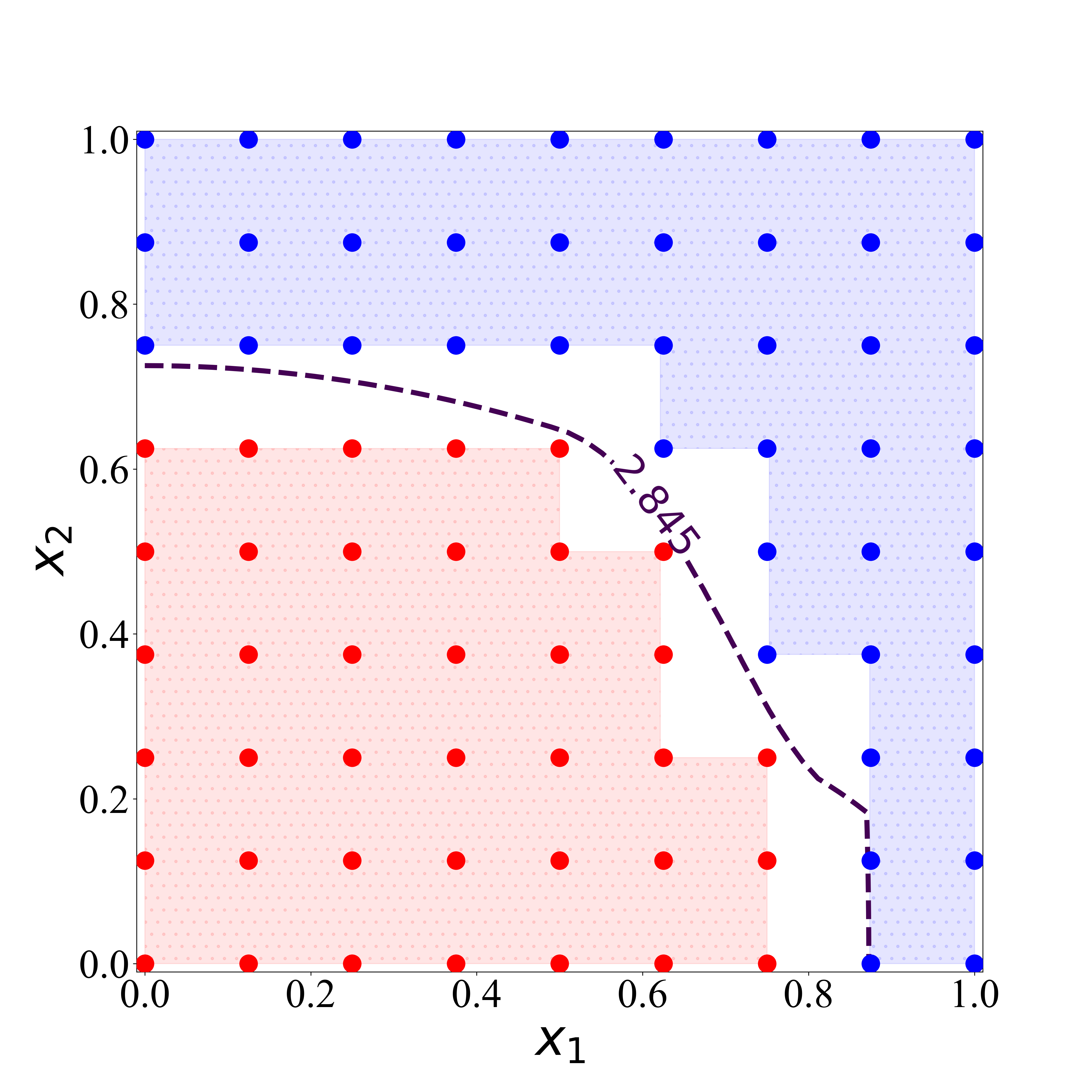}
			
			\label{illu_SG}
		\end{minipage}
	}
			%
	\subfigure[$\mathbf{D}_{\text{AG},2,8}$, $v= 0.375$.]{
		\begin{minipage}[t]{0.3\linewidth}
			\centering
			\includegraphics[width=\linewidth]{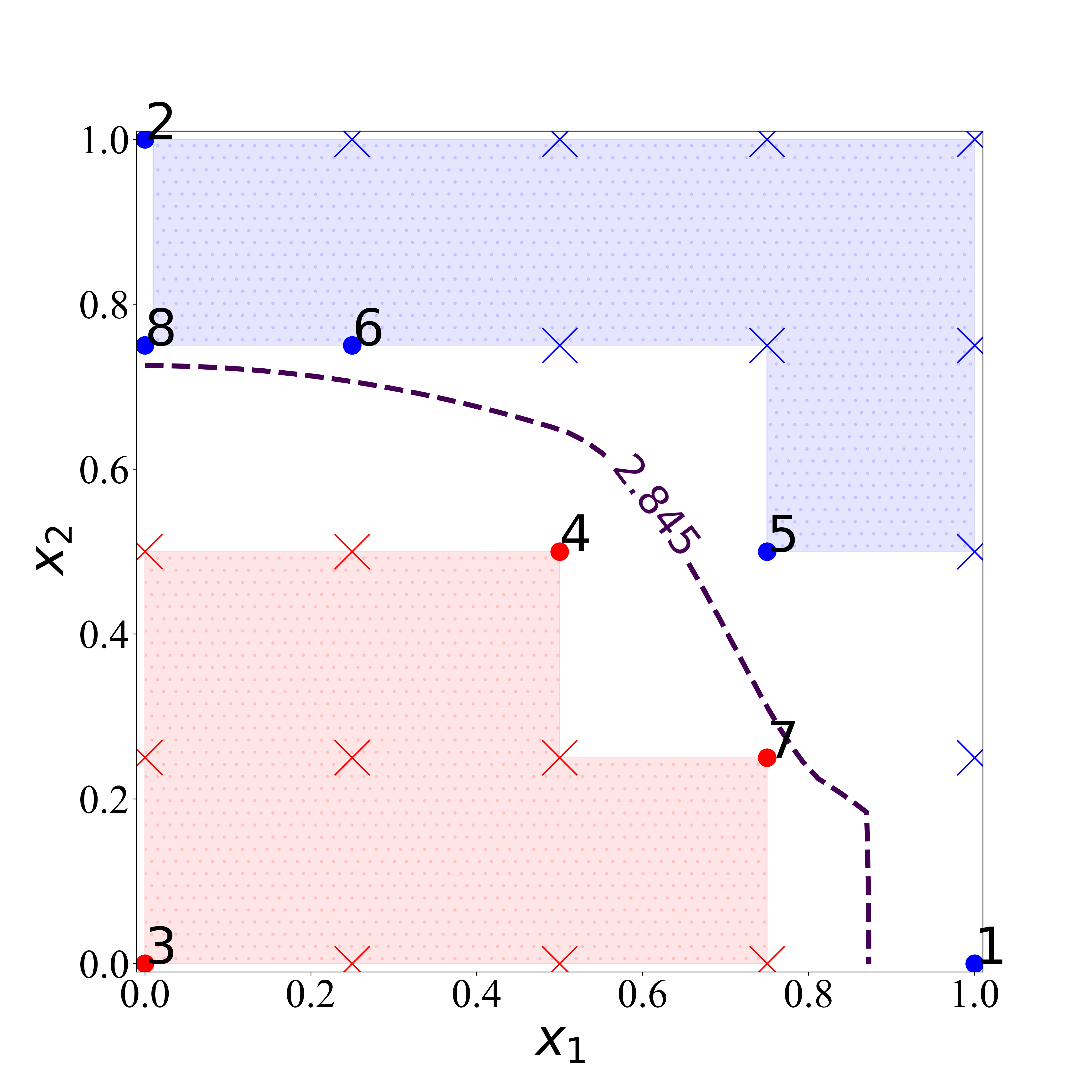}
			
			\label{illu_AG_2}
		\end{minipage}
	}
	\subfigure[$\mathbf{D}_{\text{AG},2,16}$, $v= 0.188$.]{
		\begin{minipage}[t]{0.3\linewidth}
			\centering
			\includegraphics[width=\linewidth]{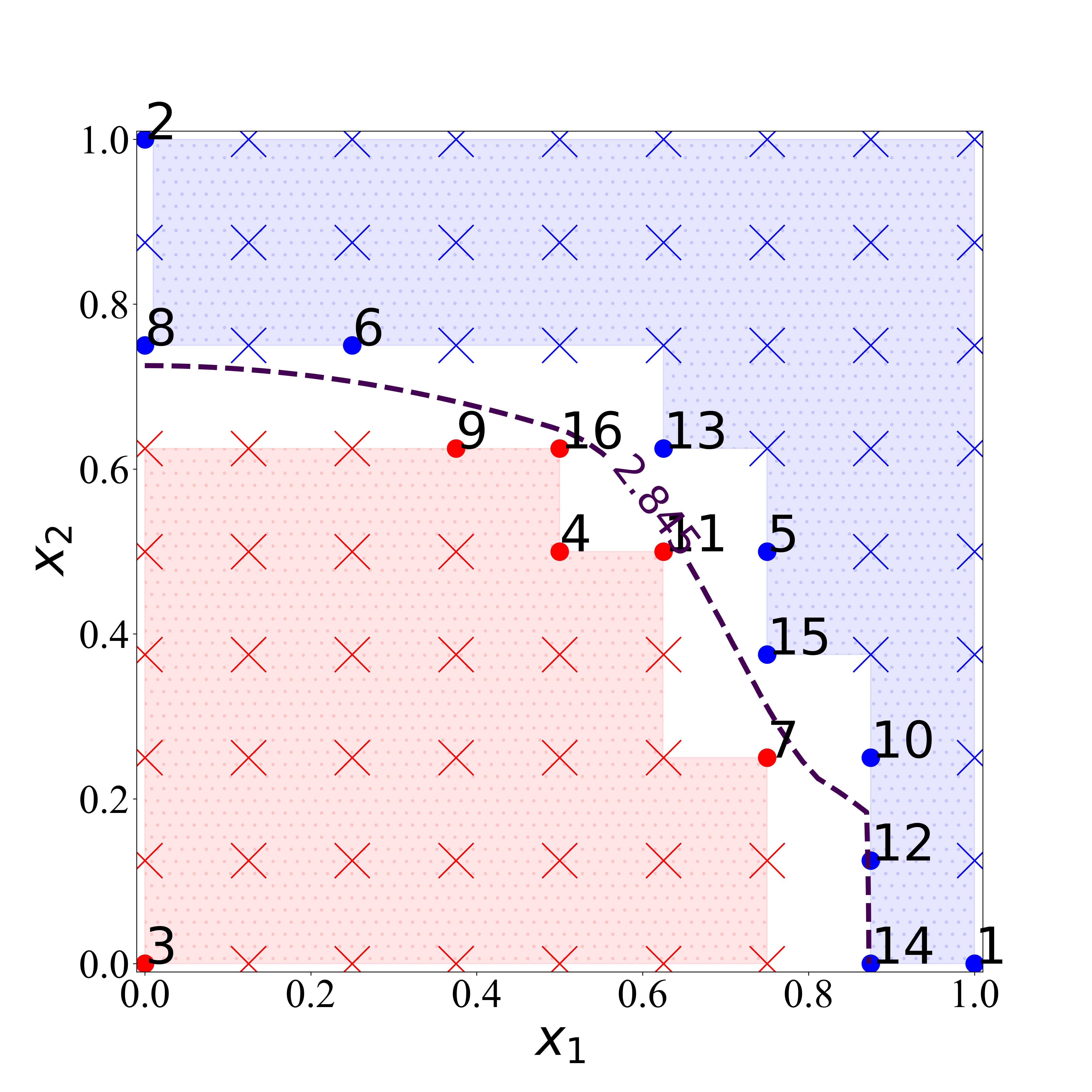}
			
			\label{illu_AG_3}
	\end{minipage}}
	\caption{SG and AG designs, 
		showing the curve separating the two regions $\mathbf{A}$ and $\mathbf{B}$ (purple dotted line), 
		the design points with negative response $\mathbf{D}\cap\mathbf{A}$ (red dots, or dots to the bottom-left of the dotted line) and positive response $\mathbf{D}\cap\mathbf{B}$ (blue dots, or dots to the top-right of the dotted line), 
		overlaid with their order in evaluation (numbers), the skipped design points (crosses), 
		the certainly negative area (red shaded area, or the bottom-left shaded area), 
		the certainly positive area (blue shaded area, or the top-right shaded area), 
		and the uncertain area $\mathbf{U}$ (white area) after evaluating the runs corresponding to $\mathbf{D}$. 
	}
	\label{fig:illu}
\end{figure}

Fig.~\ref{fig:illu} illustrates the design points $\mathbf{D}$ of the static grid design (SG) and the proposed adaptive grid design (AG), which we shall formally define later in this section, for classifying the
\begin{align}\label{eq:illustrate}
	f(\mathbf{x})= \begin{cases} 1, & x_1^2+x_2^2+15\left\{(x_1-0.5)^+\right\}^2+3\left\{(x_2-0.2)^+\right\}^{0.4} \ge 2.84,\\
		-1,& \text{otherwise}, \end{cases} 
\end{align}
where $z^+=z$ when $z\geq 0$ and $z^+=0$ otherwise.  
In the figure, alongside the design points, we delineate the certainly negative area $\cup_{\mathbf{x} \in \mathbf{D}, f(\mathbf{x})=-1} \prod_{k=1}^p [0,x_k]$, the certainly positive area $\cup_{\mathbf{x} \in \mathbf{D}, f(\mathbf{x})=1} \prod_{k=1}^p [x_k,1]$, and the uncertain area $\mathbf{U}$. 
As can be observed, the $\mathbf{U}$ for the 81-run SG is exactly the same to the $\mathbf{U}$ for the 16-run AG, 
indicating that from using the AG we can safely skip 65 runs because we can deduce their output values from the monotonic property. 
As a matter of fact, the current $\mathbf{U}$ can be concluded from only 7 runs, namely, Run-8, 16, 13, 11, 15, 7, and 14 of the AG. 
Should we know this $\mathbf{U}$ in advance, we can avoid the other 9 runs. 
Nevertheless, we will show in Section~\ref{sec:design:A} that the AG is already an excellent design because it achieves the optimal rate of converge on the number of functional evaluations needed to ensure certain level of classification accuracy. 
We will also show that the ratio on the number of functional evaluations needed for the AG divided by that for any static design will converge to zero as the $V(\mathbf{U})$ goes to zero, showing the tremendous superiority of the AG.

\subsection{Static Designs}\label{sec:design:S}

Theorem~\ref{thm:SG} below provides the asymptotic property of the SG given by 
\[ {\mathbf{D}_{\text{SG},p,n}} = \left\{0/\left(n^{1/p}-1\right),1/\left(n^{1/p}-1\right),\ldots,\left(n^{1/p}-1\right)/\left(n^{1/p}-1\right) \right\}^p, \]
when $n^{1/p}$ is an integer greater than 1, which is illustrated in Fig.~\ref{fig:illu}(a). 

\begin{theorem}\label{thm:SG}
	For the static grid design $\mathbf{D}_{\text{SG},p,n}$ with $n^{1/p}$ being an integer greater than one, the maximal possible volume of uncertain area is 
	\[\text{sup}_{f \in \mathbf{\Omega}} V\{\mathbf{U}(\mathbf{D}_{\text{SG},p,n},f)\} = 1 - (n^{1/p}-2)^p/(n^{1/p}-1)^{p}.\]
\end{theorem} 

As indicated by Theorem~\ref{thm:SG}, for large $n$, the $\text{sup}_{f \in \mathbf{\Omega}} V\{\mathbf{U}(\mathbf{D}_{\text{SG},p,n},f)\}$ converges to approximately $p n^{-1/p}$. 
Consequently, it requires approximately $p^p v^{-p}$ simulation trails to ensure that $V(\mathbf{U}) \leq v$. 
Next, Theorem~\ref{thm:static} below provides a lower bound on the required number of runs for arbitrary static designs.

\begin{theorem}\label{thm:static}
	Suppose $\mathbf{D}$ is a static design with $n$ points.  
	Then 
	\[\text{sup}_{f \in \mathbf{\Omega}} V\{\mathbf{U}(\mathbf{D},f)\} \geq (n+1)^{-1} 
	\]
	when $p=1$ and 
	\[
	\text{sup}_{f \in \mathbf{\Omega}} V\{\mathbf{U}(\mathbf{D},f)\} \geq 2^{-1/(p-1)}10^{-1/p}(p-1)!^{-1}n^{{-1/p}}\]
	when $p\geq 2$ and $n\geq 10^{p-1}p^p$. 
\end{theorem}

From Theorem~\ref{thm:static}, SG achieves the lowest rate on the number of functional evaluations. 
Similarly results hold for static inner grid design (SI), i.e., SG without boundary points, 
\[ {\mathbf{D}_{\text{SI},p,n} }=  \left\{1/\left(n^{1/p}+1\right),2/\left(n^{1/p}+1\right),\ldots,n^{1/p}/\left(n^{1/p}+1\right) \right\}^p, \]
and static Monte Carlo design (MC), for which the points are independently and uniformly generated from $[0,1]^p$. The corresponding theoretical properties are provided in the Appendix.

\subsection{Adaptive Designs}\label{sec:design:A}

Now that the $n^{-1/p}$ rate is not improvable from using static designs, attention shifts towards adaptive designs for a potentially better rate. 
The core idea is to organize the SG runs into groups and sequentially skip runs with known outcomes. 
We propose to partition the runs in $\mathbf{D}_{\text{SG},p,(2^g+1)^p}$ to groups $\mathbf{D}_{\text{SG},p,2^p}, \mathbf{D}_{\text{SG},p,3^p} \setminus \mathbf{D}_{\text{SG},p,2^p}, \mathbf{D}_{\text{SG},p,5^p} \setminus \mathbf{D}_{\text{SG},p,3^p}, \ldots, \mathbf{D}_{\text{SG},p,(2^g+1)^p} \setminus \mathbf{D}_{\text{SG},p,(2^{g-1}+1)^p}$ and carry out the experiments sequentially by groups. 
Let $\text{card}({\mathbf{S}})$ denotes the cardinality of the set ${\mathbf{S}}$. Algorithm~\ref{alg:GG} details this algorithm.

\begin{algorithm}
	\label{alg:GG}
	\DontPrintSemicolon
	
	\KwInput{Dimension $p$, simulation outcome $f(\cdot)$, and number of runs $n$}
	
	Initialize $\mathbf{D} \leftarrow \emptyset$, $\mathbf{C} \leftarrow \emptyset$, and $l \leftarrow 0$ \;
	\While{$\text{card}(\mathbf{D})<n$}
	{
		\If{$\mathbf{C} = \emptyset$}{ 
			Generate the candidate point set $\mathbf{C} \leftarrow \mathbf{D}_{\text{SG},p,(2^l+1)^p} \cap \mathbf{U}(\mathbf{D},f) $ and let $l \leftarrow l+1$
		}
		\Else{
			Randomly choose an $\mathbf{x} \in \mathbf{C}$\; 
			Evaluate $f(\mathbf{x})$, let $\mathbf{D}\leftarrow \mathbf{D}\cup\{\mathbf{x}\}$ and $\mathbf{C} \leftarrow \mathbf{C} \setminus \{\mathbf{x}\}$ \;
		}
	}
		Output the grouped-adaptive grid design $\mathbf{D}_{\text{GG},p,n} \leftarrow \mathbf{D}$\; 
		\caption{Steps of the grouped-adaptive grid design method}
	\end{algorithm}

We term the $\mathbf{D}_{\text{GG},p,n}$ as a grouped-adaptive grid design (GG). 
Let $m_{\text{GG}}(g)$ denote the $\text{card}(\mathbf{D})$ when the outcomes corresponding to the $\mathbf{D}_{\text{SG},p,(2^g+1)^p}$ are all known,  
i.e., when $\mathbf{C} = \emptyset$ and $l = g+1$ from Algorithm~\ref{alg:GG}. 
It's evident that from the $m_{\text{GG}}(g)$ functional evaluations of the GG, we gather precisely the same information as from the $(2^g+1)^p$ runs of the SG. 
To quantify the difference between $m_{\text{GG}}(g)$ and $(2^g+1)^p$, Theorem~\ref{thm:GG} below gives an upper bound of $m_{\text{GG}}(g)$. 

\begin{theorem}\label{thm:GG}
	\[	m_{\text{GG}}(g) \leq 2^p + \sum_{l=1}^g \{ p (2^l+1)^{p-1}\}.\]
\end{theorem}

Combining Theorems~\ref{thm:SG} and~\ref{thm:GG}, when $p=1$, $m_{\text{GG}}(g) \leq g+2$ and thus $\text{sup}_{f \in \mathbf{\Omega}} V\{\mathbf{U}( \allowbreak \mathbf{D}_{\text{GG},p,n},\allowbreak f)\}\allowbreak \leq 2^{-n+2}$ and it requires at most $2-\log_2 v$ functional evaluations to guarantee that $V\{\mathbf{U}\allowbreak(\mathbf{D}_{\text{GG},p,n},f)\}\allowbreak \leq v$. 
When $p\geq 2$ and $g$ is large, the $m_{\text{GG}}(g)$ is at most approximately $p 2^{p-1}(2^{p-1}-1)^{-1} 2^{g(p-1)} $ and thus $\text{sup}_{f \in \mathbf\Omega} V\{\mathbf{U}(\allowbreak\mathbf{D}_{\text{GG},p,n},f)\}$ is at most approximately $2p^{p/(p-1)}(2^{p-1}-1)^{-1/(p-1)} n^{-1/(p-1)}$ and it requires at most approximately $p^p 2^{p-1}(2^{p-1}-1)^{-1}v^{-(p-1)}$ functional evaluations to guarantee that $V\{\mathbf{U}(\mathbf{D}_{\text{GG},p,n},f)\} \leq v$. 
In both cases, the greatest possible number of functional evaluations for the GG is an order of magnitude fewer than that of the SG, demonstrating the significant superiority of the GG over the SG.

From Algorithm~\ref{alg:GG}, the $\mathbf{U}(\mathbf{D},f)$ is updated whenever a group of computer runs are completed. 
However, by updating $\mathbf{U}(\mathbf{D},f)$ after each single functional evaluation in Step~7, we may potentially skip more runs.
Let $\mathbf{A}_{\mathbf{x}}=\{\mathbf{y}\in \mathbf{C}: y_k\leq x_k, k=1,2,\ldots,p\}$ and $\mathbf{B}_{\mathbf{x}}=\{\mathbf{y}\in \mathbf{C}: y_k\geq x_k, k=1,2,\ldots,p\}$. 
Since evaluating $f(\mathbf{x})$ enables us to skip at least $\min\{\text{card}(\allowbreak \mathbf{A}_{\mathbf{x}}),\text{card}(\mathbf{B}_{\mathbf{x}})\}$ runs among those in $\mathbf{C}$, instead of selecting $\mathbf{x}$ randomly from $\mathbf{C}$, it is reasonable to prioritize evaluating the $f(\mathbf{x})$ for the $\mathbf{x} \in \mathbf{C}$ that maximizes $\min\{\text{card}(\mathbf{A}_{\mathbf{x}}),\text{card}(\mathbf{B}_{\mathbf{x}})\}$ before evaluating other runs. 
Exploiting these ideas, we propose the fully adaptive grid design (AG) method, whose steps are given in Algorithm~\ref{alg:AG}. 

\begin{algorithm}
	\label{alg:AG}
	\DontPrintSemicolon
	\KwInput{Dimension $p$, simulation outcome $f(\cdot)$, and number of runs $n$}
	Initialize $\mathbf{D} \leftarrow \emptyset$, $\mathbf{C} \leftarrow \emptyset$, and $l\leftarrow 0$ \;
	\While{$\text{card}(\mathbf{D})<n$}
	{
		\If{$\mathbf{C} \leftarrow \emptyset$}{ 
			Generate the candidate point set $\mathbf{C} = \mathbf{D}_{\text{SG},p,(2^l+1)^p} \cap \mathbf{U}(\mathbf{D},f) $ and let $l \leftarrow l+1$
		}
		\Else{
			Choose the $\mathbf{x} \in \mathbf{C}$ that maximizes $\min\{\text{card}(\mathbf{A}_{\mathbf{x}}),\text{card}(\mathbf{B}_{\mathbf{x}})\}$. If there are multiple choices of $\mathbf{x}$, select the one that also maximizes 
			$\max\{\text{card}(\mathbf{A}_{\mathbf{x}}),\text{card}(\mathbf{B}_{\mathbf{x}})\}$\; 
			Evaluate $f(\mathbf{x})$, let $\mathbf{D}\leftarrow \mathbf{D}\cup\{\mathbf{x}\}$, update $\mathbf{U}(\mathbf{D},f)$, and let $\mathbf{C} \leftarrow \mathbf{C} \cap \mathbf{U}(\mathbf{D},f)$ \;
		}
	}
	Output the full adaptive grid design $\mathbf{D}_{\text{AG},p,n} \leftarrow \mathbf{D}$\; 
	\caption{Steps of the fully adaptive grid design method}
\end{algorithm}

The AG designs are also illustrated in Figs.~\ref{fig:illu}(b-c). 
From numerical results that will be provided in Sections~\ref{sec:comp} and~\ref{sec:ice}, the AG remarkably outperforms the GG, although we are not able to establish rigorous theoretical results to verify it. 
By substituting the $\mathbf{D}_{\text{SI},p,(2^{l}-1)^p}$ for the $\mathbf{D}_{\text{SG},p,(2^l+1)^p}$ and initializing $l=1$ instead of $l=0$ in Algorithms~\ref{alg:GG}~and~\ref{alg:AG}, we obtain the grouped-adaptive inner grid design (GI) and the fully adaptive inner grid design (AI), which exhibit similar properties to GG and AG, respectively.

It is widely acknowledged that an effective design for computer simulations should have space-filling projections.
For instance,  Latin hypercube design~\citep{Mckay1979} (LHD) has the optimal univariate projections because each univariate projection of an $n$-point LHD has exactly one point in each of the $n$ equally spaced intervals $[0,1/n), \ldots, [1-1/n,1)$. 
For the SG, however, each univariate projection consists of $n^{(p-1)/p}$ many $z/(n^{1/p}-1)$ for each $z \in \{0,\ldots,n^{1/p}-1\}$. 
Having a large amount of coincident values, the SG is surely not space-filling on projections. 
However, we still consider the SG to be the best basis for adaptive design when the objective is to classify monotonic binary deterministic computer simulations and $p$ is low. 
This preference stems from the fact that using the SG as the basis design allows us to skip more runs.
Recall that a necessary condition for the run $\mathbf{x}$ to be skipped is the existence of a point $\mathbf{y}\in \mathbf{D}$ such that either $y_k\leq x_k$ for all $k$ or $x_k\leq y_k$ for all $k$.
Therefore, the more pairs of points $(\mathbf{x},\mathbf{y})$ in the design that holds the above ordered relationship, the more runs we are likely to skip. 
Among space-filling designs that we are aware of, grid points produce the most such pairs. 
For instance, given a 2-dimensional 81-run design, among the totally 3240 pairs of points, both the SG and the SI contain 1944 such ordered pairs,  whereas, on average, the MC has 1620.06 ordered pairs and the LHD has 1619.24 ordered pairs.
However, although similar conclusion holds for $p=3,4$, it becomes much more difficult to utilize the monotonicity for higher $p$. 
For instance, given a 5-dimensional 243-run design, among the totally 29403 pair of points, the SG contains only 7533 ordered pairs. Although this remains larger than the averaged 1837.43 ordered pairs for the MC, the proportion of ordered pairs drops remarkably.
We thus infer that the grid structure is less appealing for $p\geq 5$.

Finally, Theorem~\ref{thm:adaptive} below gives a lower bound on the required number of runs for arbitrary adaptive designs to ensure that $V(\mathbf{U})\leq v$. 
Remark that for an adaptive design, the $\mathbf{D}$ is subject to the $f$.

\begin{theorem}\label{thm:adaptive}
	Suppose the $\mathbf{D}$ is an adaptive design with $n$ points. Then the maximal possible volume of uncertain area, $\text{sup}_{f \in \Omega} V\{U(D,f)\}$, satisfies  
	\[	\text{sup}_{f \in \mathbf{\Omega}} V\{\mathbf{U}(\mathbf{D},f)\} \geq 2^{-n} \]
	when $p=1$ and 
	\[\text{sup}_{f \in \mathbf{\Omega}} V\{\mathbf{U}(\mathbf{D},f)\} \geq p^{1/(p-1)}2^{-(p+1)/(p-1)}(p-1)!^{-1}n^{-1/(p-1)}\]
	when $p\geq 2$ and $n\geq 4^{p-2}p^p$. 
\end{theorem}

We believe the bound in Theorem~\ref{thm:adaptive} is not tight for $p\geq 2$. 
As indicated by Theorem~\ref{thm:adaptive}, the AG exhibits the lowest rate on $\text{sup}_{f \in \mathbf{\Omega}} V\{\mathbf{U}(\mathbf{D},f)\}$, which is $2^{-n}$ when $p=1$ and $n^{-1/(p-1)}$ when $p\geq 2$. 
Accordingly, the AG has the lowest rate on the required number of functional evaluations to ensure that $V\{\mathbf{U}(\mathbf{D},f)\}\leq v$, which is $\ln(1/v)$ when $p=1$ and $v^{-(p-1)}$ when $p \geq 2$. 
Consequently, no adaptive design can significantly outperform the AG.

\section{Numerical Comparison}\label{sec:comp}
In this section, we numerically evaluate the performance of several types of static and adaptive designs for classifying monotonic binary deterministic computer simulations. 
Besides the MC, SG, GG, AG, and AI methods discussed in Section~\ref{sec:design}, we further include adaptive Monte Carlo design in which points are independently generated one-by-one from the uniform distribution on $[0,1]^p$ but only the runs within the uncertain area are conducted~\citep{DEROCQUIGNY2009363} (AMC), the active learning design that sequentially supplement an initial design using the point $\mathbf{x}^\star$ that maximizes the entropy~\citep{Lewis1994} (ALE), maximin-distance Latin hypercube designs~\citep{Shan2015} (OLH), minimum energy designs~\citep{Roshan2015} (MED), partitioned active learning Cohn designs with ordinary Gaussian process models  \citep{Lee2023} (PALC), and contrastive active learning  \citep{Margatina2021} (CAL) in the comparison. For MED, PALC, ALE, and CAL, the initial designs are OLH with $10p$ points.
We have also tried the margin sampling method~\citep{margin2002} but omit its results because they are very similar to that of the ALE. 
We consider the test function $f(\mathbf{x})$ defined as follows
\begin{align}\label{eqn:countour}
	f(\mathbf{x})= \begin{cases}
		1, & \sum_{i=1}^p \arctan\left\{5(p+1-i)x_i/(p+1)\right\} \ge \mu,\\
		-1, & \text{otherwise}.\end{cases}
\end{align} 
From choosing $\mu$ in [0.92,2.10], [1.53,2.95], [2.14,3.75], [2.76,4.59], and [3.43,5.40] when $p=2,3,4,5,6$, respectively, the $V(\mathbf{A})$ is between 10\% to 90\% for any $p$. 
We use the support vector classification (SVC) method~\citep{Corinna1995} as the classifier hereafter.
Details about the model and the fitting procedure are provided in the Appendix.
We use $V(\mathbf{U})$ and classification accuracy, i.e., the proportion of correct predictions for $10^5$ independent testing samples, to measure the performance of the designs. 
For each given $p$ and $n$, we independently sample 100 $\mu$'s from the uniform distribution within the specified range. Subsequently, we construct different types of designs and make predictions for each $\mu$.

\begin{figure}
	\centering
	\subfigure{
		\begin{minipage}{0.4\linewidth}
			\centering
			\includegraphics[width=\linewidth]{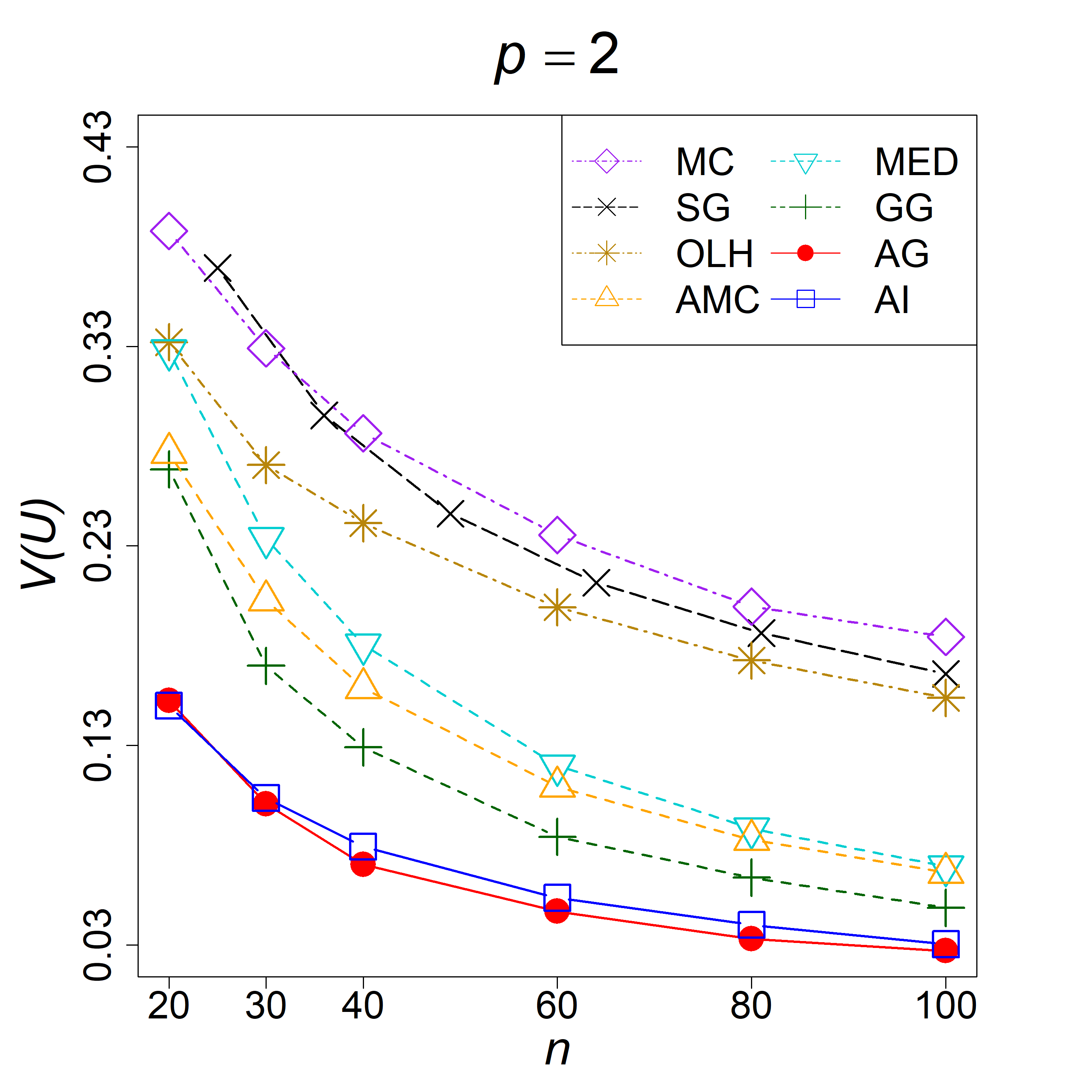}
		\end{minipage}
	}
	\subfigure{
		\begin{minipage}{0.4\linewidth}
			\centering
			\includegraphics[width=\linewidth]{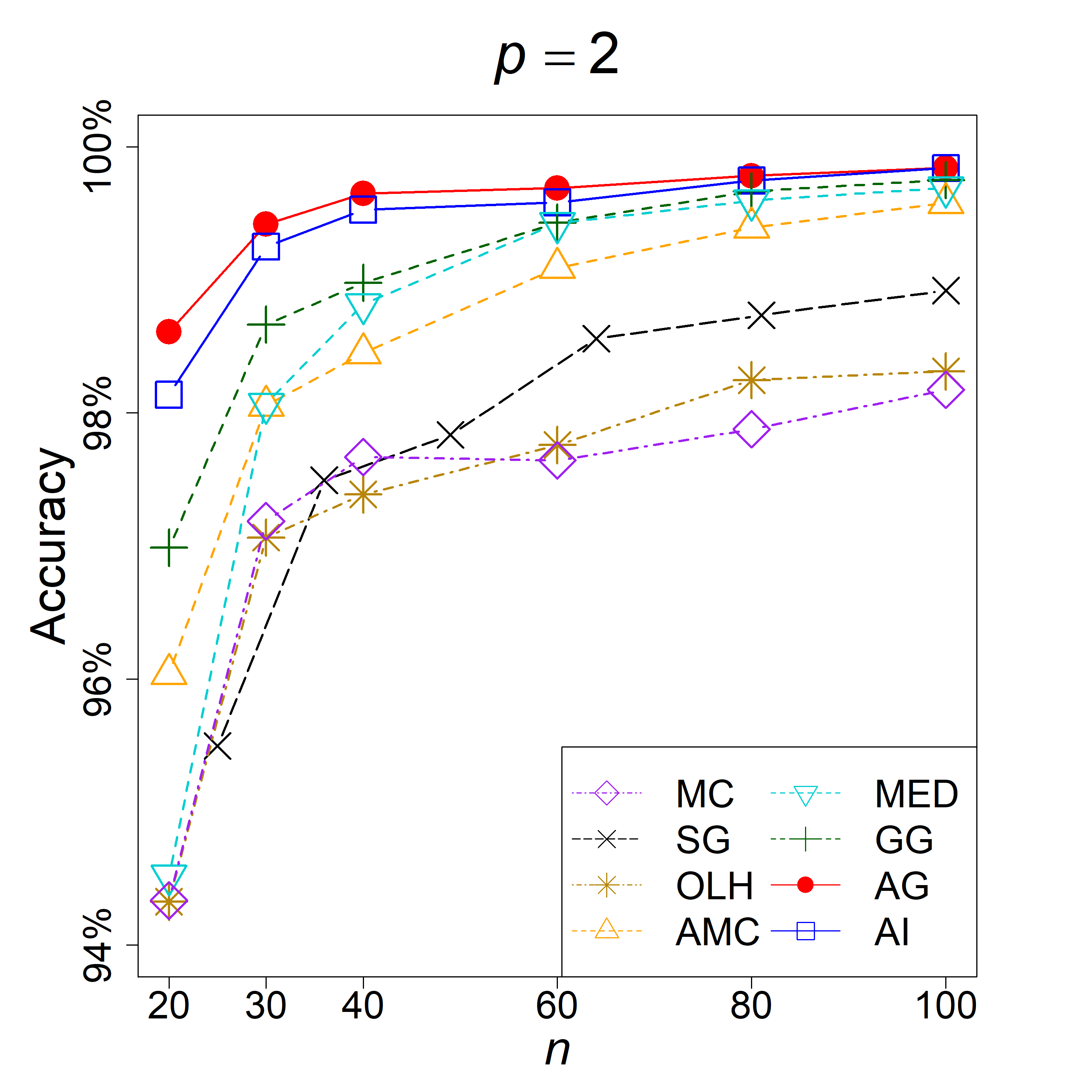}
		\end{minipage}
	}
	\subfigure{
		\begin{minipage}{0.4\linewidth}
			\centering
			\includegraphics[width=\linewidth]{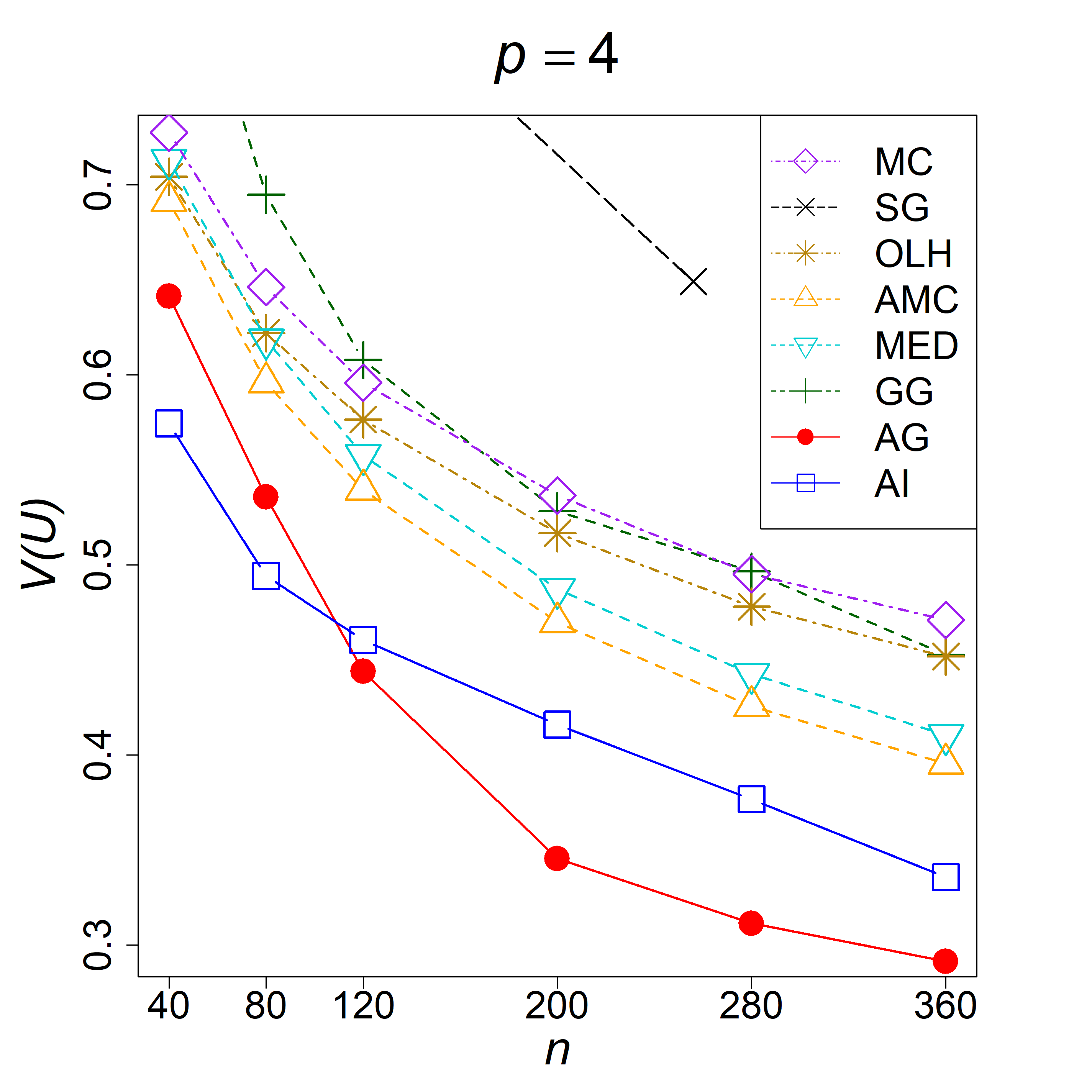}
		\end{minipage}
	}
	\subfigure{
		\begin{minipage}{0.4\linewidth}
			\centering
			\includegraphics[width=\linewidth]{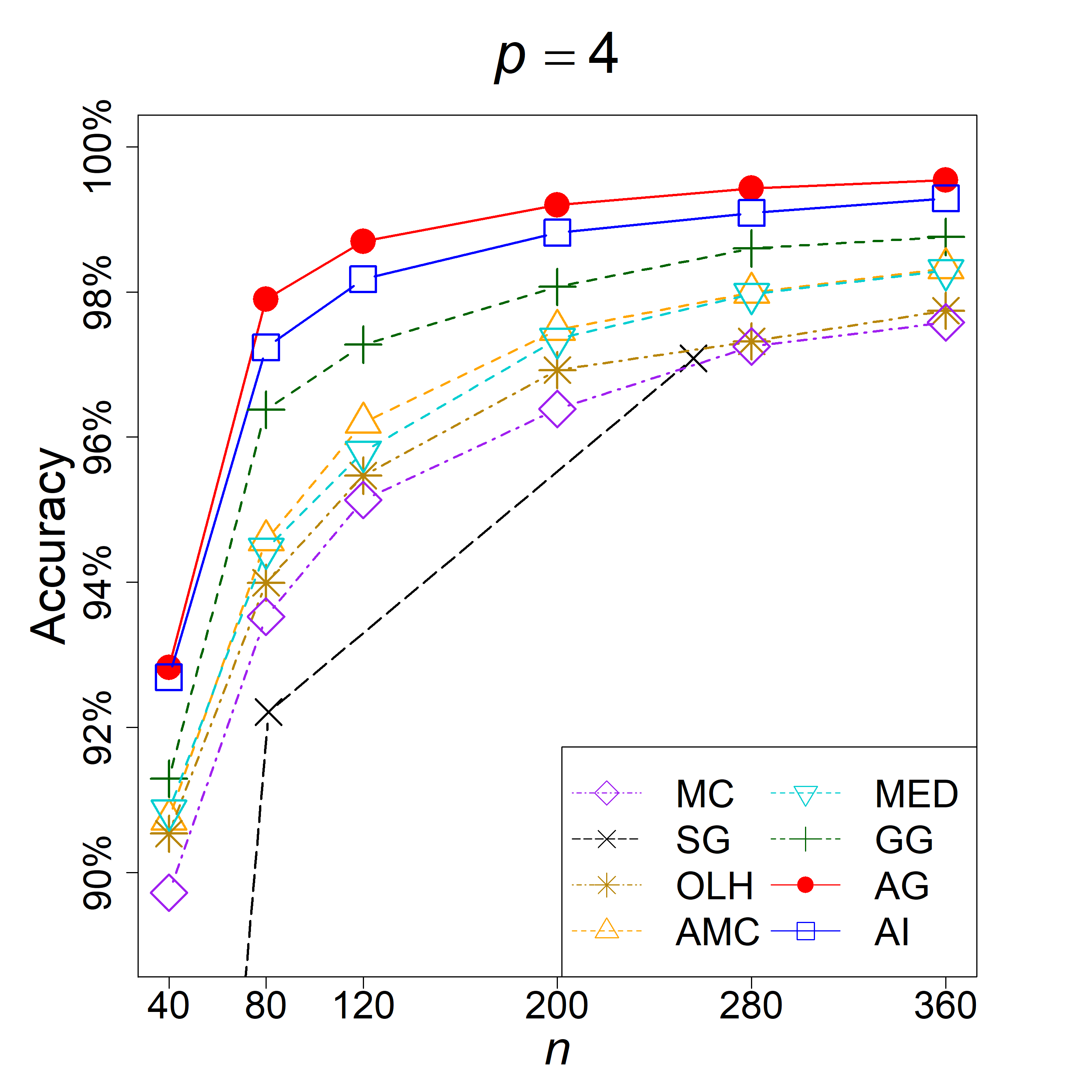}
		\end{minipage}
	}
	\subfigure{
		\begin{minipage}{0.4\linewidth}
			\centering
			\includegraphics[width=\linewidth]{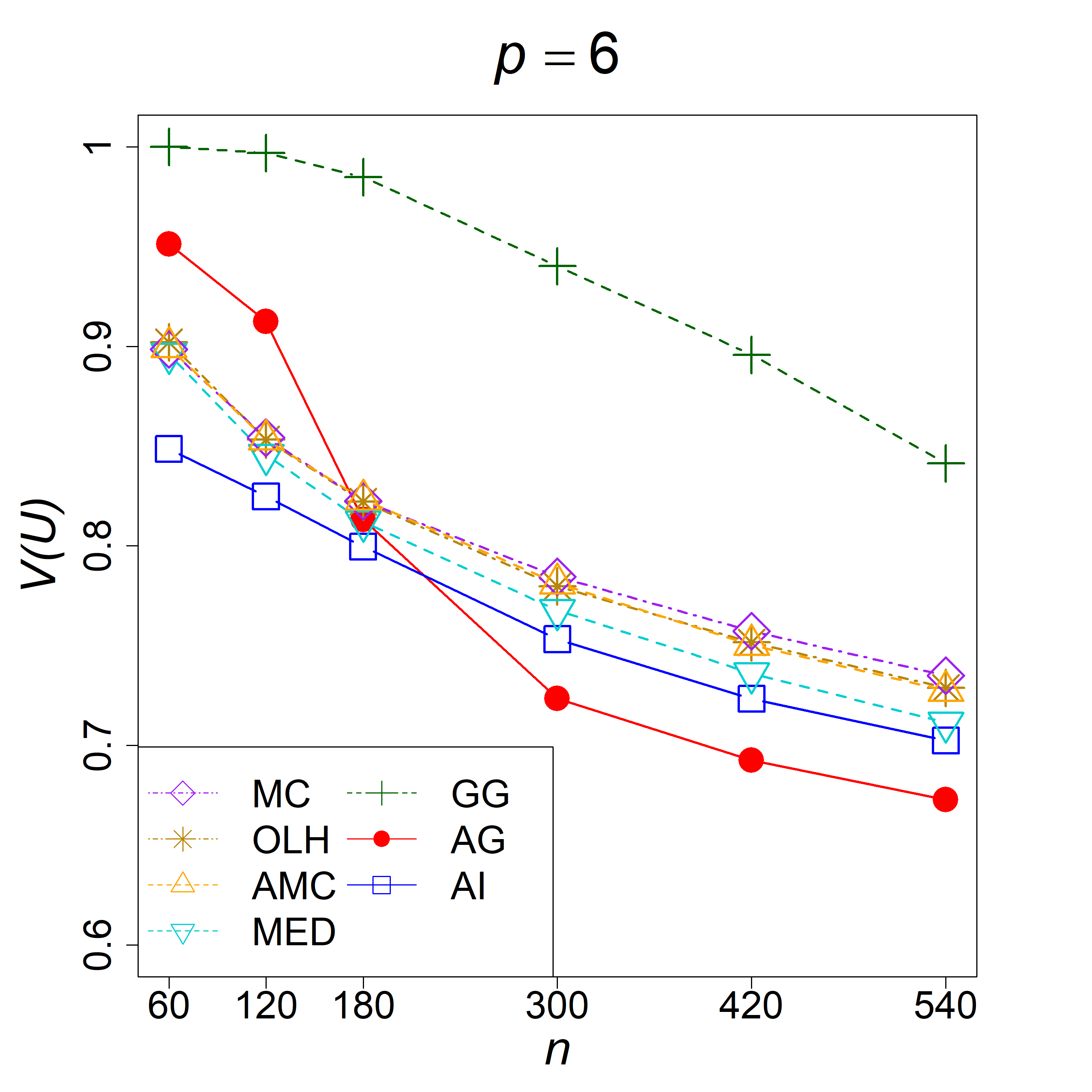}
		\end{minipage}
	}
	\subfigure{
		\begin{minipage}{0.4\linewidth}
			\centering
			\includegraphics[width=\linewidth]{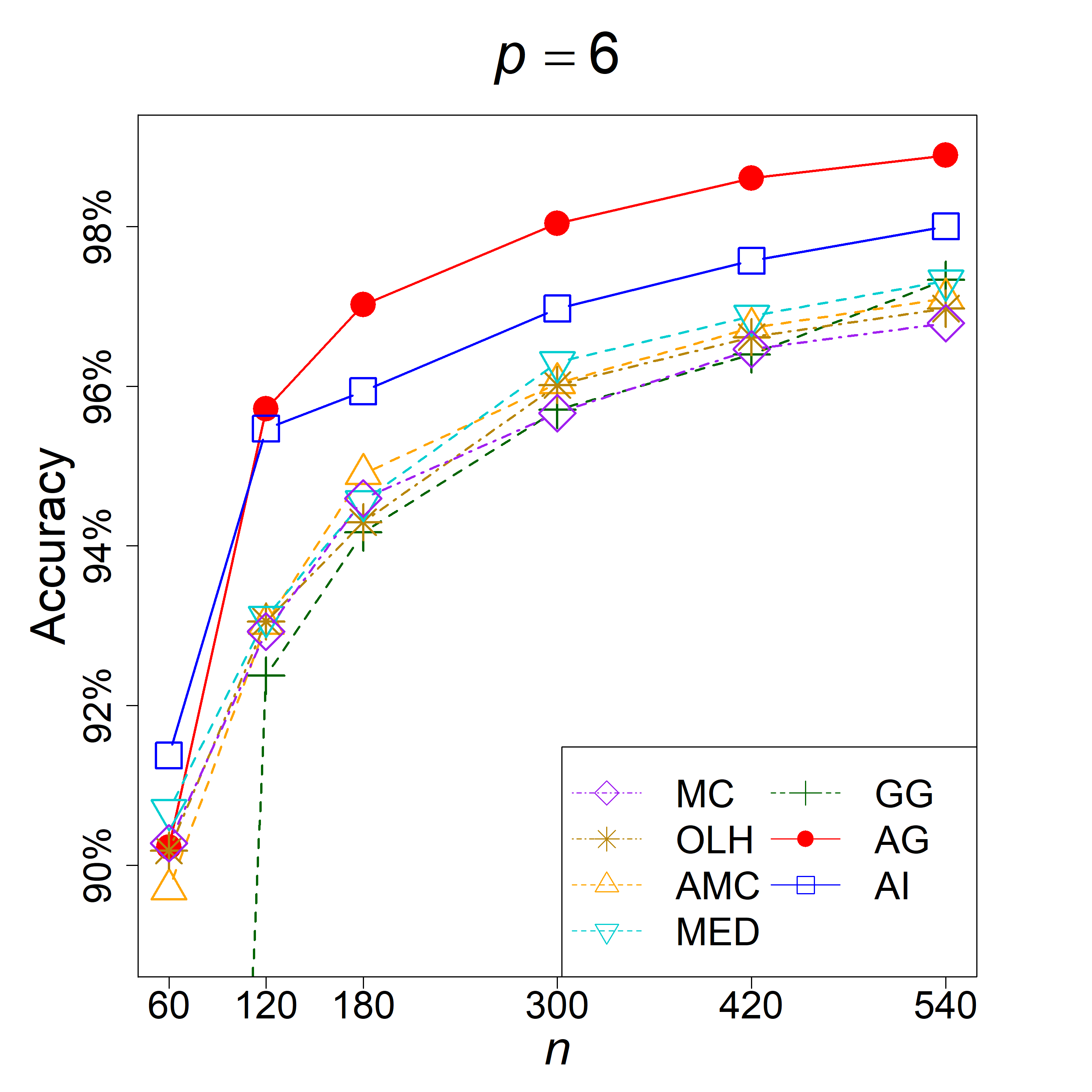}
		\end{minipage}
	}
	\caption{Averaged volume of uncertain area $V(\mathbf{U})$ (left) and classification accuracy (right) for the test function in \eqref{eqn:countour} with $p=2$ (top), $p=4$ (middle), and $p=6$ (bottom).}
	\label{fig:gridresult}
\end{figure}

Firstly, we compare methods MC, AMC, OLH, MED, SG, GG, AG, and AI. 
The averaged $V(\mathbf{U})$ and accuracies for $p=2,4,6$ with various $n$ are displayed in Fig.~\ref{fig:gridresult}. 
As can be observed, the AG and AI are the best methods. 
In all cases, the AG outperforms the GG significantly, while the GG exhibits notably better performance than the SG. 
From results not shown here, the AI outperforms the GI, while the GI outperforms the SI.
Similarly, the AMC demonstrates remarkable superiority over the MC. 
To summarize, adaptive designs show considerable advantages over their static counterparts, whereas the GG and GI can be regarded as halfly adaptive. 
Besides, the GG is better than the AMC when $p\leq 3$ and the conclusion is reversed when $p\geq 4$.
This aligns closely with our theoretical results provided in Section~\ref{sec:design} and Appendix. 

\begin{figure}
	\centering
	\subfigure{
		\begin{minipage}{0.4\linewidth}
			\centering
			\includegraphics[width=\linewidth]{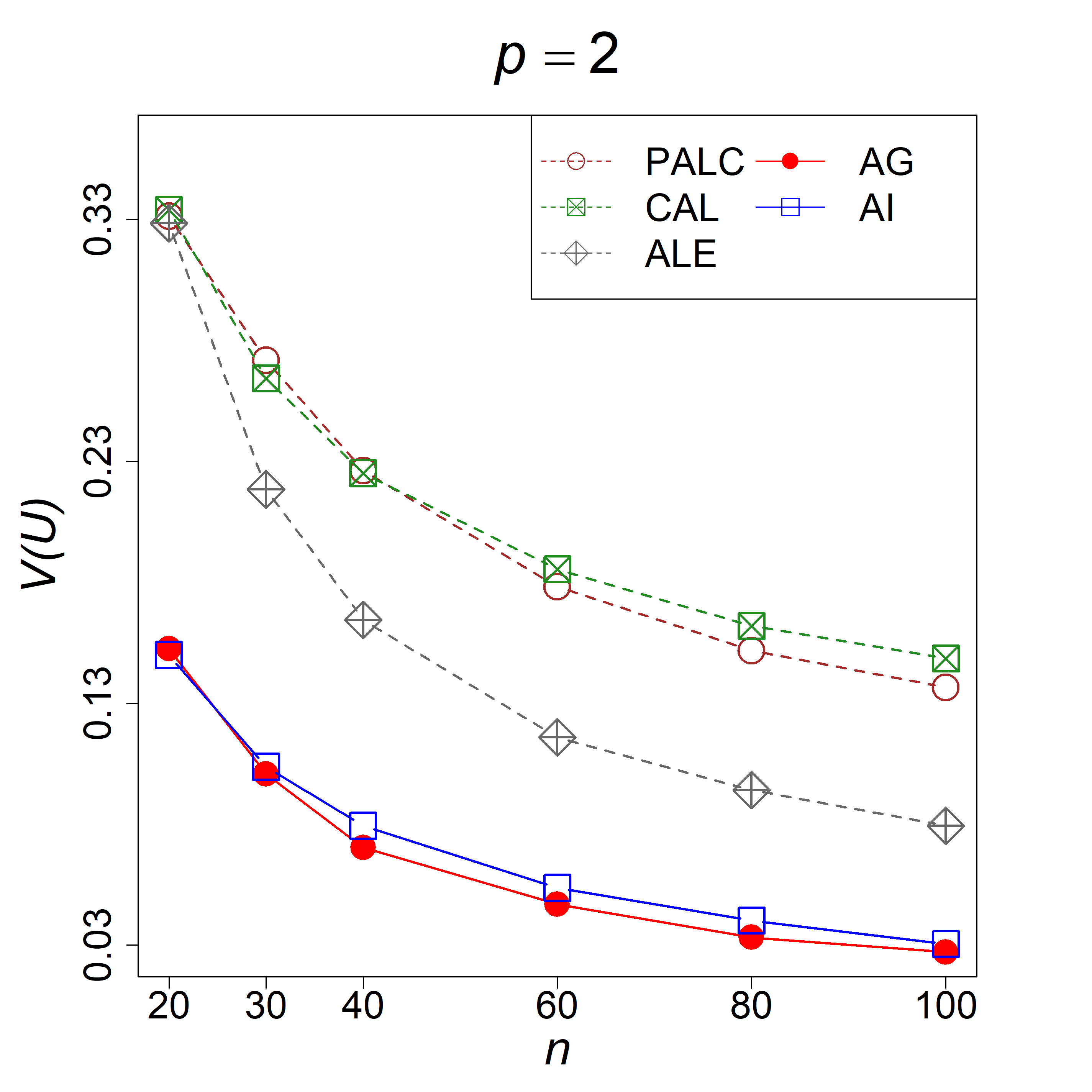}
		\end{minipage}
	}
	\subfigure{
		\begin{minipage}{0.4\linewidth}
			\centering
			\includegraphics[width=\linewidth]{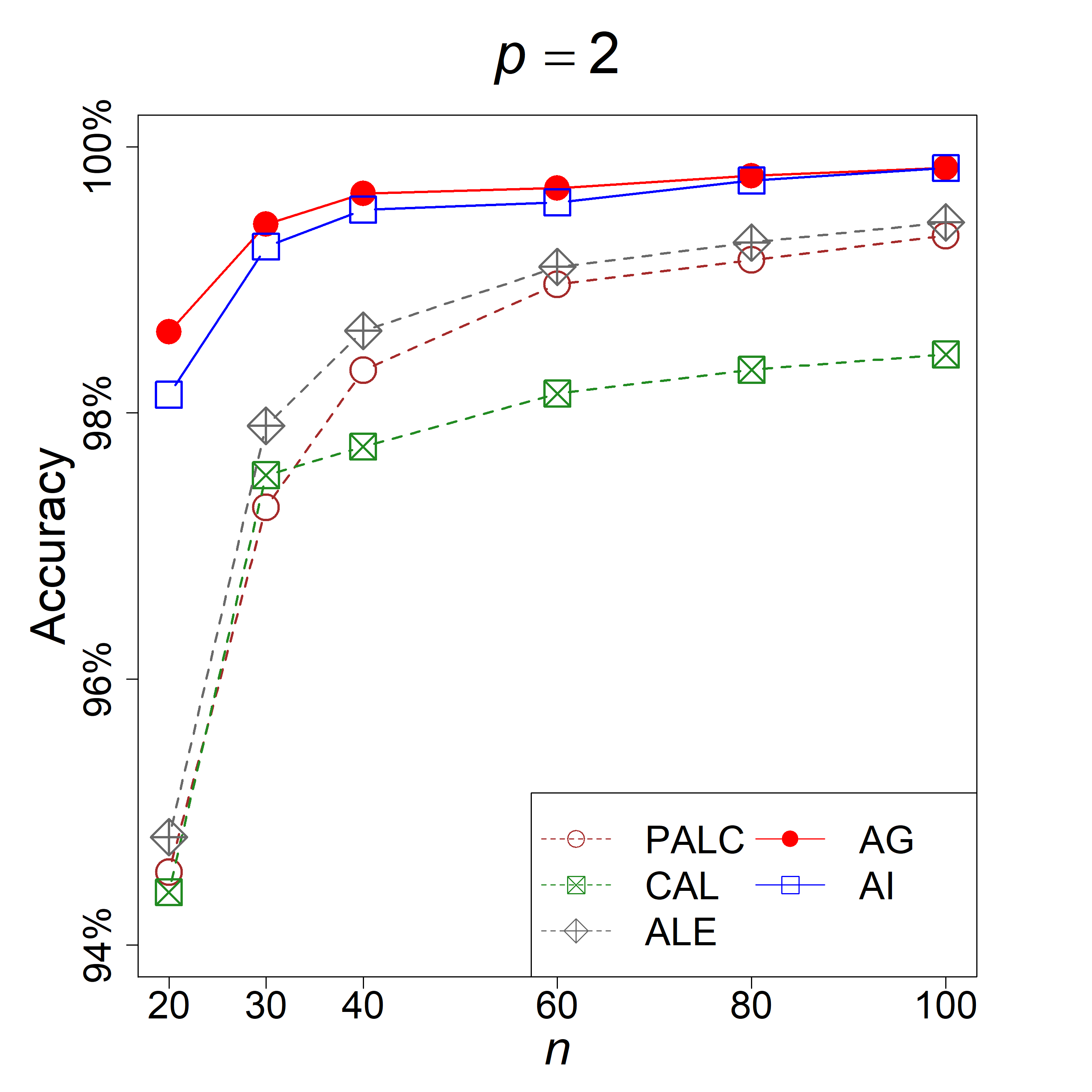}
		\end{minipage}
	}
	
	\subfigure{
		\begin{minipage}{0.4\linewidth}
			\centering
			\includegraphics[width=\linewidth]{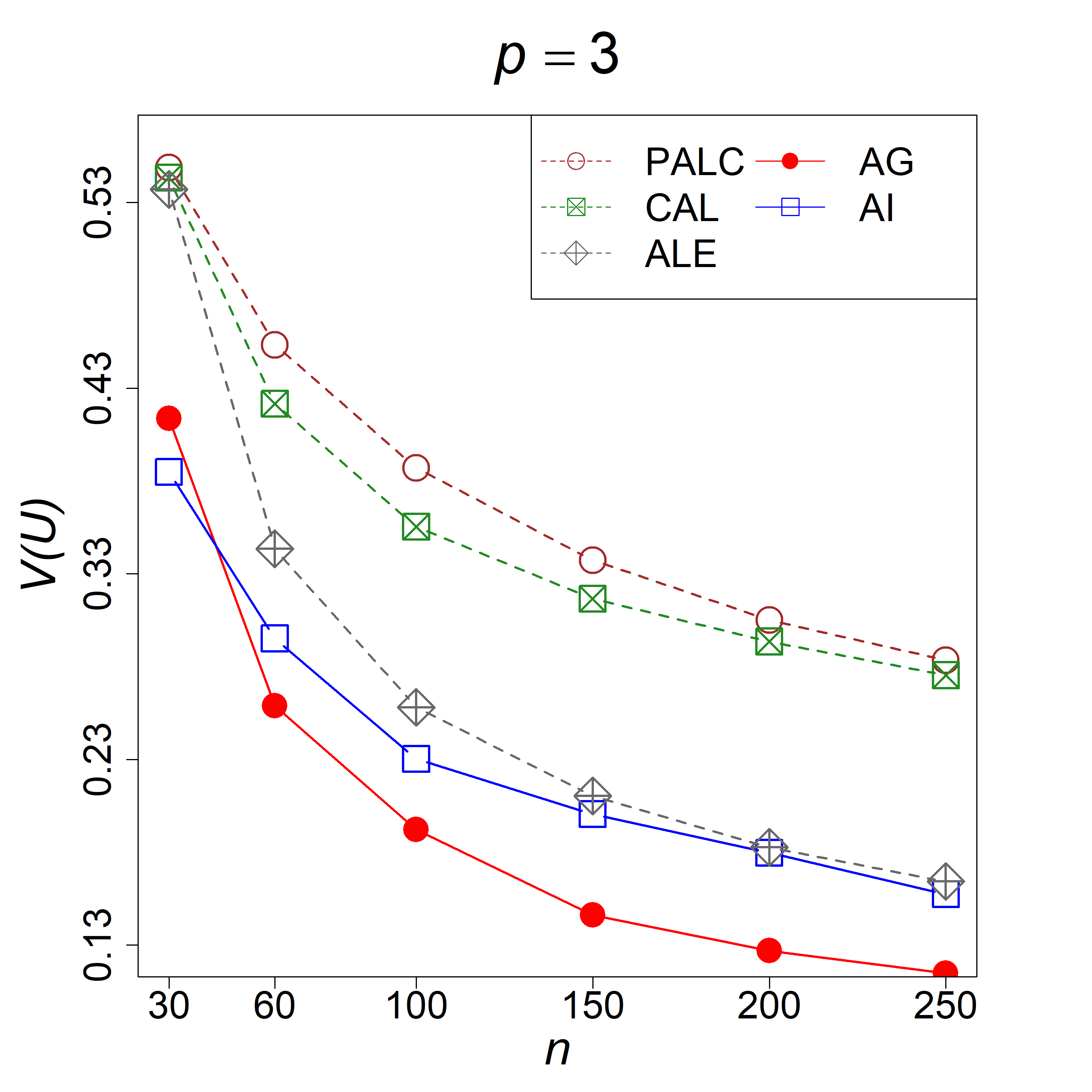}
		\end{minipage}
	}
	\subfigure{
		\begin{minipage}{0.4\linewidth}
			\centering
			\includegraphics[width=\linewidth]{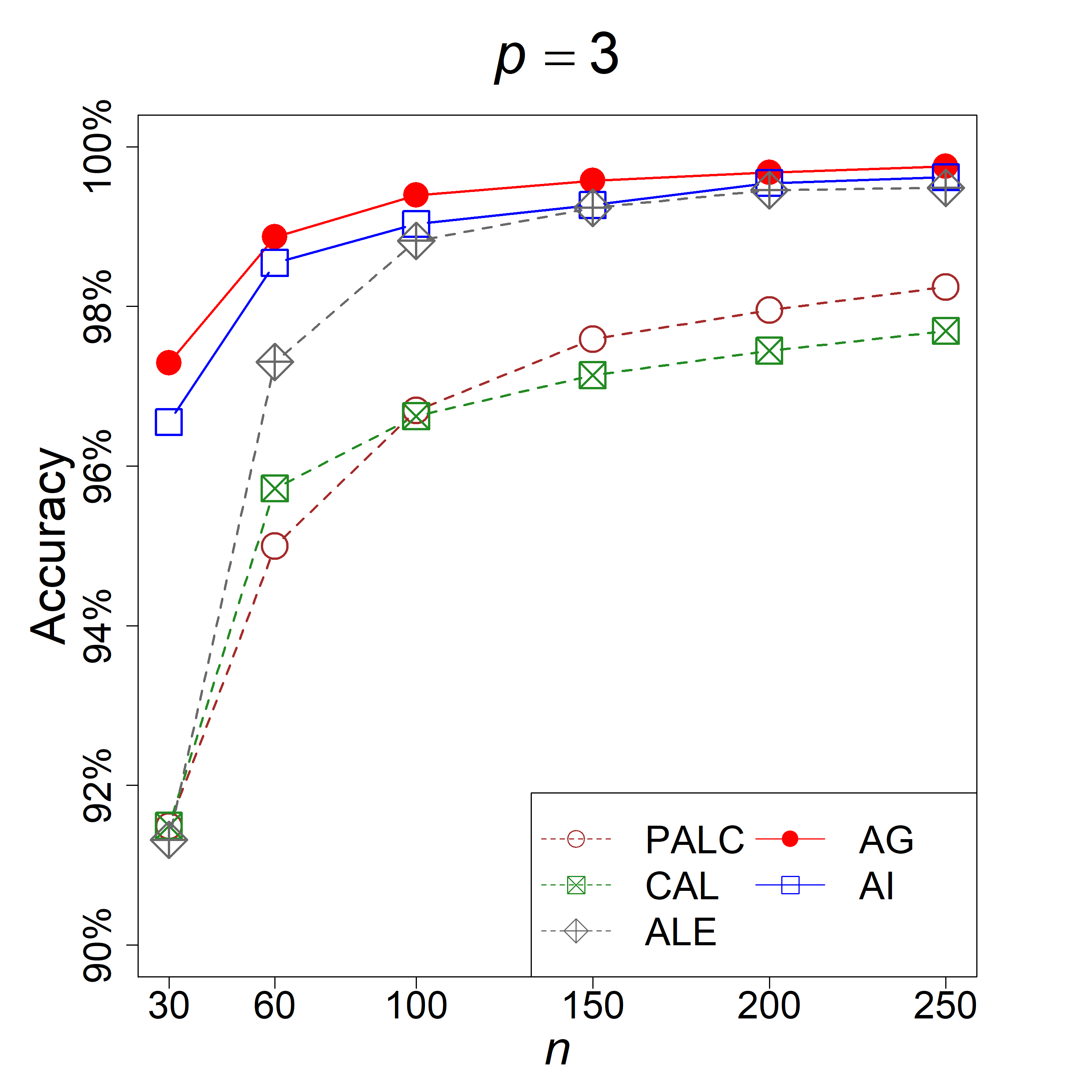}
		\end{minipage}
	}
	
	\subfigure{
		\begin{minipage}{0.4\linewidth}
			\centering
			\includegraphics[width=\linewidth]{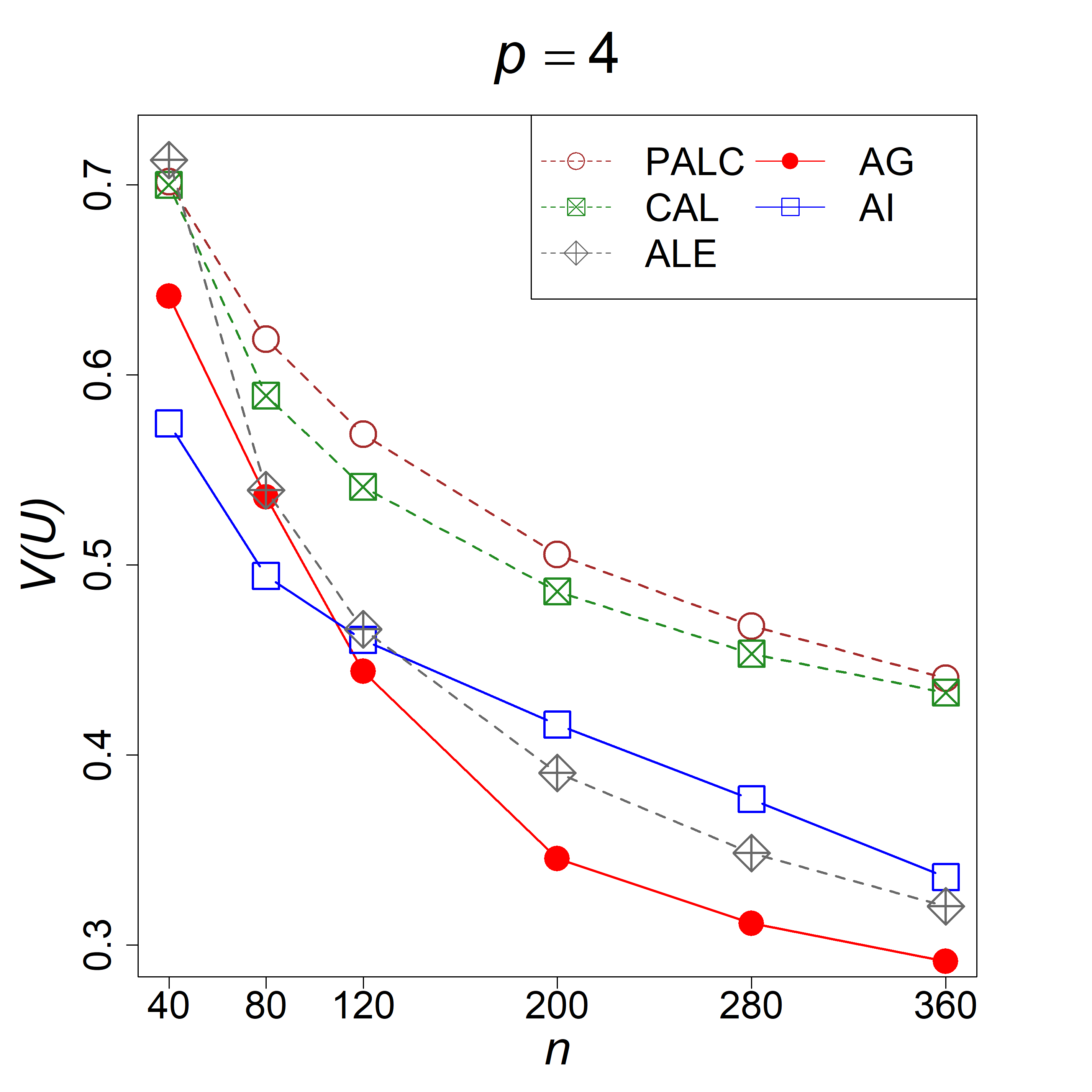}
		\end{minipage}
	}
	\subfigure{
		\begin{minipage}{0.4\linewidth}
			\centering
			\includegraphics[width=\linewidth]{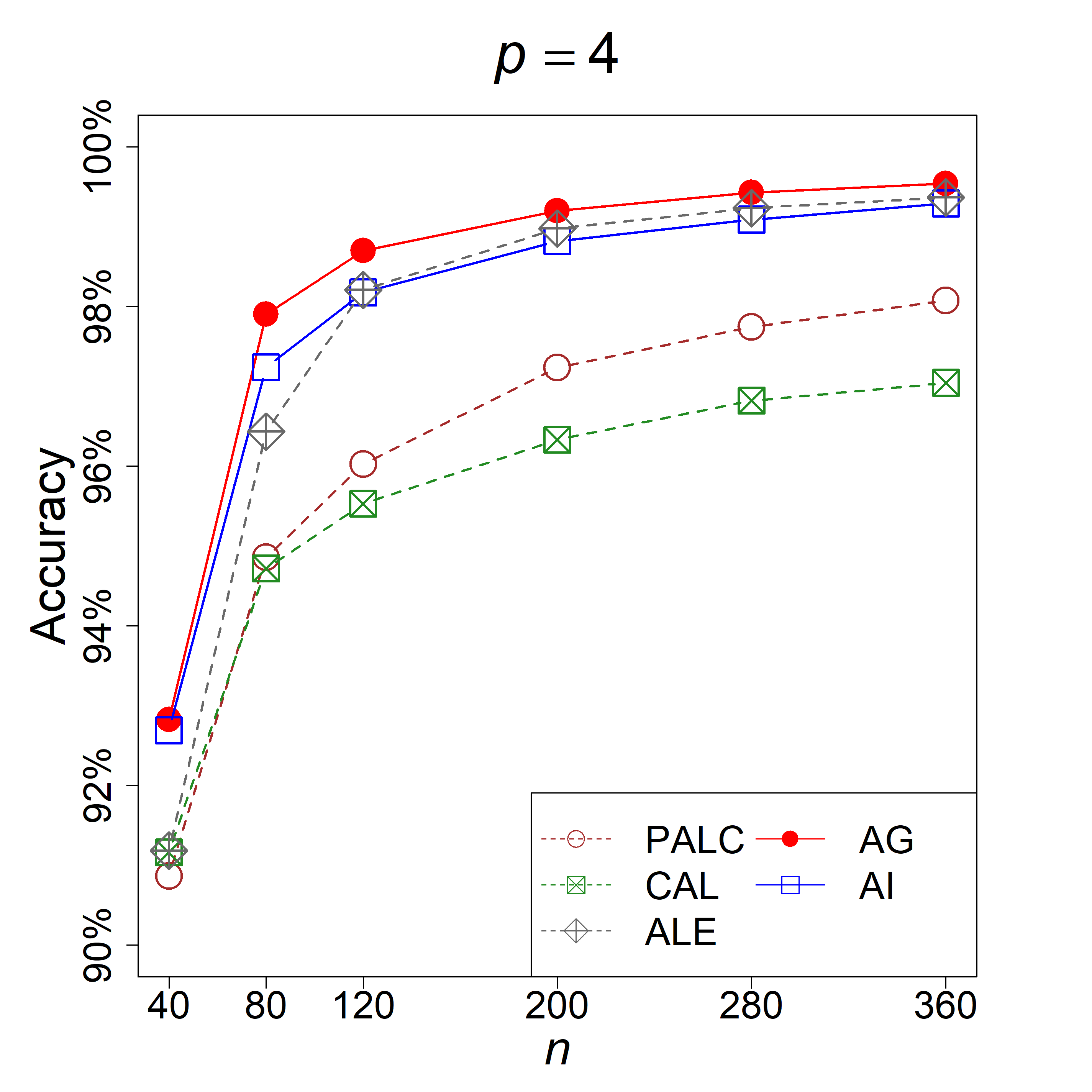}
		\end{minipage}
	}
	\caption{Averaged volume of uncertain area $V(\mathbf{U})$ (left) and classification accuracy (right) for the test function in \eqref{eqn:countour} with $p=2, 3, 4$.}
	\label{fig:p=234}
\end{figure}

\begin{figure}
	\centering
	\subfigure{
		\begin{minipage}{0.4\linewidth}
			\centering
			\includegraphics[width=\linewidth]{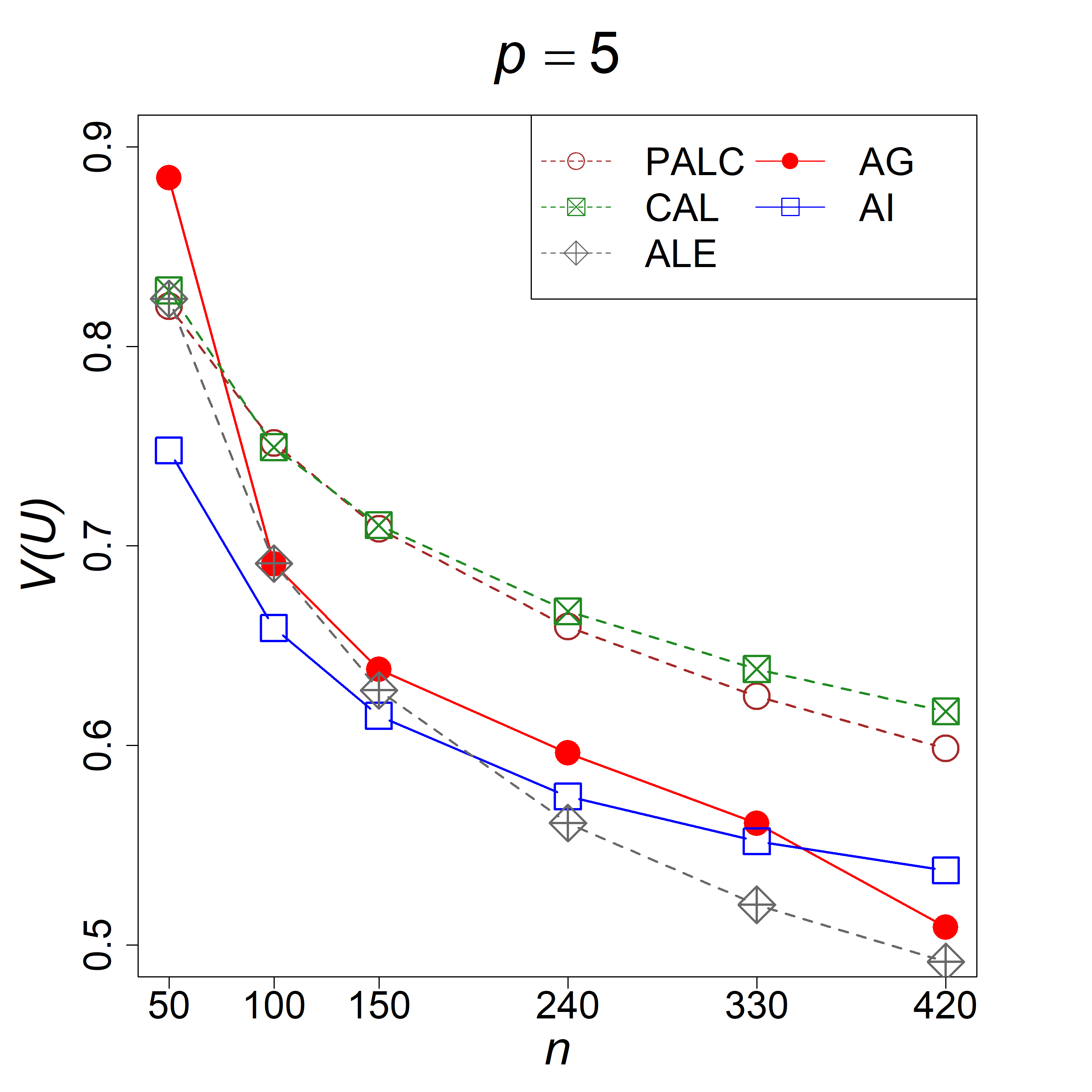}
		\end{minipage}
	}
	\subfigure{
		\begin{minipage}{0.4\linewidth}
			\centering
			\includegraphics[width=\linewidth]{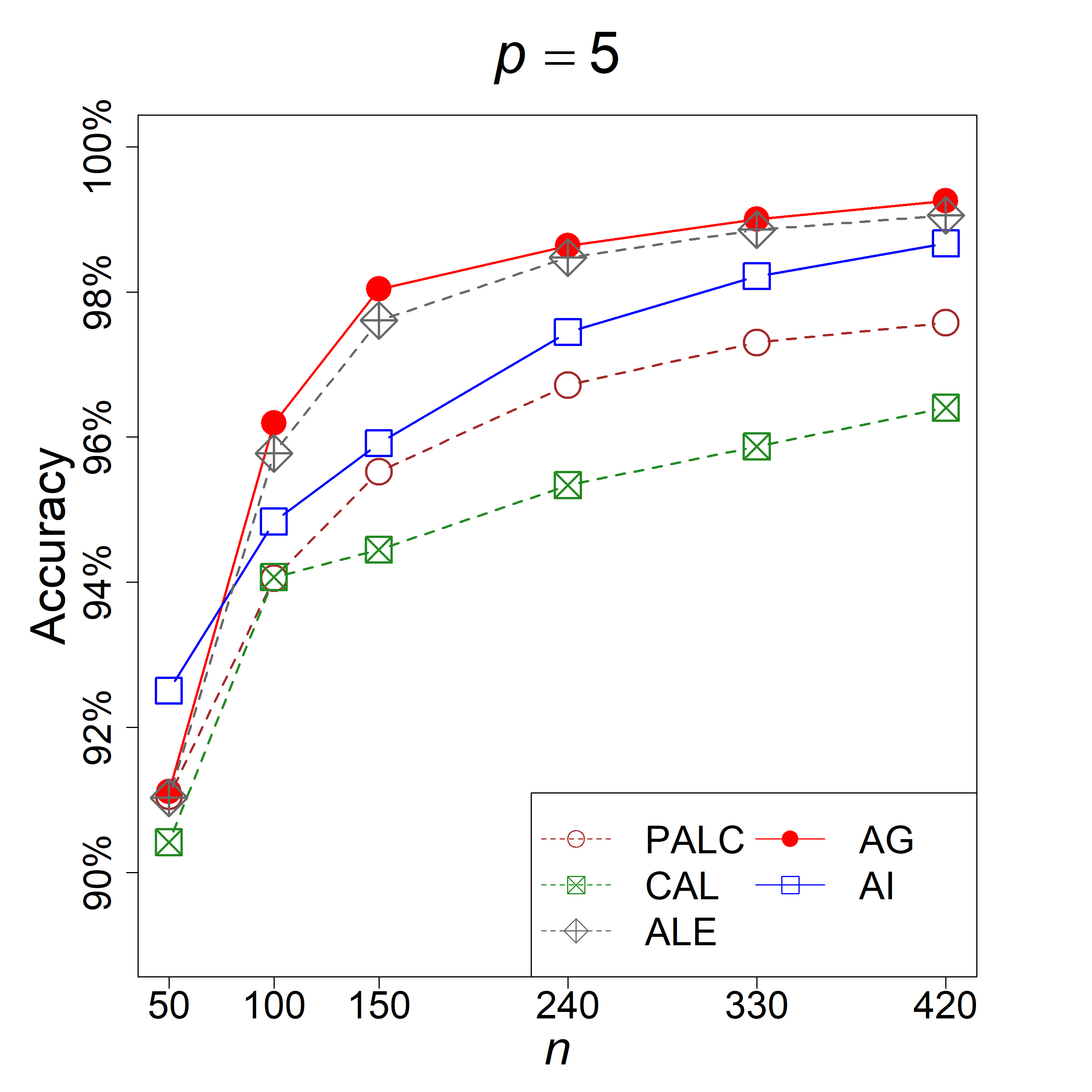}
		\end{minipage}
	}
	
	\subfigure{
		\begin{minipage}{0.4\linewidth}
			\centering
			\includegraphics[width=\linewidth]{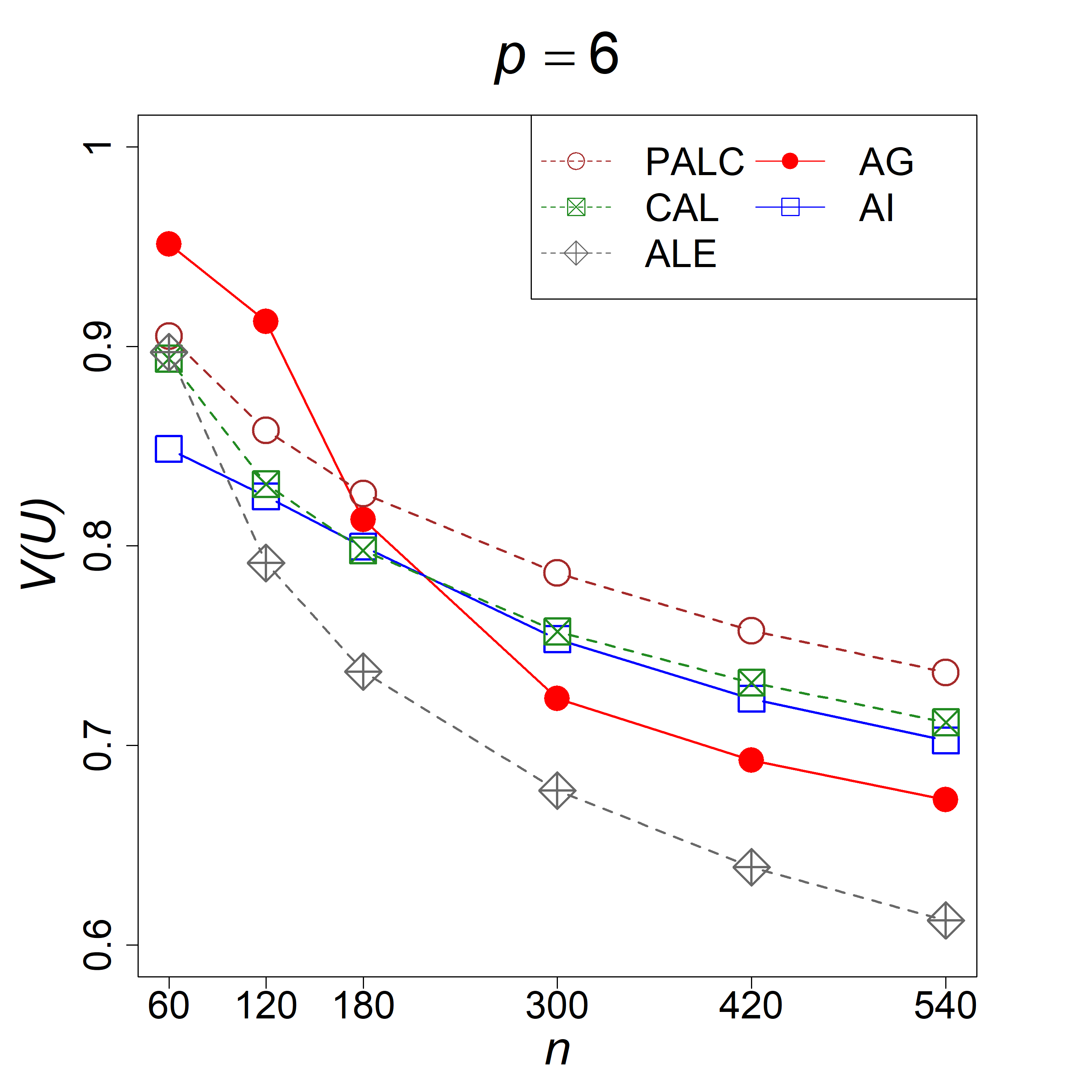}
		\end{minipage}
	}
	\subfigure{
		\begin{minipage}{0.4\linewidth}
			\centering
			\includegraphics[width=\linewidth]{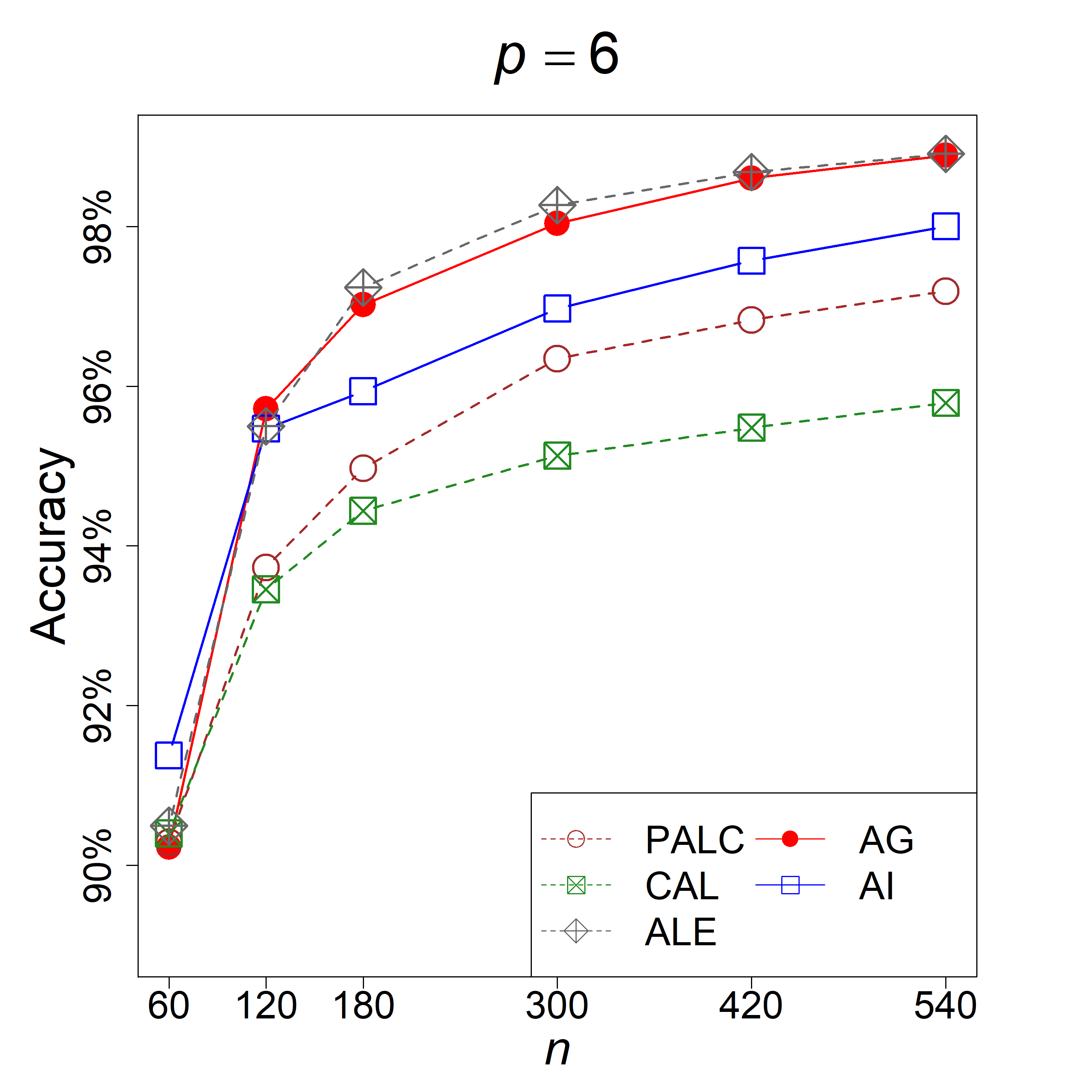}
		\end{minipage}
	}
	\caption{Averaged volume of uncertain area $V(\mathbf{U})$ (left) and classification accuracy (right) for the test function in \eqref{eqn:countour} with $p=5, 6$.}
	\label{fig:p=56}
\end{figure}

In the second batch of comparison, we include the AG and AI as well as active learning methods ALE, PALC, and CAL. 
The results for $2\leq p\leq 6$ are shown in Figs.~\ref{fig:p=234}--\ref{fig:p=56}.
We have several notable observations from the results. 
Firstly, the AG and AI outperform the ALE, PALC, and CAL significantly for $2\leq p\leq 4$. 
We conjecture that this is because the grid structure facilitates the skipping of computer runs for low $p$, as discussed in Section~\ref{sec:design}, while the active learning methods ALE, PALC, and CAL do not leverage monotonic information.
We thus recommend the use of the AG or AI for classifying monotonic binary deterministic computer simulations for low-dimensional problems. 
When $n$ is small, 
we recommend the AI because it surpasses the AG in performance. Otherwise we recommend the AG. 
However, compared to the ALE, the performance of the AG and AI deteriorates as $p$ grows. 
In particular, they are inferior to the ALE for $p=6$. 
This is presumbly because the grid structure is less compelling for higher $p$ cases, as discussed in Section~\ref{sec:design}.
The PALC and the CAL perform the worst among adaptive designs, likely because the PALC relies on ordinary Gaussian process models, which are not suited for binary outcomes, and the CAL fails to choose points near the SVC classification boundary in our problems.

\begin{figure}
	\centering
	\subfigure{
		\begin{minipage}{0.4\linewidth}
			\centering
			\includegraphics[width=\linewidth]{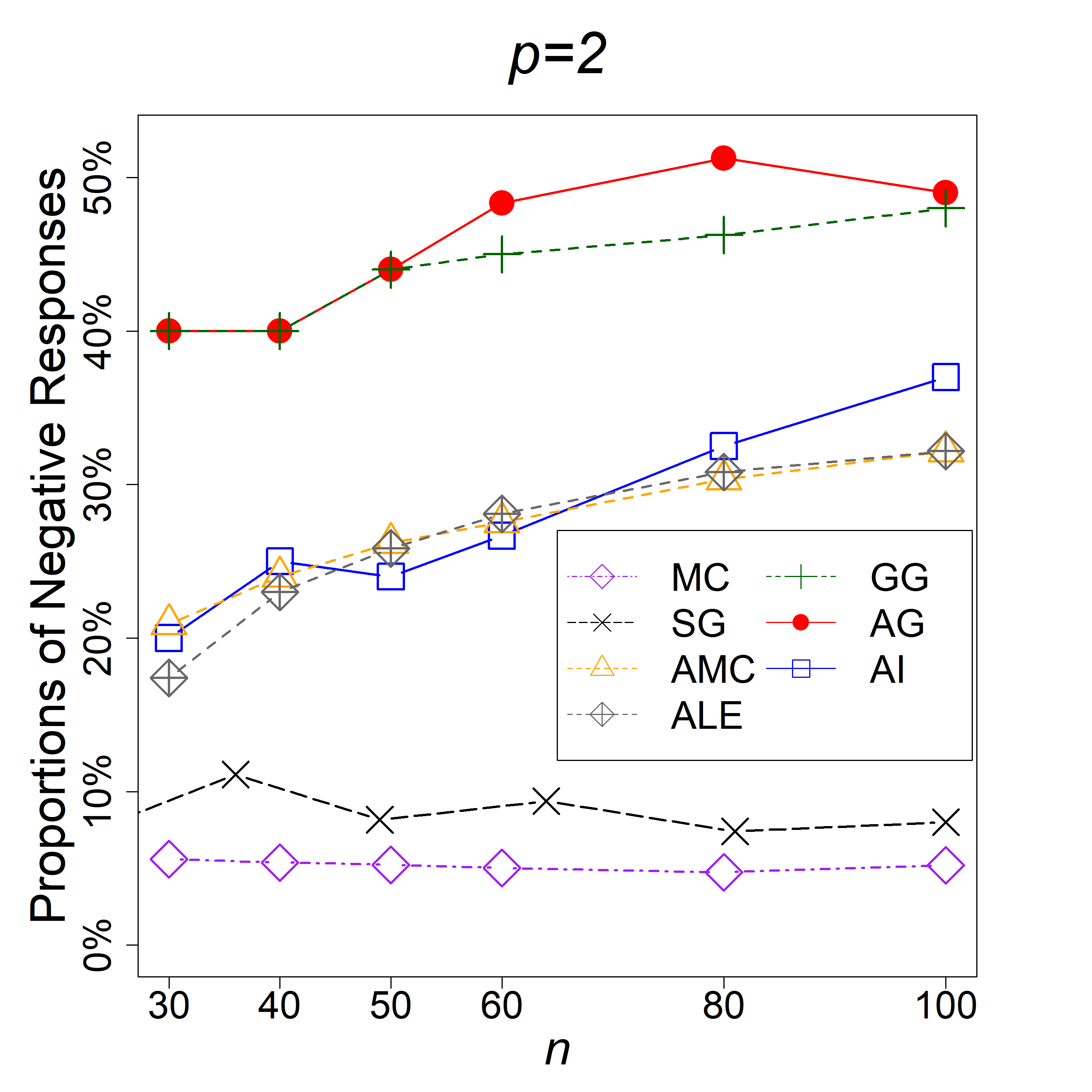}
		\end{minipage}
	}
	\quad
	\subfigure{
		\begin{minipage}{0.4\linewidth}
			\centering
			\includegraphics[width=\linewidth]{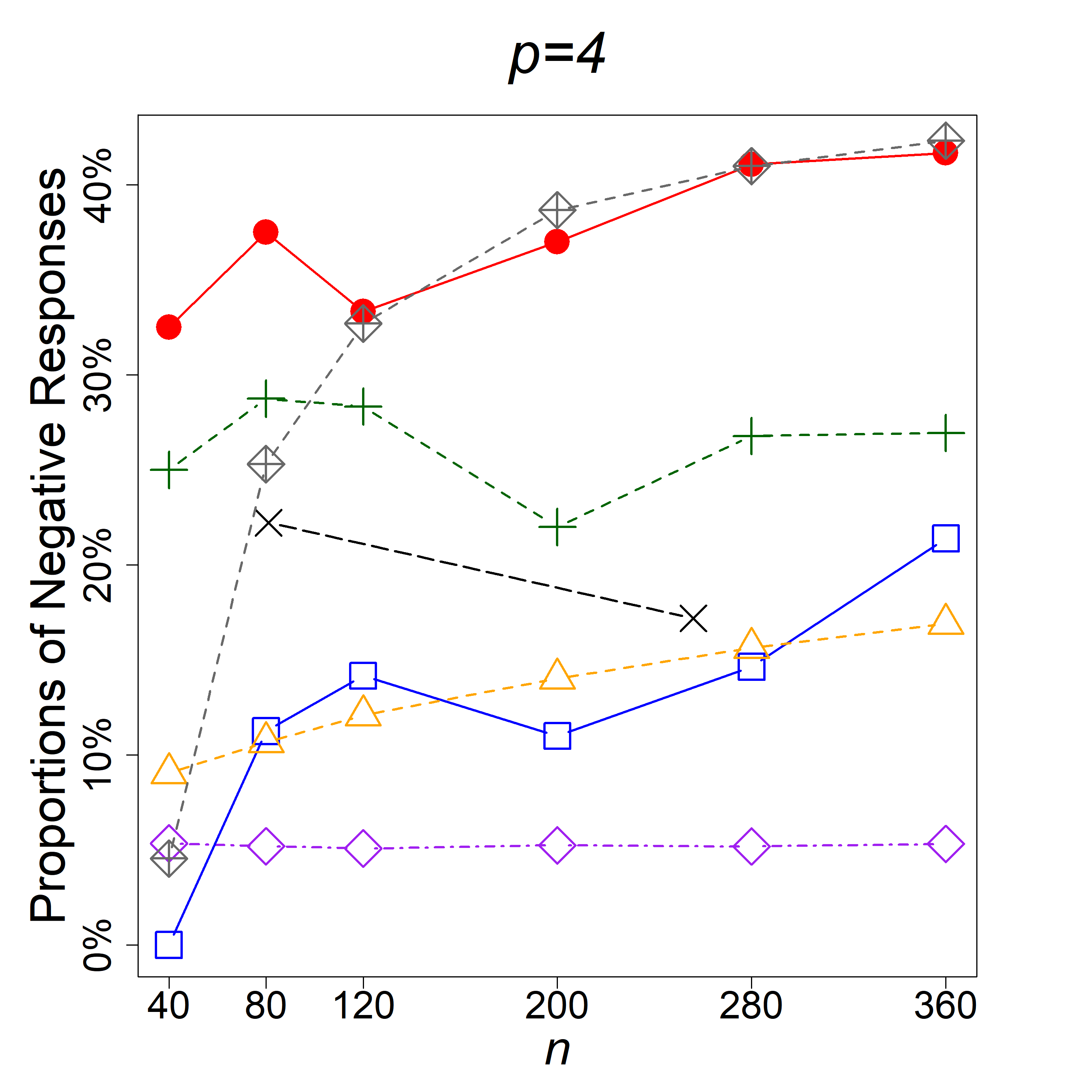}
		\end{minipage}
	}
	\caption{Proportions of negative responses for $p=2$ (left) and $p=4$ (right).}
	\label{fig:extreme}
\end{figure}

\begin{figure}
	\centering
	\subfigure[$\mathbf{D}_{\text{GG},2,30}$, $v=0.038$.]{
		\begin{minipage}{0.4\linewidth}
			\centering
			\includegraphics[width=\linewidth]{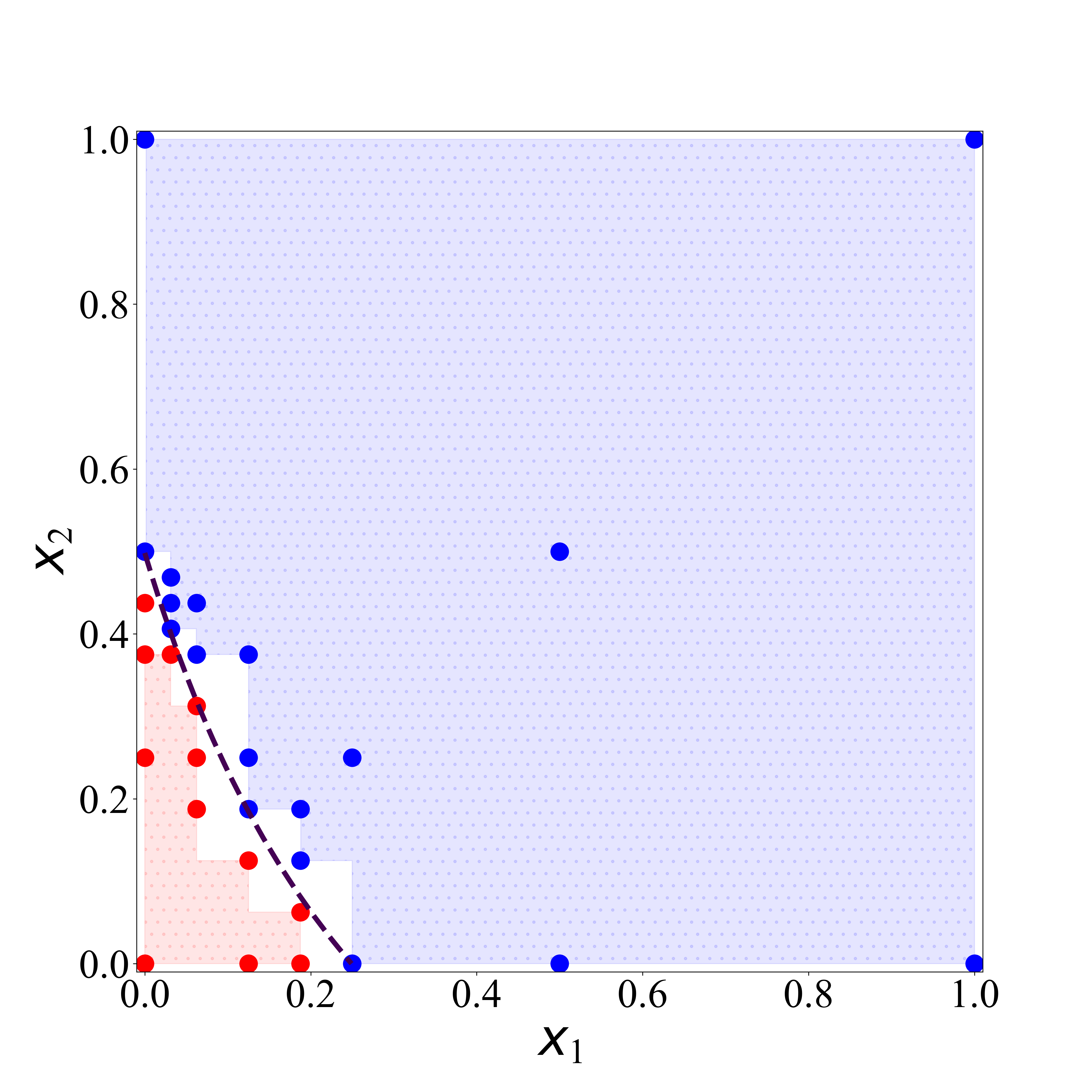}
			\label{extr_GG}
		\end{minipage}
	}
	\subfigure[$\mathbf{D}_{\text{AG},2,30}$, $v=0.023$.]{
		\begin{minipage}{0.4\linewidth}
			\centering
			\includegraphics[width=\linewidth]{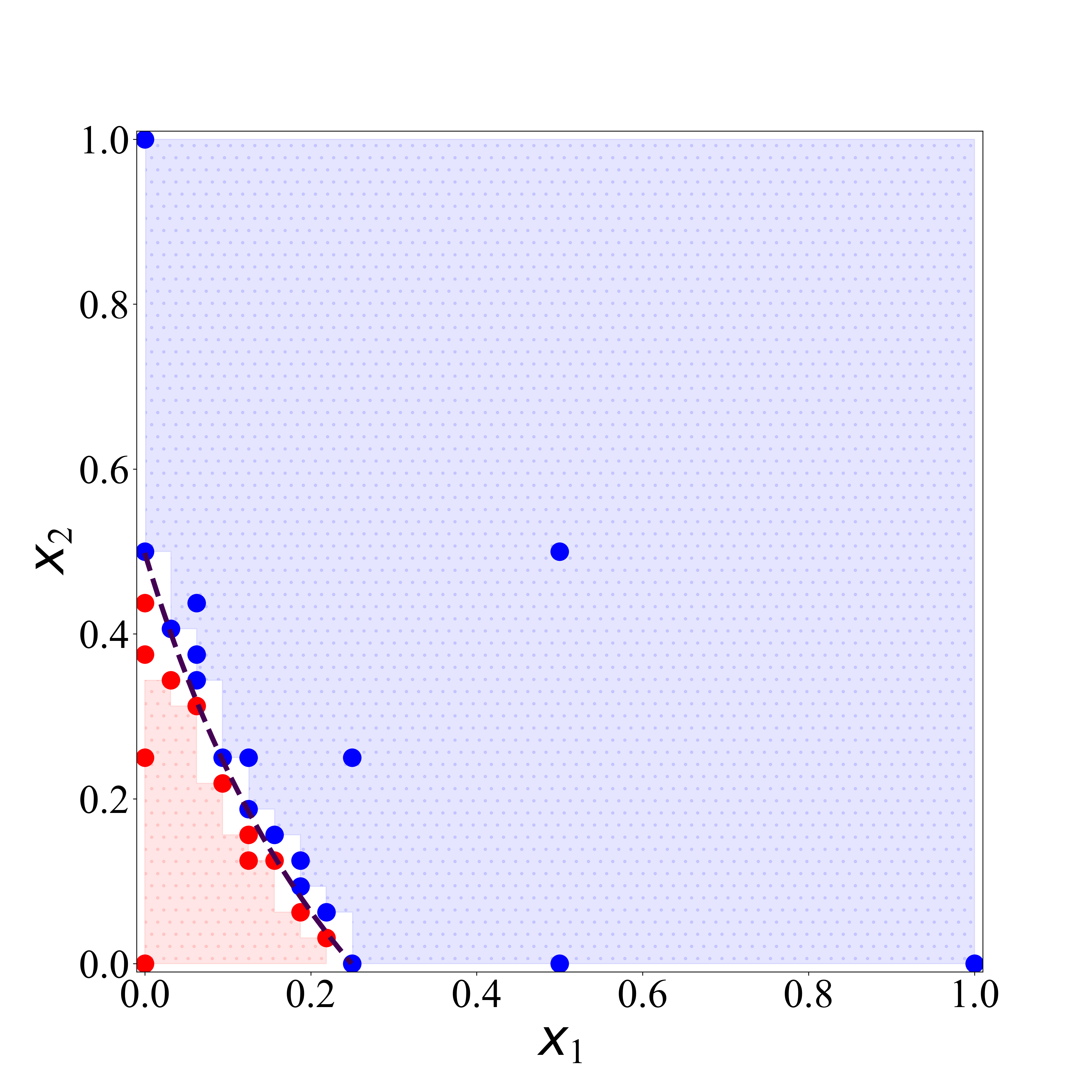}
			\label{extr_AG}
		\end{minipage}
	}
	\qquad
	\subfigure[$\mathbf{D}_{\text{AMC},2,30}$, $v=0.074$.]{
		\begin{minipage}{0.4\linewidth}
			\centering
			\includegraphics[width=\linewidth]{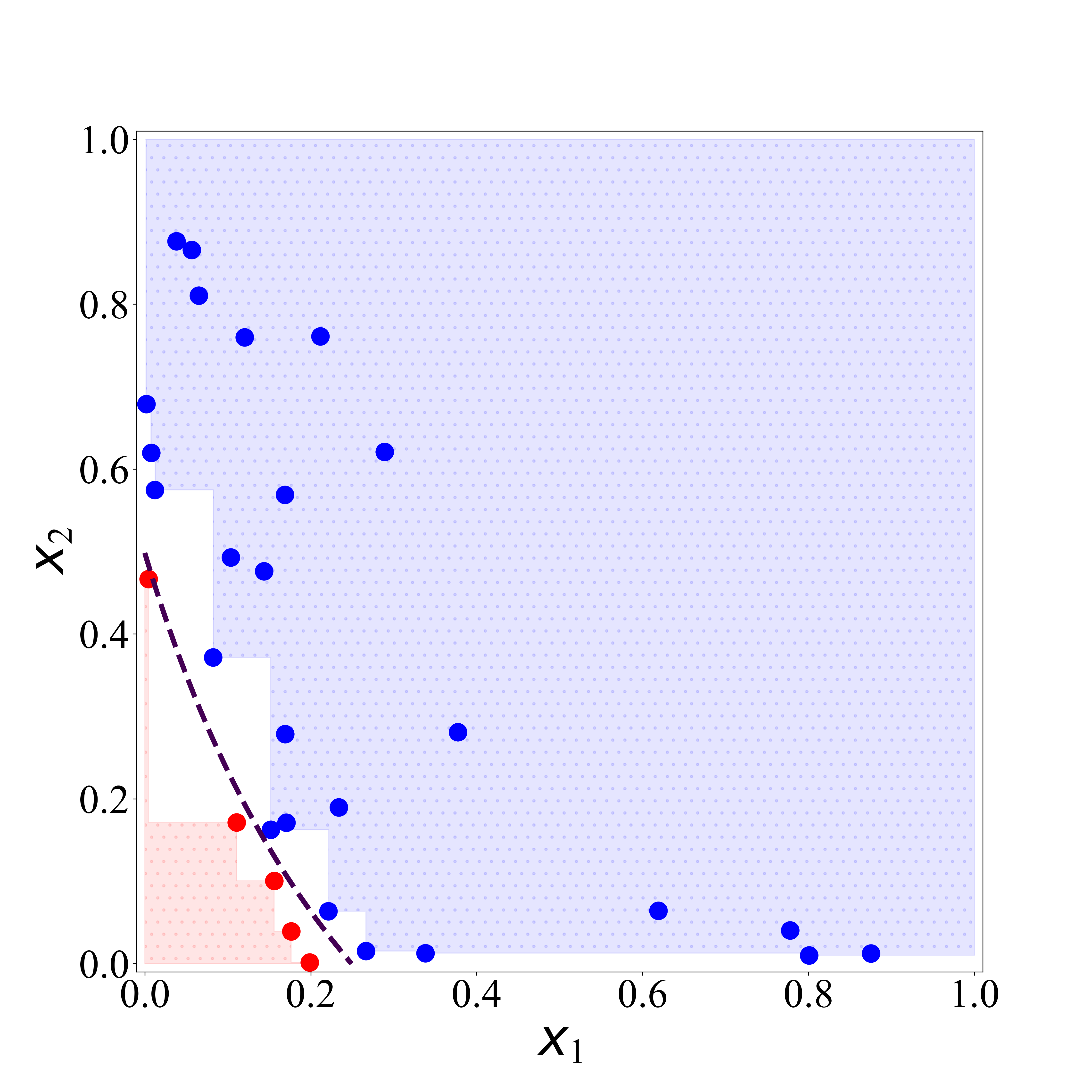}
			\label{extr_AMC}
		\end{minipage}
	}
	\subfigure[$\mathbf{D}_{\text{AI},2,30}$, $v=0.069$.]{
		\begin{minipage}{0.4\linewidth}
			\centering
			\includegraphics[width=\linewidth]{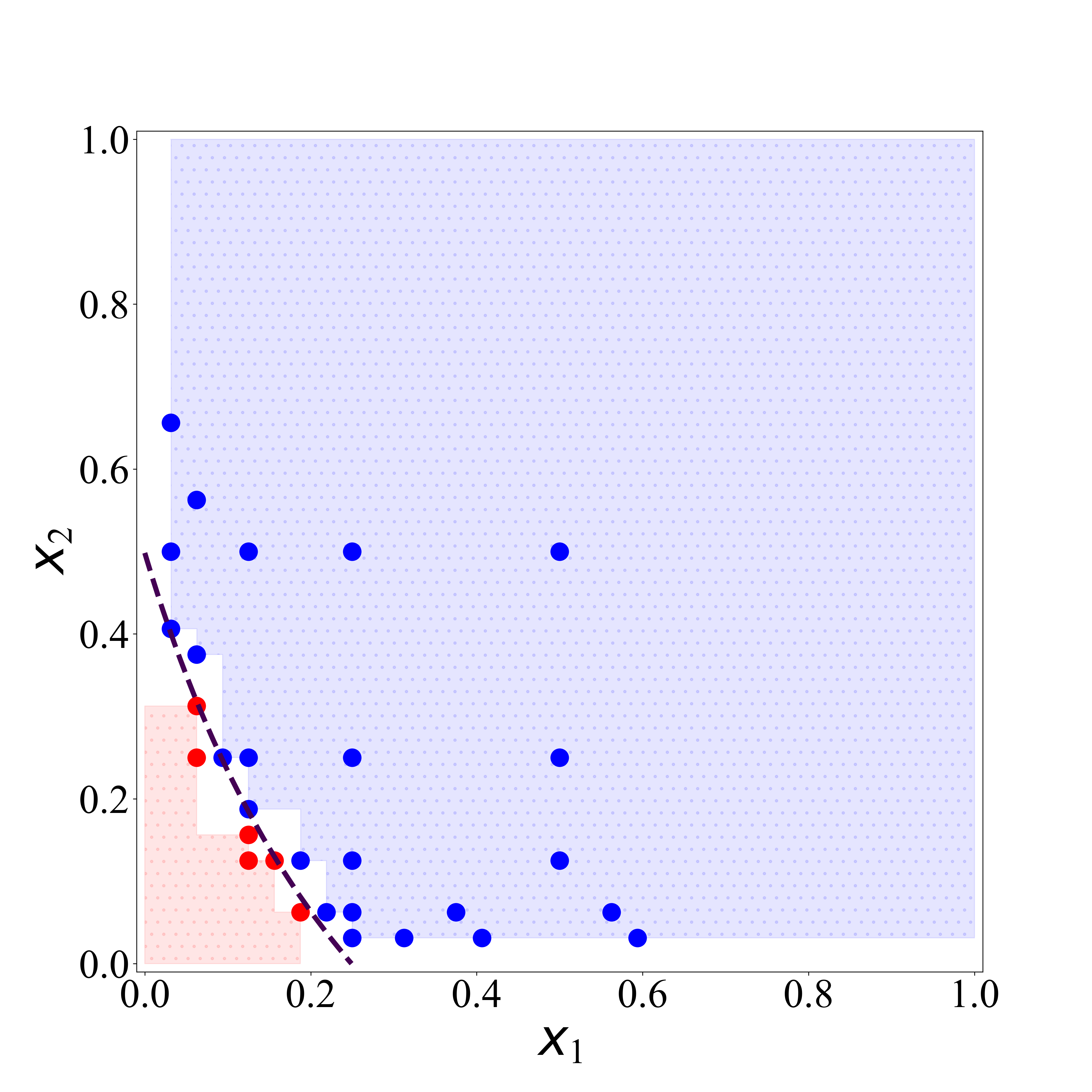}
			\label{extr_AI}
		\end{minipage}
	}
	\caption{Four designs in $p=2$, 
		showing the design points with negative response (red dots, or dots to the bottom-left of the dotted line) and positive response (blue dots, or dots to the top-right of the dotted line), 
		the curve separating the two regions $\mathbf{A}$ and $\mathbf{B}$ (purple dotted line), the certainly negative area 
		(red shaded area, or the bottom-left shaded area), the certainly positive area 
		(blue shaded area, or the top-right shaded area), and the uncertain area (white area) after evaluating the runs corresponding to the designs. 
	}
	\label{fig:illu:extreme}
\end{figure}

To further assess the performance of design methods when one outcome dominates the majority of the input space, we employ the same test function in \eqref{eqn:countour}, but set $\mu$ to be 0.69 and 1.88 when $p=2$ and $p=4$, respectively, resulting in $V(\mathbf{A}) = 5\%$. 
For each design with various $n$, we record the proportions of negative responses for the first $n$ runs. 
Fig.~\ref{fig:extreme} presents these proportions. 
From the results, the MC maintains stable proportions around 5\% due to its uniform distribution of points across the input space.
In contrast, the proportions for the AG, AI, GG, AMC, and ALE are notably higher and tend to increase as $n$ grows, suggesting that these adaptive designs tend to scatter most of the runs around the boundary line separating the two regions when $n$ is big. 
The ALE performs poorly when $n$ is small, suggesting that it is not robust in extreme cases with small sample sizes.
Furthermore, the AG and GG exhibit higher proportions compared to the AMC and AI. 
To uncover its cause, Fig.~\ref{fig:illu:extreme} depicts the $\mathbf{D}_{\text{GG},2,30}$, $\mathbf{D}_{\text{AG},2,30}$, $\mathbf{D}_{\text{AMC},2,30}$, and $\mathbf{D}_{\text{AI},2,30}$. 
As can be observed from the figure, from using the GG or AG, $\mathbf{U} \subset [0,1/4] \times [0,1/2]$ because both designs contain $(1/4,0)$ and $(0,1/2)$. 
Conversely, with no boundary point, from using the AMC or AI the $\mathbf{U} \cap [1/4,1] \times [0,1]$ and $\mathbf{U} \cap [0,1] \times [1/2,1]$ are non-empty, thereby yielding a larger $V(\mathbf{U})$.
Consequently, the AG is appealing when a single type of response dominants the input space. 

\section{Road Crash Simulation}\label{sec:road}

In this section, we provide numerical comparison on road crash simulation. 
Safety systems have been innovatively designed to enhance traffic safety, demonstrating significant potential in avoiding crashes. 
To gauge the outcomes with and without the implementation of a specific safety system, a glance-and-deceleration crash-causation model was employed, yielding a deterministic binary outcome, indicating whether the crash occurs under specific off-road glance duration and deceleration rate~\citep{seyedi2021safety}. 
The Volvo Car Corporation has reconstructed 40 rear-end crashes in Sweden~\citep{coelingh2007collision} using the specific pre-crash kinematics, focusing on two main causes: drivers not keeping their eyes on the forward roadway and drivers not braking at the maximum level.
For each scenario, they simulated outcomes indicating whether a crash occurred for different off-road glance durations and decelerations, with $-1$ indicating the avoidance and 1 indicating the crash.
The off-road glance duration spans 67 equally-spaced levels from $0.0$ to $6.6$ seconds and the deceleration is taken to be the 15 equally-spaced levels from $-10.3$ to $-3.3$ meters per squared seconds. 
They have provided the outcomes for all 40 occasions, covering all $67 \times 15 = 1005$ combinations of inputs. 
The simulation is monotonic because the longer the off-road glance duration and the lower the deceleration, the more likely of a crash. 
The data set, along with additional details regarding these simulations, can be found in \citep{imberg2022active}.

\begin{figure}
	\centering
	\subfigure[$\mathbf{D}_{\text{AG},2,10}$, $v= 0.258$.]{
		\begin{minipage}{0.3\linewidth}
			\centering
			\includegraphics[width=\linewidth]{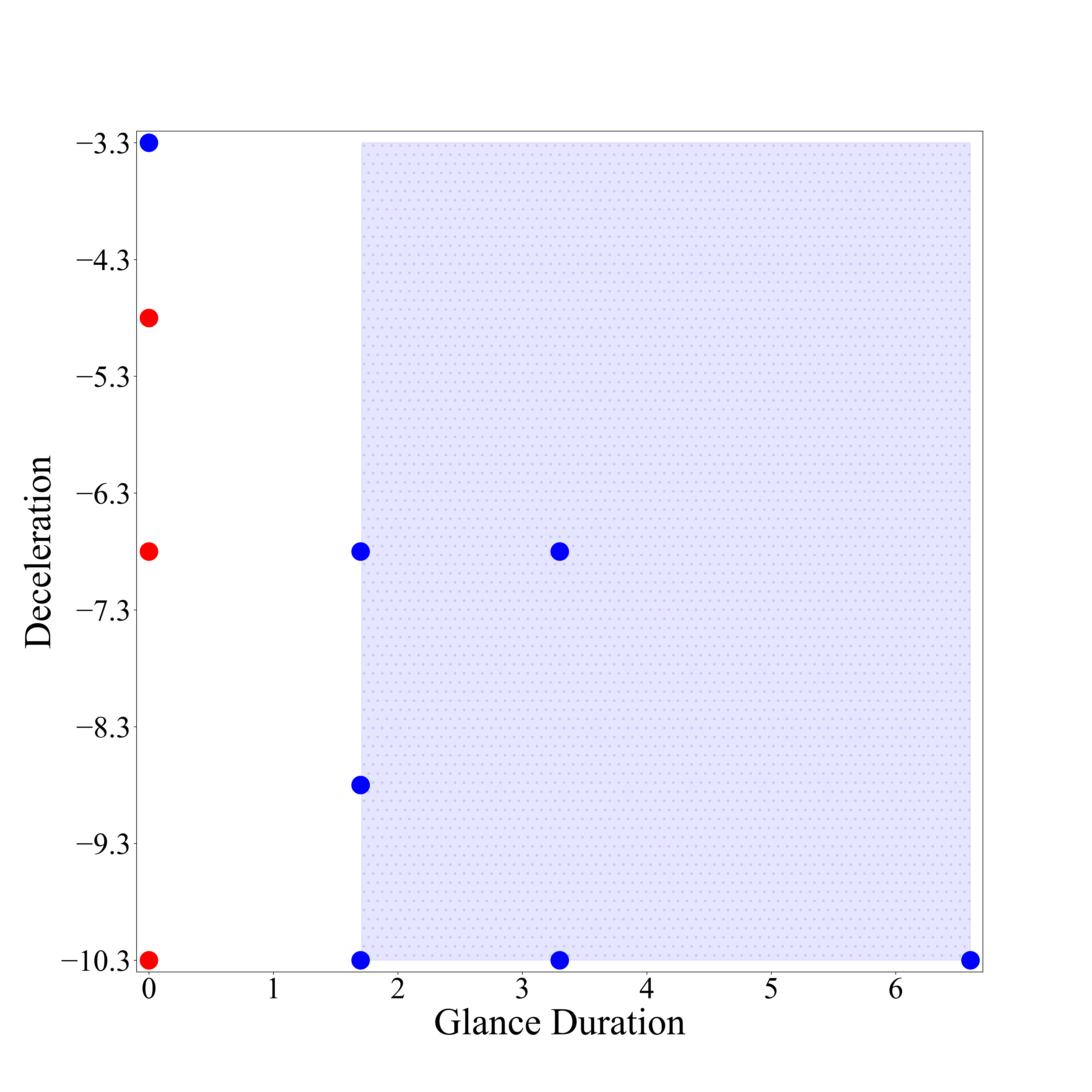}
			\label{illu_crash10}
		\end{minipage}
	}
	\subfigure[$\mathbf{D}_{\text{AG},2,20}$, $v= 0.085$.]{
		\begin{minipage}{0.3\linewidth}
			\centering
			\includegraphics[width=\linewidth]{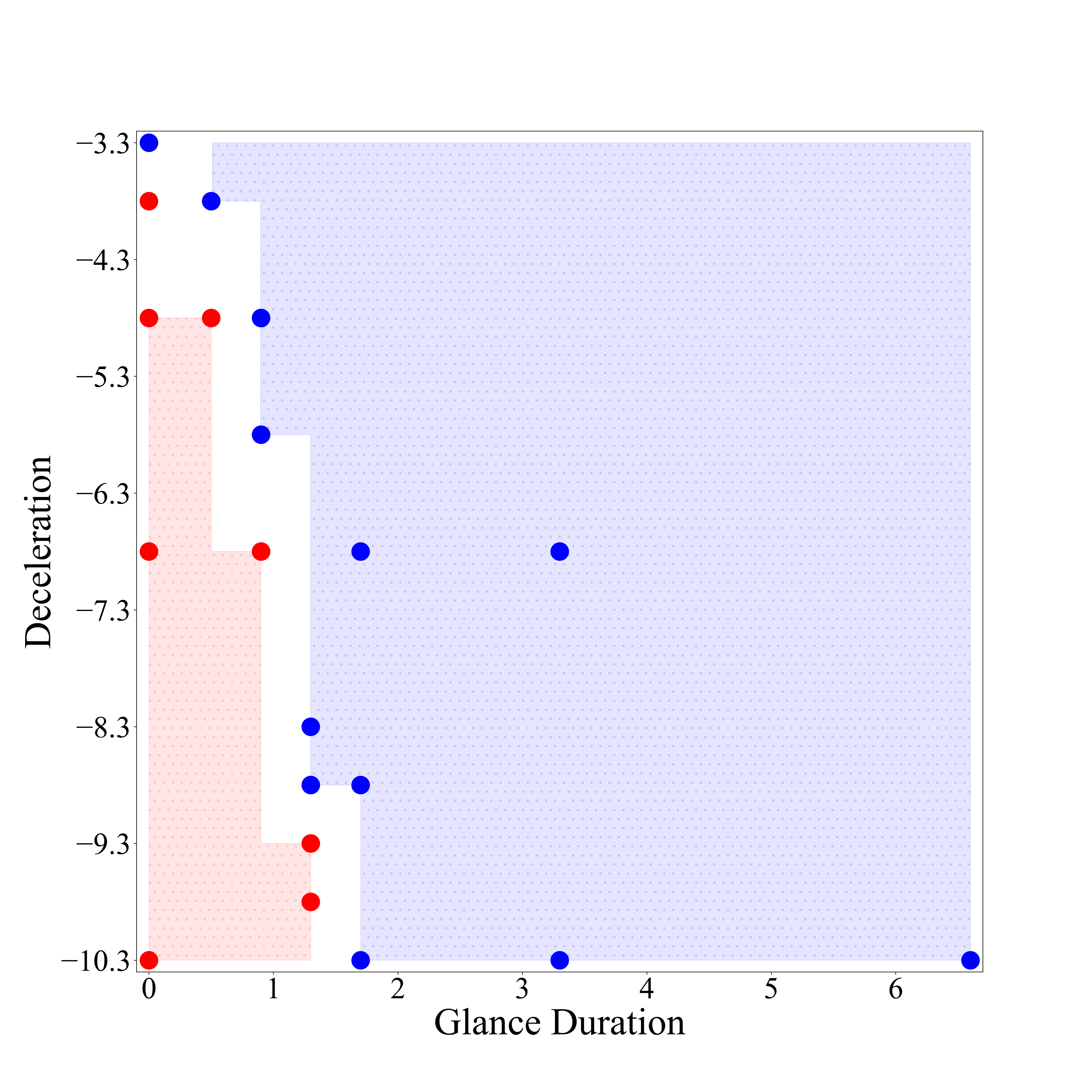}
			\label{illu_crash20}
		\end{minipage}
	}
	\subfigure[$\mathbf{D}_{\text{AG},2,30}$, $v= 0.048$.]{
		\begin{minipage}{0.3\linewidth}
			\centering
			\includegraphics[width=\linewidth]{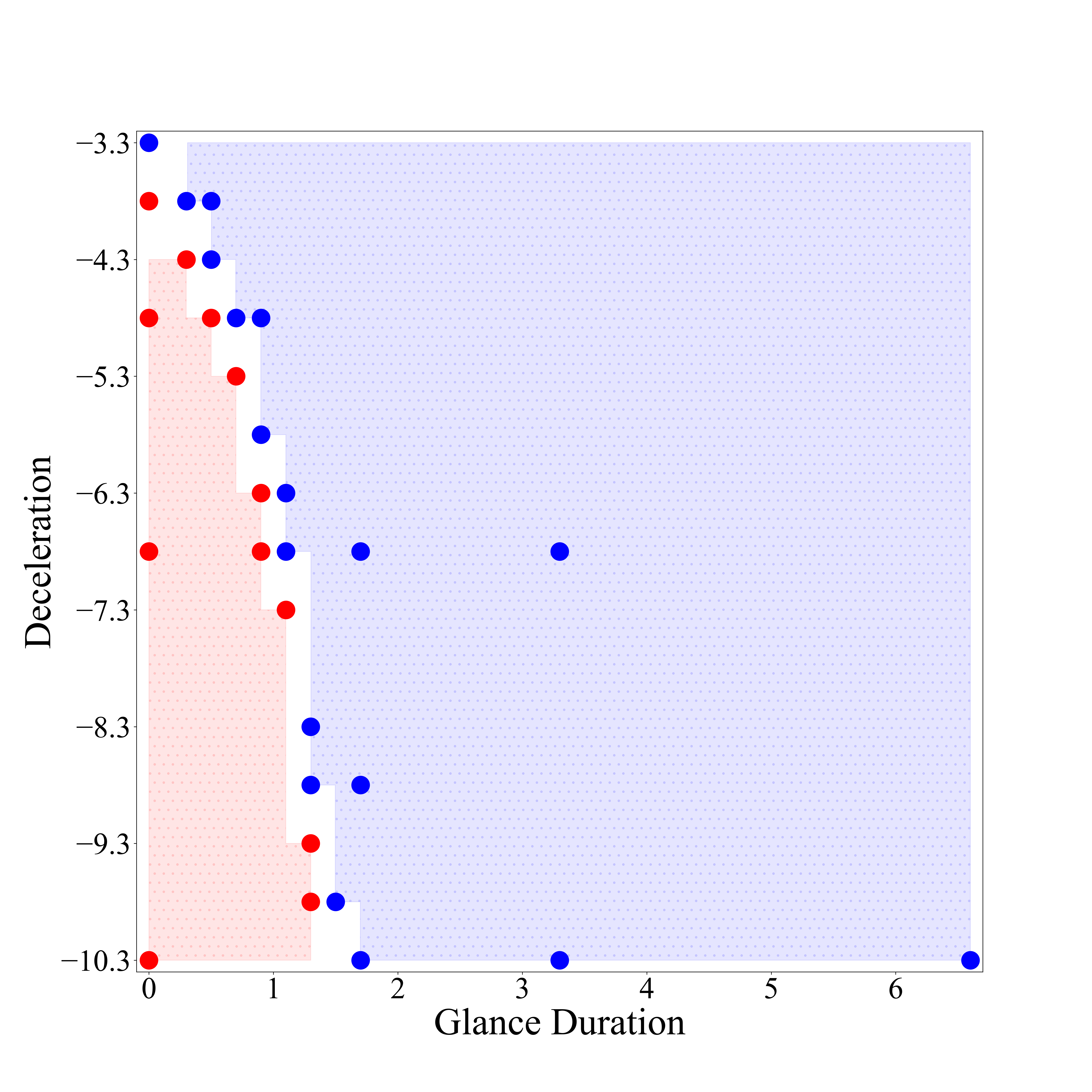}
			\label{illu_crash30}
		\end{minipage}
	}
	\caption{The $\mathbf{D}_{\text{AG},2,10}$, $\mathbf{D}_{\text{AG},2,20}$, and $\mathbf{D}_{\text{AG},2,30}$ for the 17th crash occasion,
		showing the design points with crash avoidance (red dots) and crash (blue dots), the certainly crash avoidance area 
		(red shaded area, or the bottom-left shaded area), the certainly crash area (blue shaded area, or the top-right shaded area), and the uncertain area (white area) after evaluating the runs corresponding to the designs.}
	\label{fig:crash}
\end{figure}

We use this data set to compare the designs. 
Since the outcomes are available for only the 1005 combinations, we assume the computer simulation can be carried out only on these 1005 combinations. We aim to assess how the designs perform on problems with ordinal input variables.  
To adapt to the discrete input space, for the MC, AMC, and ALE, we sample $\mathbf{x}$ uniformly from the 1005 possible combinations instead of generating the $\mathbf{x}$ from the uniform distribution in $[0,1]^2$. 
For grid designs, we map design points in $[0,1]^2$ to the discrete input space. 
To be specific, let $x_\text{o}$ and $x_\text{d}$ denote the off-road glance duration in seconds and the deceleration in meters per squared seconds, respectively, and $\tilde f(x_\text{o},x_\text{d})$ denote the functional relationship between $(x_\text{o},x_\text{d})$ and the outcome. 
For the GG and AG, we employ the following nonlinear mappings 
\begin{equation*}
	h_{\text{GG,o}}(x) = \begin{cases}
		0, & x=0,\\
		0.1, & x=1/128,\\
		6.4x_1+0.1, & 1/64 \leq x \leq 63/64, \\
		6.5, & x=127/128,\\
		6.6, & x=1, 
	\end{cases}
\end{equation*}
and
\begin{equation*}
	h_{\text{GG,d}}(x) = \begin{cases}
		-10.3, & x=0,\\
		8x-10.8, & 1/8\leq x\leq 7/8, \\
		-3.3, & x=1.
	\end{cases}
\end{equation*}
Then $f(x_1,x_2) = \tilde f\{h_{\text{GG,o}}(x_1),h_{\text{GG,d}}(x_2)\}$ 
is monotonic non-decreasing in the discrete input space $[0,1/128,1/64,2/64,\ldots,63/64,127/128,1] \times [0, 2/16,3/16, \ldots, 14/16, 1] \allowbreak\subset [0,1]^2$.  
Using the $f(\mathbf{x})$ in place of $\tilde f$, we can apply the GG and AG to the road crash problem. 
Remark that here we use a nonlinear mapping so that the lawful $x_1$ and $x_2$ values are rational whose denominator is a power of two. 
Similarly, for the AI, we employ the following nonlinear mappings
\begin{equation*}
	\begin{aligned}
		h_{\text{AI,o}}(x) = \begin{cases}
			0, & x=1/256,\\
			0.1, & x=1/128,\\
			6.4x_1+0.1, & 1/64 \leq x \leq 63/64, \\
			6.5, & x=127/128,\\
			6.6, & x=255/256, 
		\end{cases}
		\qquad
		h_{\text{AI,d}}(x) = 8x-10.8.
	\end{aligned}
\end{equation*}
We believe these mappings show examples on how the grid design methods can be applied to input spaces other than $[0,1]^p$. 
For illustration, Fig.~\ref{fig:crash} depicts the $\mathbf{D}_{\text{AG},2,10}$, $\mathbf{D}_{\text{AG},2,20}$, and $\mathbf{D}_{\text{AG},2,30}$ for the 17th crash occasion. 

\begin{figure}
	\centering
	\subfigure{
		\begin{minipage}{0.4\linewidth}
			\centering
			\includegraphics[width=\linewidth]{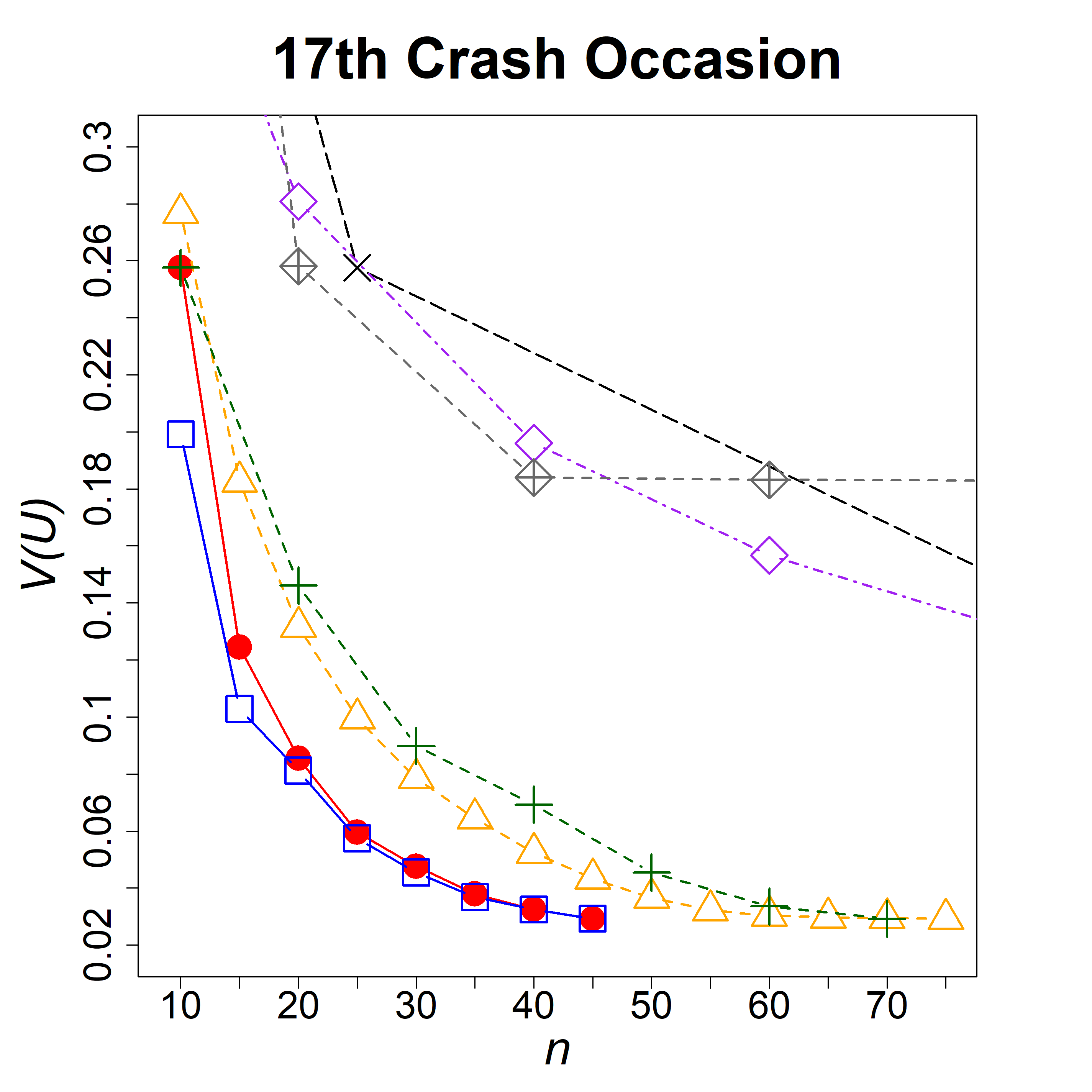}
			\label{crash_v}
		\end{minipage}
	}
	\subfigure{
		\begin{minipage}{0.4\linewidth}
			\centering
			\includegraphics[width=\linewidth]{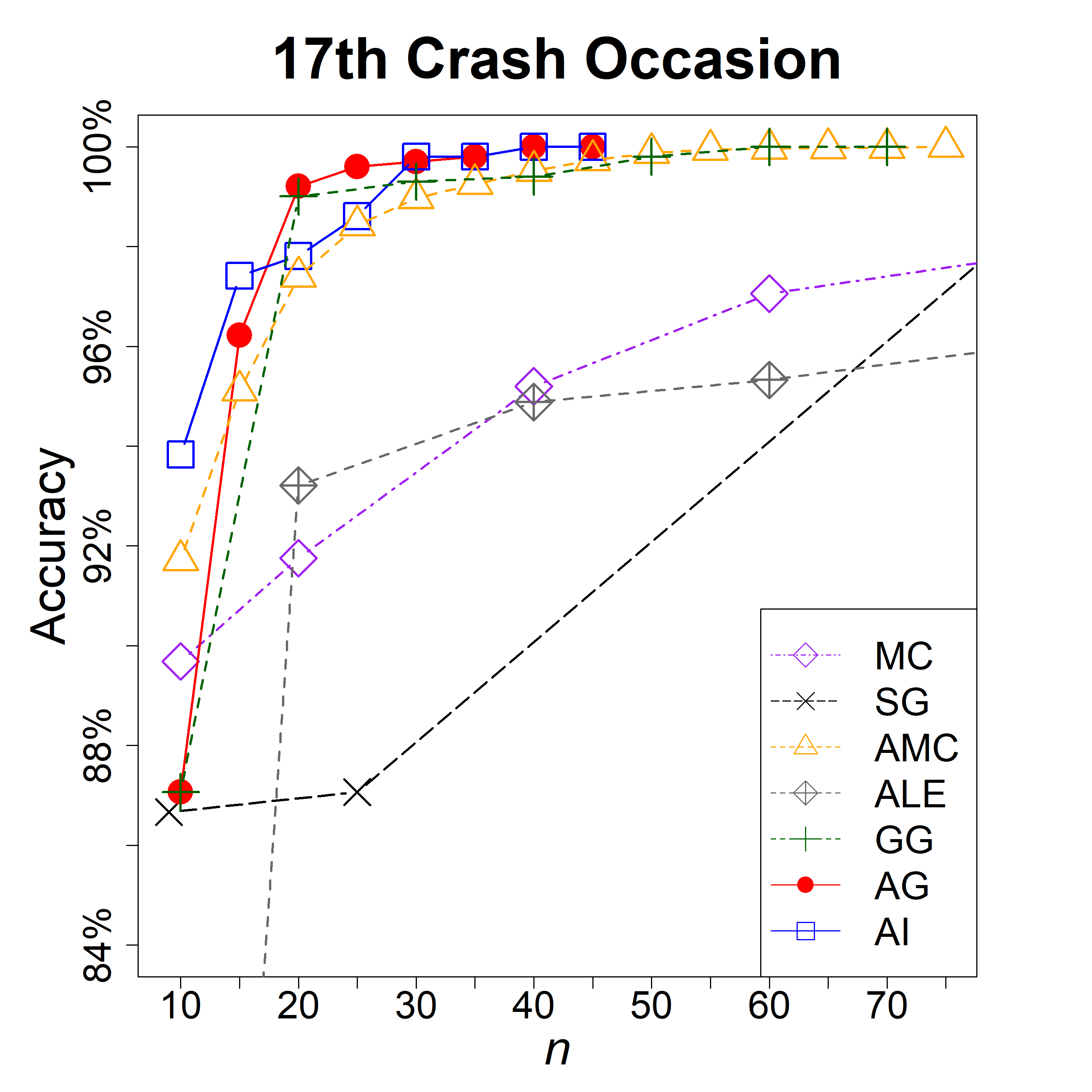}
			\label{crash_clf}
		\end{minipage}
	}
	\caption{Averaged volume of uncertain area $V(\mathbf{U})$ (left) and classification accuracy (right) for the 17th crash occasion.}
	\label{fig:crash:17}
\end{figure}

\begin{figure}
	\centering
	\subfigure{
		\begin{minipage}{0.4\linewidth}
			\centering
			\includegraphics[width=\linewidth]{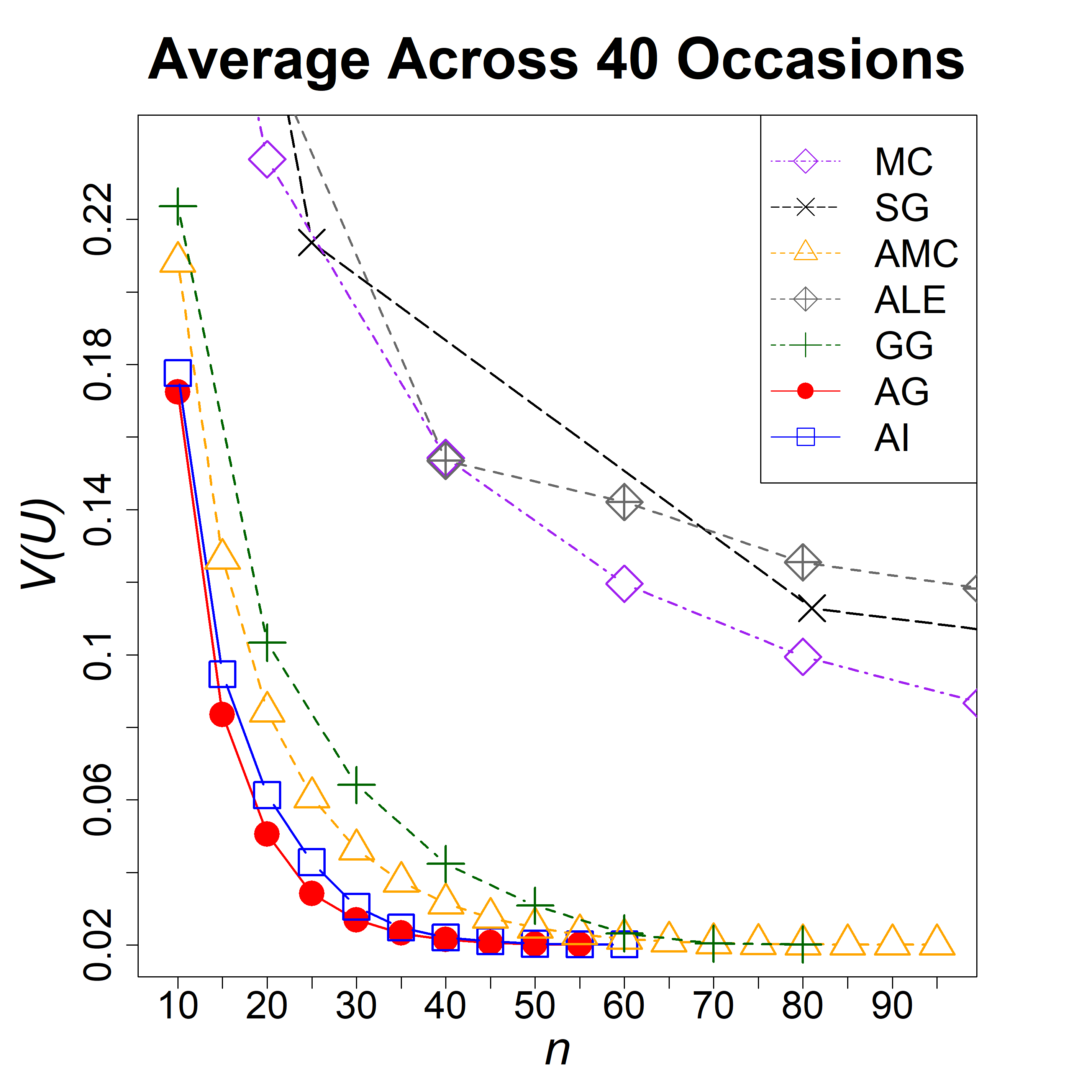}
			\label{crash_v}
		\end{minipage}
	}
	\subfigure{
		\begin{minipage}{0.4\linewidth}
			\centering
			\includegraphics[width=\linewidth]{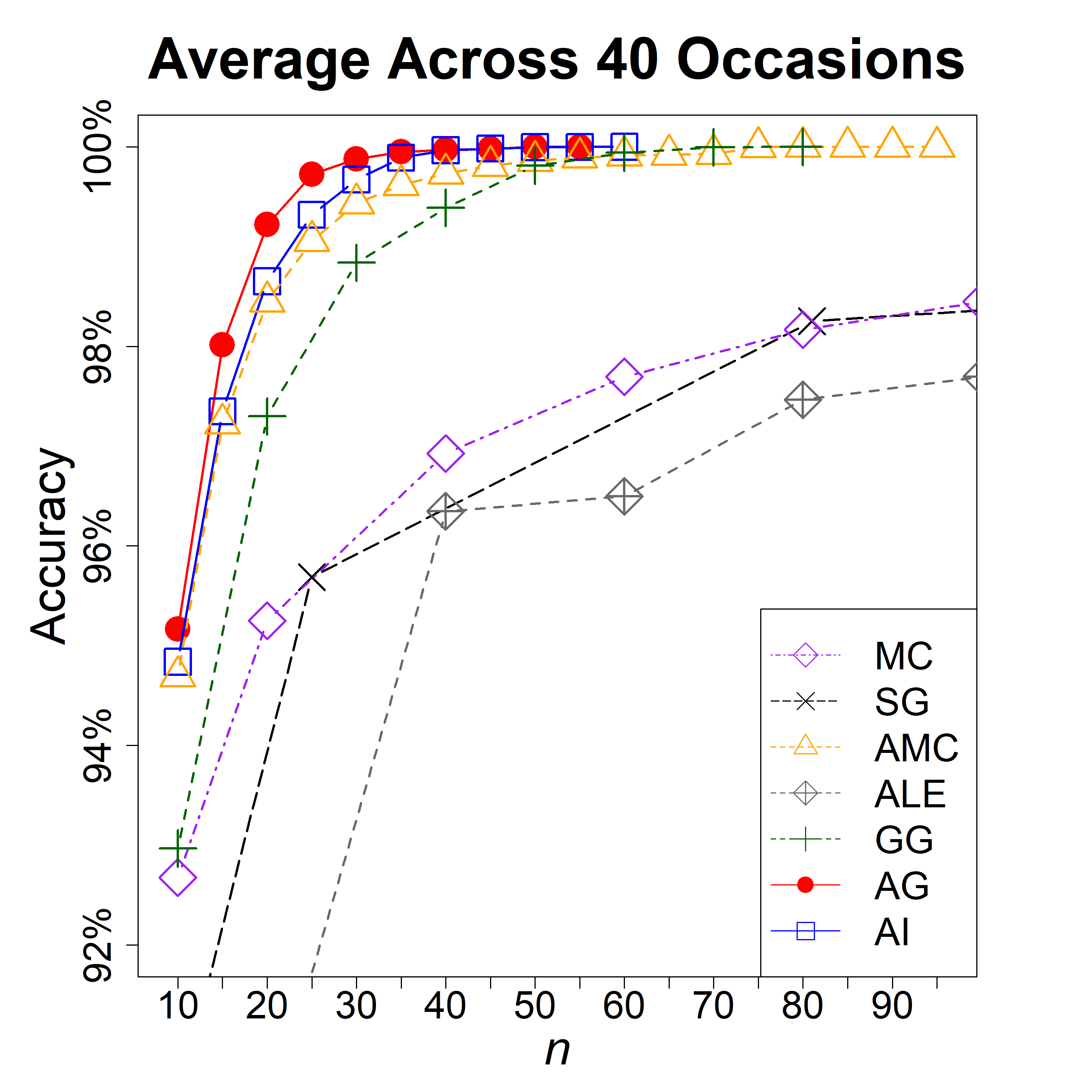}
			\label{crash_clf}
		\end{minipage}
	}
	\caption{Averaged volume of uncertain area $V(\mathbf{U})$ (left) and classification accuracy (right) across all 40 crash occasions.}
	\label{fig:crash:all}
\end{figure}

Similar to the comparison in Section~\ref{sec:comp} of the paper, we use the $V(\mathbf{U})$ and the classification accuracy to measure the performance of different types of designs. 
However, for computing the accuracy, the testing inputs are the 1005 combinations. 
Fig.~\ref{fig:crash:17} gives the results for the 17th occasion while Fig.~\ref{fig:crash:all} gives the averaged result across all 40 crash occasions. 
From the results, the AG and AI outperform other methods by requiring fewer samples to obtain the exist boundary line. 
This aligns with our theoretical and previous numerical results. 

\section{Ice-breaking Simulation}\label{sec:ice}

Finally, we apply our recommended AG method to the ice-breaking simulation. 
In the simulation, the sphere with a diameter of $0.0254$ meters, a density of $3900$ kilograms per cubic meters, and initial velocity $x_{\text{v}}$ meters per second impacts a cuboid ice sheet with $0.15$ meters long and wide and $x_{\text{t}}$ millimeters thick. 
The ice sheet is assumed to has a Poisson ratio of $0.33$ and an elasticity modulus $x_{\text{e}}$ gigapascals. 
We use the finite element method to simulate the ice-breaking dynamics. 
We label the outcome ``1'' if the sphere breaks the ice sheet and ``-1'' otherwise. 
According to expert guidance, the ranges of inputs are $x_{\text{v}} \in [5,40]$, $x_{\text{t}} \in [5,15]$, and $x_{\text{e}} \in [1,5]$. 
Based on domain knowledge, the higher the $x_{\text{v}}$, the lower the $x_{\text{t}}$, and the lower the $x_{\text{e}}$, the likely the sphere breaks the ice sheet. 

Here, the input space is not $[0,1]^3$ and the outcome is monotonic non-increasing on two inputs. 
To tackle this problem, we apply linear transformations $h_{\text{v}}(x) = 35x + 5$, $h_{\text{t}}(x) = -10x + 15$, and $h_{\text{e}}(x) = -4x + 5$. 
Let $\tilde f$ denote the functional relationship between the outcome and the $(x_{\text{v}},x_{\text{t}},x_{\text{e}})$. 
Then $f(x_1,x_2,x_3) = \tilde f\{h_{\text{v}}(x_1),h_{\text{t}}(x_2),h_{\text{e}}(x_3)\}$ is a three-dimensional monotonic non-decreasing function supported on $[0,1]^3$. 
Using the $f(\mathbf{x})$ to replace the $\tilde f$, we can apply any design we have discussed to the ice-breaking problem. 

\begin{figure}
	\centering
	\subfigure[$x_{\text{e}}=5$]{
		\begin{minipage}{0.28\linewidth}
			\centering
			\includegraphics[width=\linewidth]{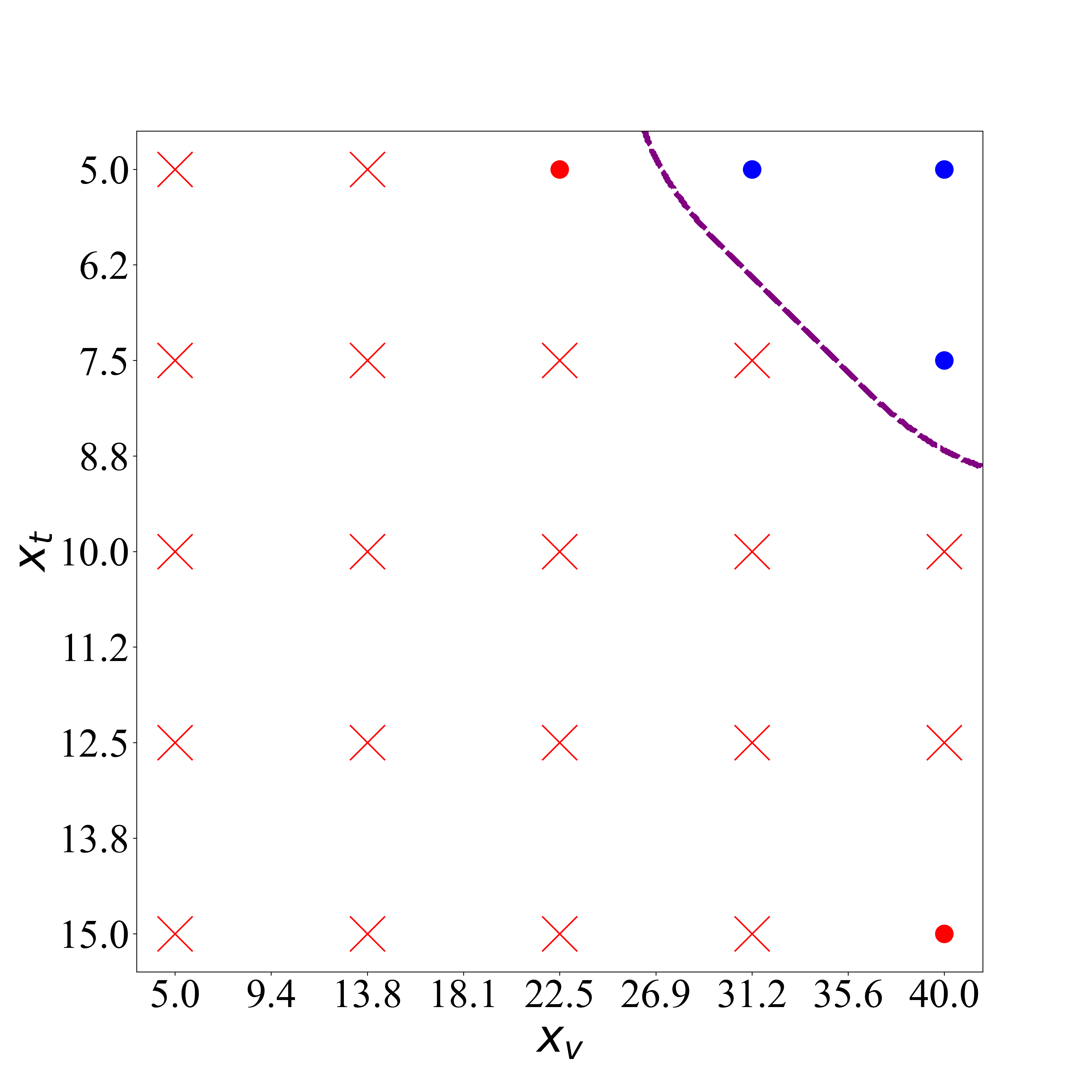}
		\end{minipage}
	}
	\subfigure[$x_{\text{e}}=4$]{
		\begin{minipage}{0.28\linewidth}
			\centering
			\includegraphics[width=\linewidth]{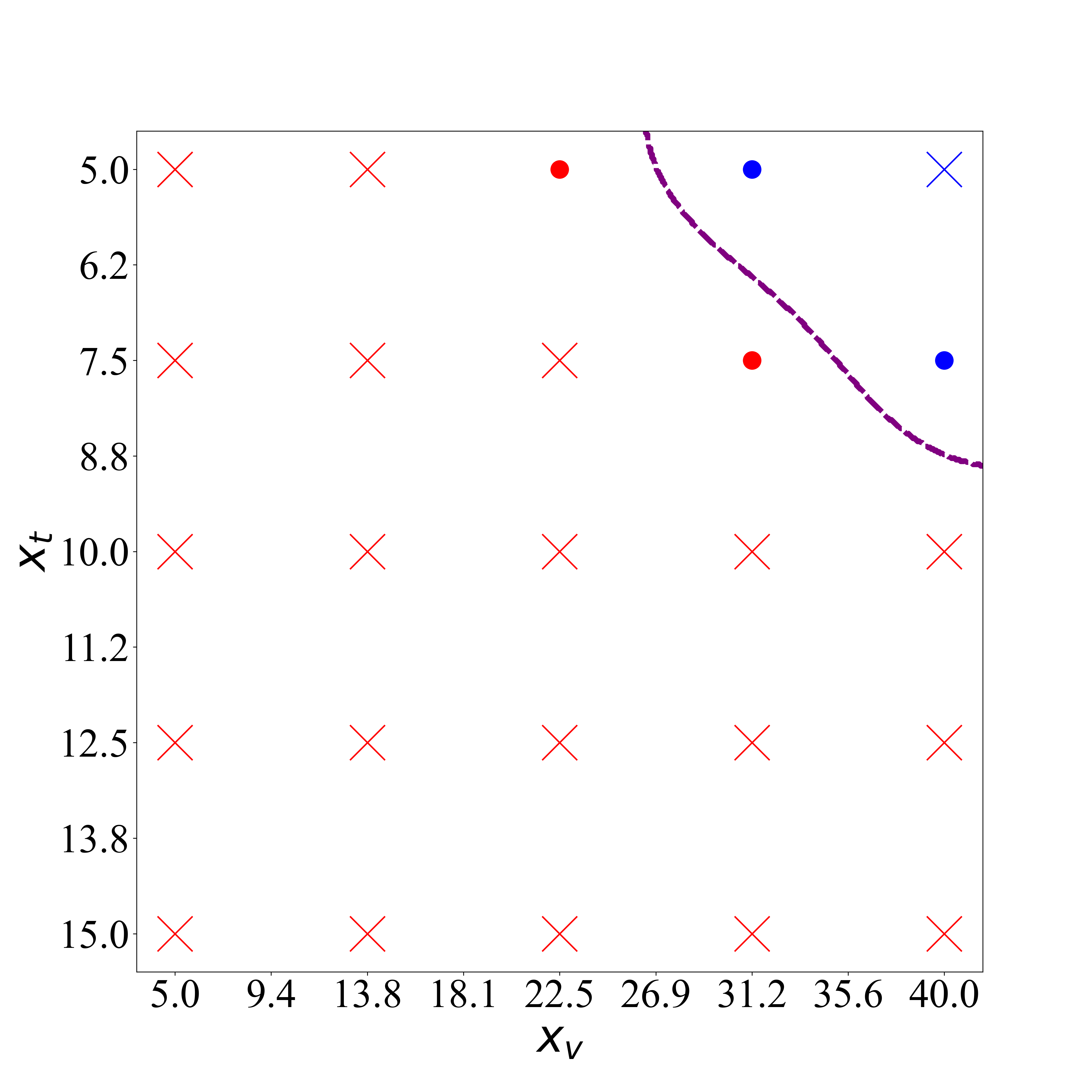}
		\end{minipage}
	}
	\subfigure[$x_{\text{e}}=3$]{
		\begin{minipage}{0.28\linewidth}
			\centering
			\includegraphics[width=\linewidth]{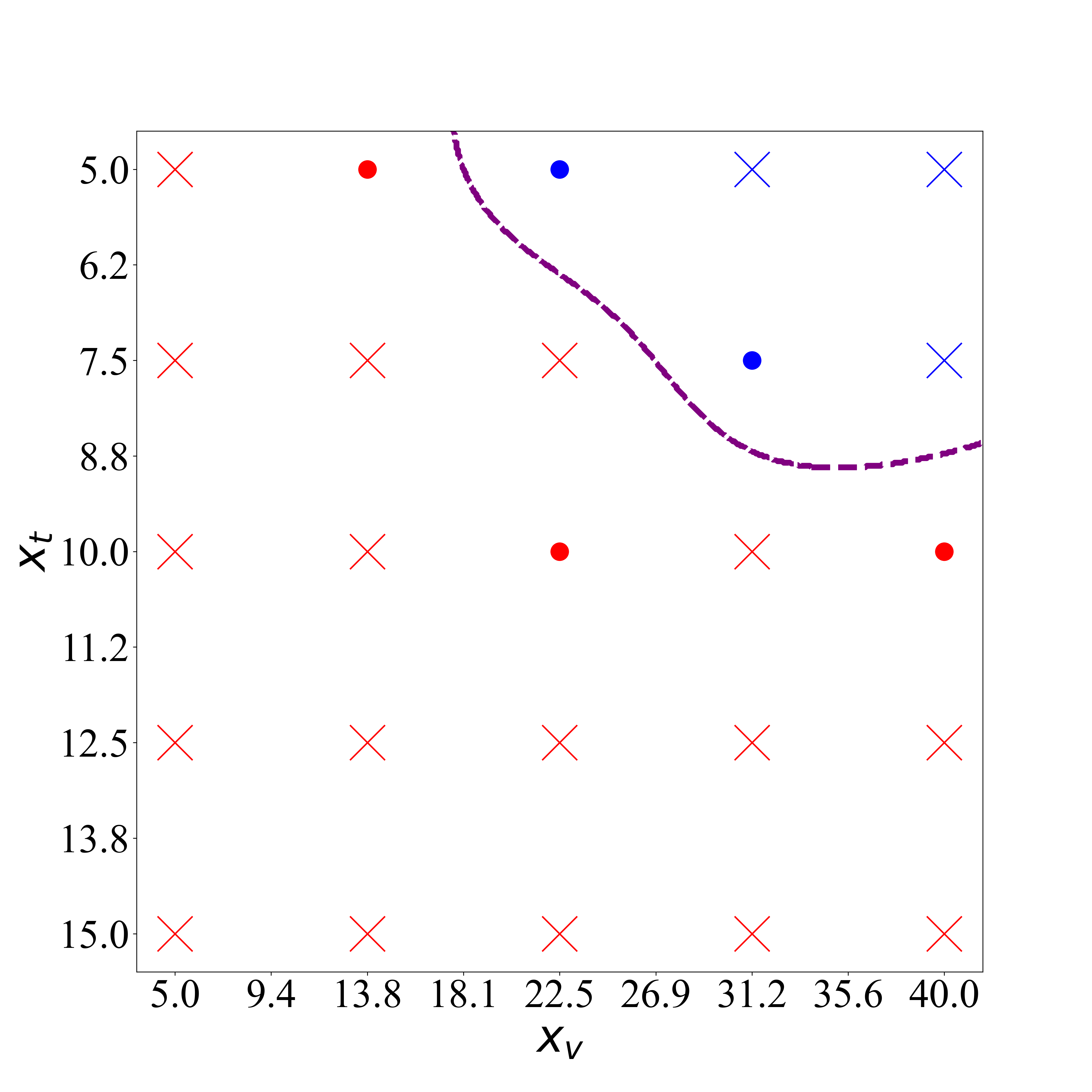}
		\end{minipage}
	}
	\qquad
	\subfigure[$x_{\text{e}}=2$]{
		\begin{minipage}{0.28\linewidth}
			\centering
			\includegraphics[width=\linewidth]{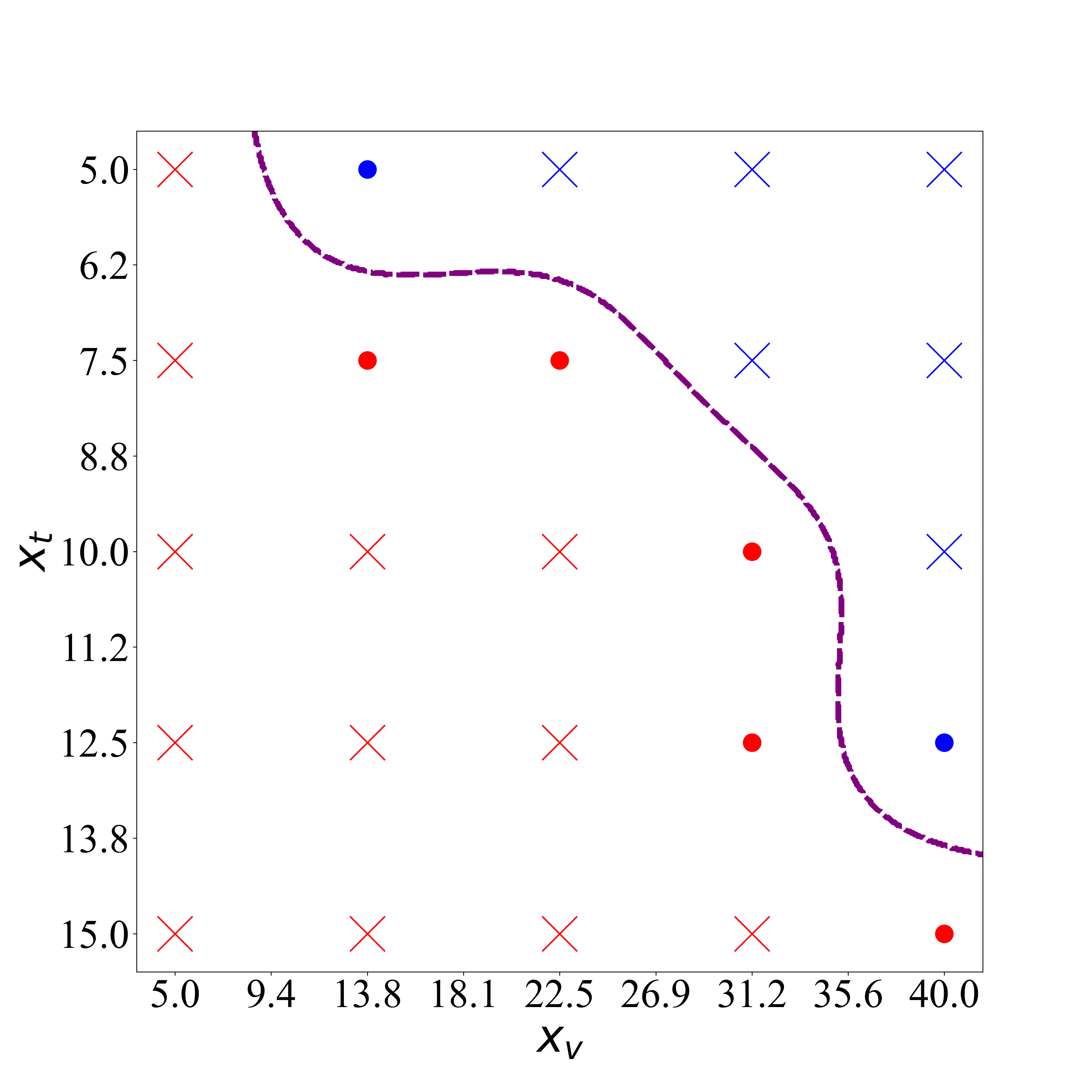}
		\end{minipage}
	}
	\qquad
	\subfigure[$x_{\text{e}}=1$]{
		\begin{minipage}{0.28\linewidth}
			\centering
			\includegraphics[width=\linewidth]{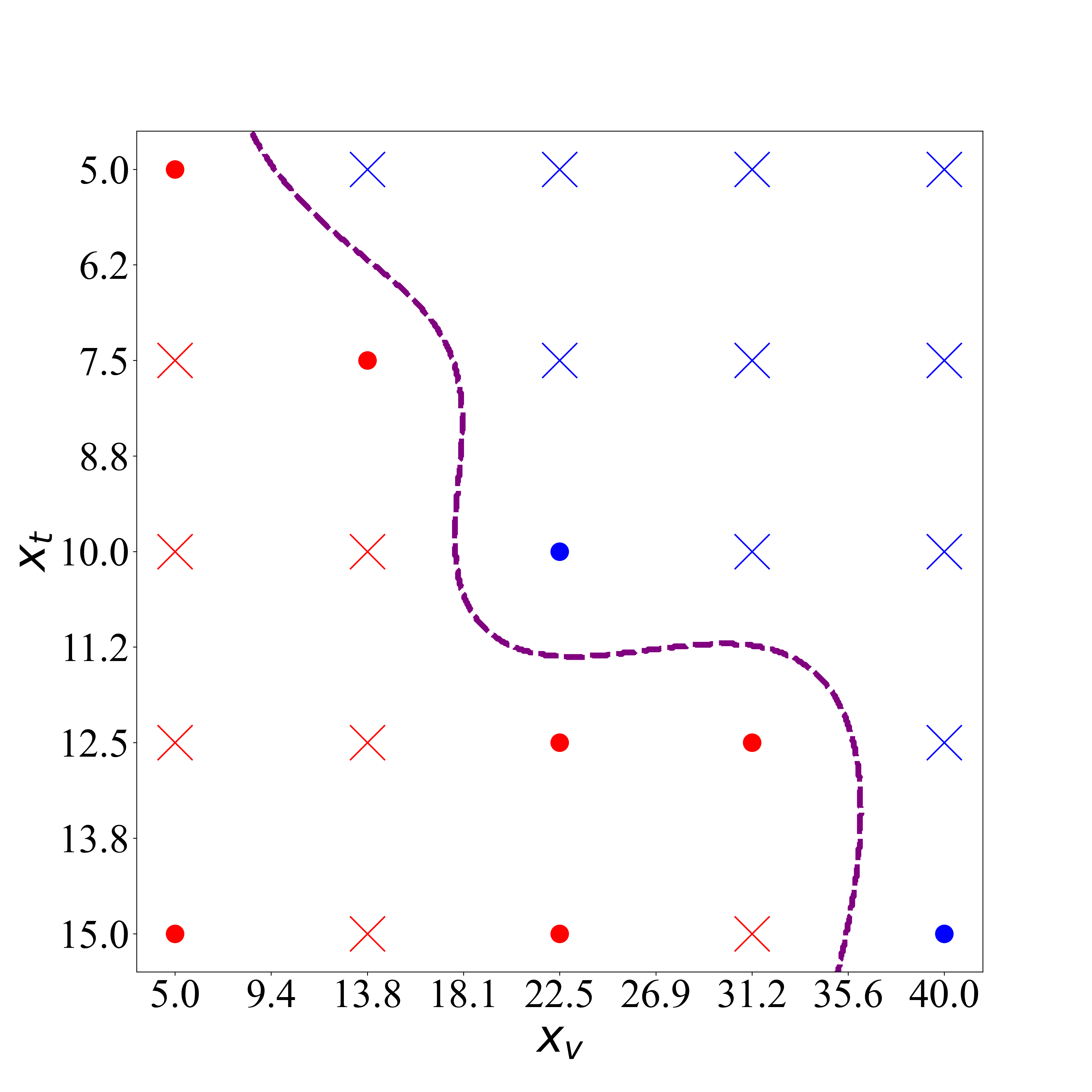}
		\end{minipage}
	}
	\caption{The $\mathbf{D}_{\text{AG},3,29}$ for all five $x_{\text{e}}$ values, 
		showing the design points with negative response (red dots, or dots to the bottom-left of the dotted line) and positive response (blue dots, or dots to the top-right of the dotted line), the skipped design points (crosses), and the estimated boundary (purple dotted line). }
	\label{fig:2dplot}
\end{figure}

Due to the high cost of evaluations, we only carry out the AG method to up to 29 simulation runs. 
Fig.~\ref{fig:2dplot} shows the points of $\mathbf{D}_{\text{AG},3,29}$ as well as the estimated boundary. 
The AG points naturally cluster around the critical energy threshold separating the subcritical and supercritical impacts, forming a narrow, band-like frontier, akin to what is observed in fracture boundary modeling or transition-state search in material mechanics. 
This concentration of points highlights AG’s inherent capacity to exploit the underlying physical transition surface, focusing uncertainty where the interplay between impact energy and material strength is most delicate.

\begin{figure}
	\centering
	\includegraphics[width=0.5\linewidth]{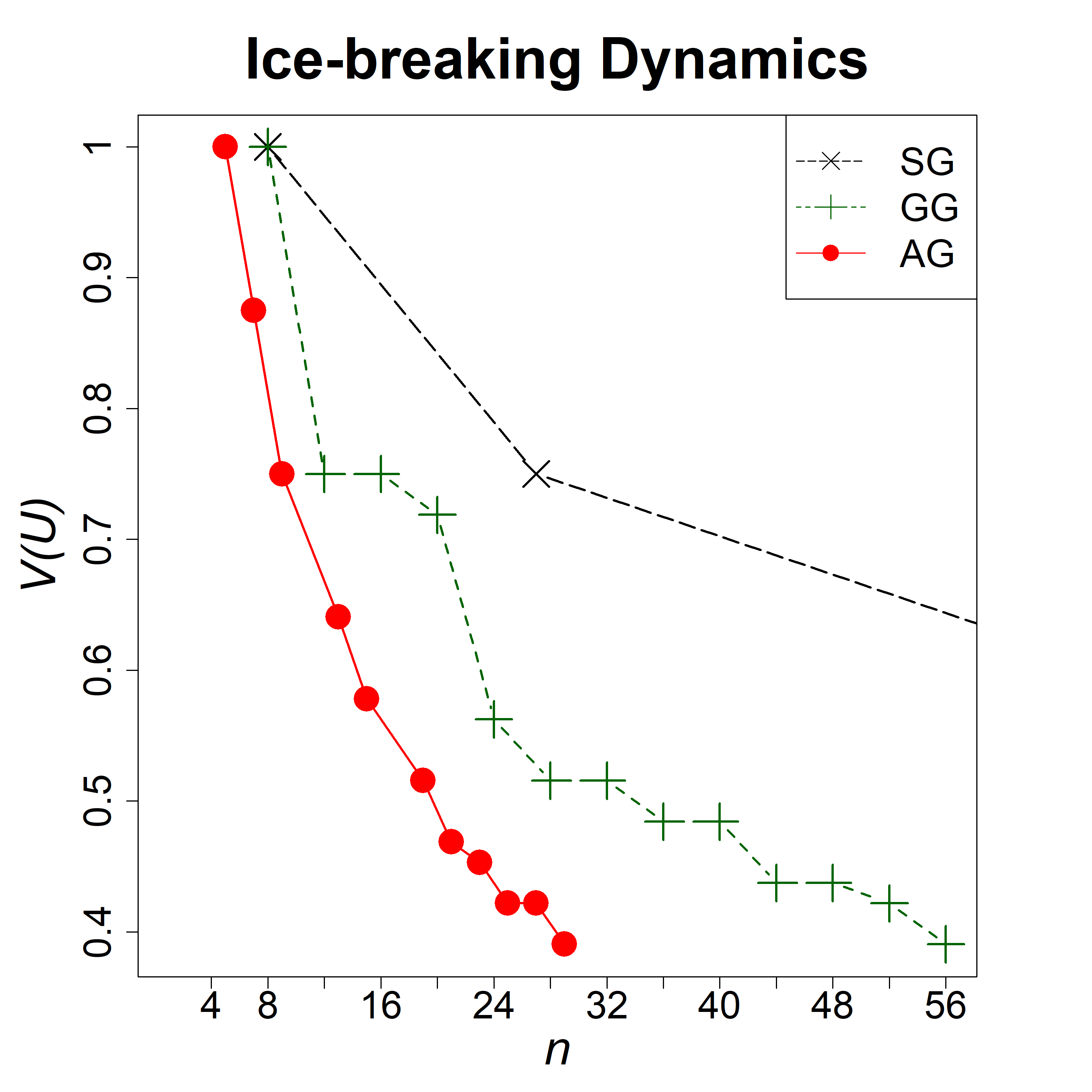}
	\caption{Volume of uncertain area $V(\mathbf{U})$ for the ice-breaking dynamics.}
	\label{fig:ice_v}
\end{figure}

With all outcome information collected for $\mathbf{D}_{\text{SG},3,125}$, we can tell the uncertain area of the $\mathbf{D}_{\text{SG},3,8}$, the $\mathbf{D}_{\text{SG},3,27}$, the $\mathbf{D}_{\text{SG},3,125}$, as well as the $\mathbf{D}_{\text{GG},3,n}$ for $1\leq n \leq 56$. 
For instance, the uncertainty area for the $\mathbf{D}_{\text{AG},3,29}$ is the same to that of the $\mathbf{D}_{\text{SG},3,125}$ and the $\mathbf{D}_{\text{GG},3,56}$. 
Fig.~\ref{fig:ice_v} shows the $V(\mathbf{U})$ for the three types of designs. 
From the results, using the AG instead of the SG or GG allows for skipping a significant proportion of simulation runs.
To be specific, evaluating the $\mathbf{D}_{\text{AG},3,29}$ requires approximately 12 days of simulation, whereas utilizing the $\mathbf{D}_{\text{SG},3,125}$ would demand nearly 2 months.

\section{Conclusions and Discussion}\label{sec:conc}

Motivated by the ice-breaking simulation problem, in this paper, we propose a novel class of adaptive designs called the adaptive grid design. We provide asymptotic theoretical results on this and several other types of designs for classifying monotonic binary deterministic computer simulations. 
We gauge design efficiency based on the required number of functional evaluations to ensure that the volume of the uncertain area remains below a prespecified value $v$.
Our findings reveal that for a $p$-dimensional problem with $p\geq 2$, any static design necessitates at least a constant multiplying $v^{-p}$ simulation trials.
In contrast, the adaptive grid design only requires a constant multiplying $v^{-(p-1)}$ simulation trials, matching the efficiency of the best possible adaptive designs.
This underscores the considerable advantage of the adaptive grid design over static designs.
Our numerical comparisons further demonstrate the superiority of the adaptive grid design over the adaptive Monte Carlo design for $2\leq p\leq 6$ and over the active learning methods when $2\leq p\leq 4$, highlighting the gird structure's strong ability to skip computer runs in low-dimensional problems.

The adaptive grid design, the adaptive Monte Carlo design, and the active learning with entropy method all require simulation runs to be carried out one after another. 
However, in some practical scenarios, there may be a preference for parallel execution of multiple runs.
It is thus a compelling research problem to adapt adaptive design methods to facilitate parallel computing. 
Finally, our study assumes deterministic, binary, and monotonic outputs.
We plan to investigate on situations in which one or two of these assumptions are violated.

\section*{Fundings}
Xu He was supported by National Key R\&D Program of China 2021YFA 1000300, 2021YFA 1000301. 
Dianpeng Wang was supported by National Natural Science Foundation of China, Grant/Award Numbers:12171033.

\section*{Conflict of Interest}
The authors report there are no competing interests to declare.

\appendix

\section*{Appendix}
In this appendix, we provide additional design construction methods as well as theoretical properties and proofs of theorems.

\section{Support Vector Classification}\label{sec:SVC}

In this section, we briefly review the support vector classification method~\cite{Corinna1995} that we use in Section~\ref{sec:comp}-\ref{sec:ice} of the paper to predict outcomes. 
Based on outcomes corresponding to the design ${\mathbf{D}}$, the prediction at site ${\mathbf{x}}$ from the SVC method with the Gaussian kernel and the tuning parameter $\gamma>0$ is 
\[\hat{f}(\mathbf{x})=\text{sgn}\left[\sum_{\mathbf{y} \in \mathbf{D}} \left\{\alpha(\mathbf{y}) f(\mathbf{y}) \exp(-\gamma\|\mathbf{x}-\mathbf{y}\|^2)\right\} +b\right],\]
where $\text{sgn}(z)$ gives the sign of $z$, 
\begin{align*}
	b = \sum_{\mathbf{x}\in \mathbf{D}, \alpha(\mathbf{x})>0} [ f(\mathbf{x}) &- \sum_{\mathbf{y} \in \mathbf{D}, \alpha(\mathbf{y})>0} \{ \alpha(\mathbf{y}) f(\mathbf{y}) \exp(-\gamma\|\mathbf{x}-\mathbf{y}\|^2) \} ] \\
	&/ {\text{card}(\{\mathbf{y} \in \mathbf{D}, \alpha(\mathbf{y})>0\})},
\end{align*}
$\text{card}({\mathbf{S}})$ denotes the cardinality of the set ${\mathbf{S}}$, 
and the values $\{ \alpha(\mathbf{y}) : \mathbf{y} \in \mathbf{D}\} $ are chosen to minimize 
\( \sum_{\mathbf{x},\mathbf{y}\in \mathbf{D}} \{ \alpha(\mathbf{x})\alpha(\mathbf{y}) f(\mathbf{x})f(\mathbf{y}) \exp(-\gamma\|\mathbf{x}-\mathbf{y}\|^2) /2 \} -\sum_{\mathbf{y}\in \mathbf{D}} \alpha(\mathbf{y}), \) 
subjecting to $\sum_{\mathbf{y} \in \mathbf{D}} \{ \alpha(\mathbf{y}) f(\mathbf{y}) \} = 0$ and $\alpha(\mathbf{y}) \geq 0$ for all $\mathbf{y}\in \mathbf{D}$.  
When the data set contains at least 5 negative observations as well as 5 positive observations, we select the tuning parameter $\gamma$ that minimizes the 5-fold cross-validated prediction error. 
Otherwise, we simply assign all predictions to be the majority class among observations. 

\section{Further Constructions and Properties of Designs}
In this section, we provide further constructions and properties of various types of designs discussed in the paper, including the MC, SI, AMC~\cite{DEROCQUIGNY2009363}, GI, and ALE~\cite{Lewis1994}.

\subsection{Static Designs}
We begin by introducing several static designs, starting with the simple Monte Carlo (MC) design, denoted by $\mathbf{D}_{\text{MC},p,n}$, in which the points are independently and uniformly generated from $[0,1]^p$. 
Let $\tilde f$ denote the function that yields $-1$ if and only if $\sum_{k=1}^p x_k < p/2$, $\Gamma(\cdot)$ denote the Gamma function, and 
\begin{equation}\label{eqn:g} 
	g_p = \sum_{0\leq k< p/2} \left\{ (-1)^{k} \binom{p}{k} (p/2-k)^{p-1} \right\}. 
\end{equation}
Theorem~\ref{thm:MC} below gives the efficiency of the $\mathbf{D}_{\text{MC},p,n}$. 

\begin{theorem}\label{thm:MC}
	For the simple Monte Carlo design $\mathbf{D}_{\text{MC},p,n}$, 
	\begin{align}\label{eqn:MC:1}
		\text{sup}_{f \in \mathbf{\Omega}} \text{E} \left[ V\left\{\mathbf{U}\left(\mathbf{D}_{\text{MC},p,n},f\right)\right\} \right]  = \left(2 - 2^{-n}\right) /\left(n+1\right)
	\end{align}
	when $p=1$ and 
	\begin{align*}
		\lim_{n\to \infty} \left\{ \text{E} \left[ V\left\{\mathbf{U}\left(\mathbf{D}_{\text{MC},p,n},\tilde f\right)\right\} \right] / n^{-1/p} \right \} = 2{p!}^{1/p-1}\Gamma\left(1/p\right)g_p  
	\end{align*}
	when $p\geq 2$. 
\end{theorem} 

As indicated by Theorem~\ref{thm:MC}, for $p=1$ and large $n$, it requires approximately $2/v$ MC runs to ensure that $V(\mathbf{U}) \leq v$.  
For $p\geq 2$ and large $n$, it requires at least approximately $\{2p!^{1/p-1}\Gamma(1/p)g_p\}^p v^{-p}$ MC runs to ensure that $V(\mathbf{U}) \leq v$. 
We have demonstrated in the proof that for $p=1$, the function $\tilde f$ is one of the functions that necessitates the most expected simulation runs.
We conjecture that the same applies to the $p\geq 2$ cases. 
If our conjecture holds, then 
\[ \lim_{n\to \infty} \left\{ \sup_{f \in \mathbf{\Omega}} \text{E} \left[ V\left\{\mathbf{U}\left(\mathbf{D}_{\text{MC},p,n},f\right)\right\} \right] / n^{-1/p} \right \} = 2{p!}^{1/p-1}\Gamma\left(1/p\right)g_p  \]
for $p \geq 2$. 
However, if our conjecture is incorrect, the limit may be higher.

A design similar to the SG is the static inner grid design (SI) given by 
\[ {\mathbf{D}_{\text{SI},p,n} }=  \left\{1/\left(n^{1/p}+1\right),2/\left(n^{1/p}+1\right),\ldots,n^{1/p}/\left(n^{1/p}+1\right) \right\}^p \]
when $n^{1/p}$ is a positive integer. 
The $\mathbf{D}_{\text{SI},p,m^p}$ can be seen as the $\mathbf{D}_{\text{SG},p,(m+2)^p}$ without the boundary points.  
Theorem~\ref{thm:SI} below provides the efficiency of the $\mathbf{D}_{\text{SI},p,n}$.

\begin{theorem}\label{thm:SI}
	For the static inner grid design $\mathbf{D}_{\text{SI},p,n}$ with $n^{1/p}$ being a positive integer, 
	\begin{align}\label{eqn:SI}
		V\left\{\mathbf{U}\left(\mathbf{D}_{\text{SI},p,n},f\right)\right\} = 1 - n/\left(n^{1/p}+1\right)^{p}. 
	\end{align}
\end{theorem} 

As indicated by Theorem~\ref{thm:SI}, the $V\{\mathbf{U}(\mathbf{D}_{\text{SI},p,n},f)\}$ is entirely uncorrelated with $f(\cdot)$.
For large $n$, the $V\{\mathbf{U}(\mathbf{D}_{\text{SI},p,n},f)\}$ converges to approximately $p n^{-1/p}$. 
Considering the fact that the $V\{\mathbf{U}(\mathbf{D}_{\text{SG},p,n},f)\}$ also converges to approximately $p n^{-1/p}$ for the most unfavorable $f(\cdot)$, 
the SG may surpass the SI for most functions when $n$ is large. 
However, for small $n$, the $V\{\mathbf{U}(\mathbf{D}_{\text{SI},p,n},f)\}$ is substantially smaller than the $\text{sup}_{f \in \mathbf{\Omega}} V\{\mathbf{U}(\mathbf{D}_{\text{SG},p,n},f)\}$. 
Specifically, the $V\{\mathbf{U}(\mathbf{D}_{\text{SG},p,2^p})\}$ remains at one after conducting $2^p$ functional evaluations,
while the $V\{\mathbf{U}(\mathbf{D}_{\text{SI},p,1})\}=1-2^{-p}<1$ despite only one functional evaluation.
This implies that the SI surpasses the SG when $n$ is small. 
Additionally, in scenarios where simulations on the boundary of the input space are infeasible, SI becomes particularly advantageous.

While the $V(\mathbf{U})$ is of the order $n^{-1/p}$ for all discussed static designs, the constant varies.
To compare the efficiencies of the SG, MC, and SI, we summarize results obtained in Theorems~\ref{thm:SG},~\ref{thm:MC}, and~\ref{thm:SI} for $1\leq p\leq 6$ in Table \ref{tab:coeff}. 
From the results, we observe that for all MC, SG, and SI, the $V(\mathbf{U})$ is of the order $n^{-1/p}$. 
However, the SG and SI outperform the MC in the constant for $p\leq 2$. 
Conversely, for $p\geq 3$, the MC exhibits better performance at least for classifying $\tilde f$. 
Furthermore, when $p=1$, from comparing Theorem~\ref{thm:static} and Theorem~\ref{thm:SI}, we are certain that the SI is optimal and the SG is nearly optimal. 
We believe the bound provided in Theorem~\ref{thm:static} is not tight for $p\geq 2$. 
Therefore, although none of the aforementioned static designs achieves the lower bound for $p\geq 2$, 
it remains unclear if there exists a static design superior to those discussed.
However, even if such a design exists, the improvement is unlikely to be dramatic, as the $V(\mathbf{U})$ for no static design can surpass the rate of $n^{-1/p}$.

\begin{table}
	\centering
	\caption{Volume of uncertainty area for static designs in 1 to 6 dimensions. }
	\vspace{10pt}
	\resizebox{\textwidth}{!}{
		\begin{tabular}{c|ccc}
			\hline
			$p$   & 1    & 2    & 3      \\ \hline
			$ \text{E} [ V\{\mathbf{U}(\mathbf{D}_{\text{MC},p,n},\tilde f)\} ] $ & $2n^{-1}+o(n^{-1})$ & $2.51n^{-1/2}+o(n^{-1/2})$ & $2.43n^{-1/3}+o(n^{-1/3})$ \\ 
			$\text{sup}_{f \in \mathbf{\Omega}} V\{\mathbf{U}(\mathbf{D}_{\text{SG},p,n},f)\} $             & $n^{-1}+o(n^{-1})$    & $2n^{-1/2}+o(n^{-1/2})$    & $3n^{-1/3}+o(n^{-1/3})$   \\ 
			$V\{\mathbf{U}(\mathbf{D}_{\text{SI},p,n},f)\} $             & $n^{-1}+o(n^{-1})$    & $2n^{-1/2}+o(n^{-1/2})$    & $3n^{-1/3}+o(n^{-1/3})$   \\ \hline
			$p$  & 4    & 5    & 6    \\ \hline
			$ \text{E} [ V\{\mathbf{U}(\mathbf{D}_{\text{MC},p,n},\tilde f)\} ] $ & $2.67n^{-1/4}+o(n^{-1/4})$ & $2.87n^{-1/5}+o(n^{-1/5})$ & $3.06n^{-1/6}+o(n^{-1/6})$  \\
			$\text{sup}_{f \in \mathbf{\Omega}} V\{\mathbf{U}(\mathbf{D}_{\text{SG},p,n},f)\}$  & $4n^{-1/4}+o(n^{-1/4})$    & $5n^{-1/5}+o(n^{-1/5})$               & $6n^{-1/6}+o(n^{-1/6})$      \\ 
			$V\{\mathbf{U}(\mathbf{D}_{\text{SI},p,n},f)\}$  & $4n^{-1/4}+o(n^{-1/4})$    & $5n^{-1/5}+o(n^{-1/5})$               & $6n^{-1/6}+o(n^{-1/6})$      \\ \hline
		\end{tabular}
	}
	\label{tab:coeff}
\end{table}

\begin{figure}
	\centering
	\subfigure[$\mathbf{D}_{\text{LHD},2,81}$, $v=0.248$.]{
		\begin{minipage}{0.4\linewidth}
			\centering
			\includegraphics[width=\linewidth]{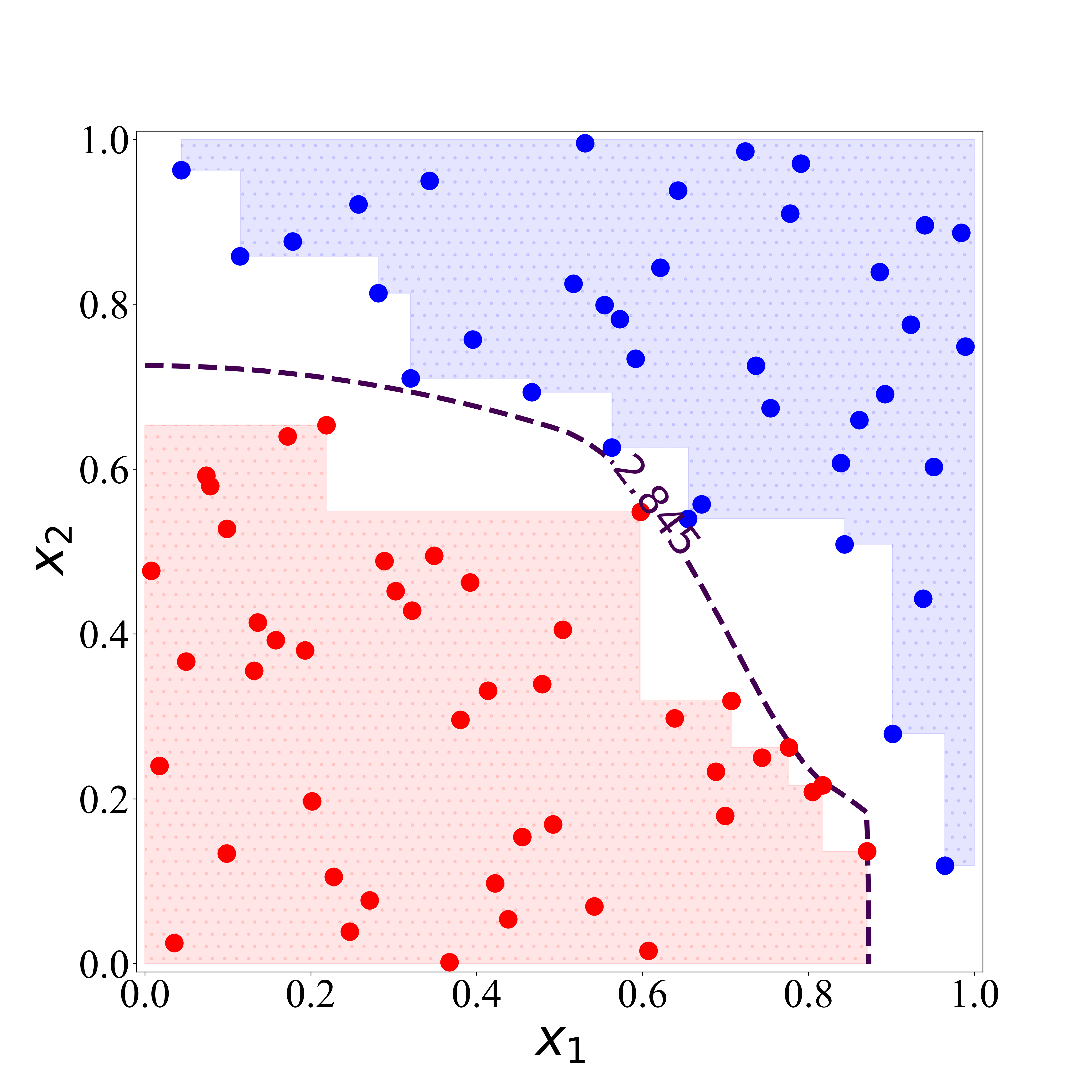}
			\label{illu_LHD}
		\end{minipage}
	}
	\subfigure[$\mathbf{D}_{\text{MC},2,81}$, $v=0.247$.]{
		\begin{minipage}{0.4\linewidth}
			\centering
			\includegraphics[width=\linewidth]{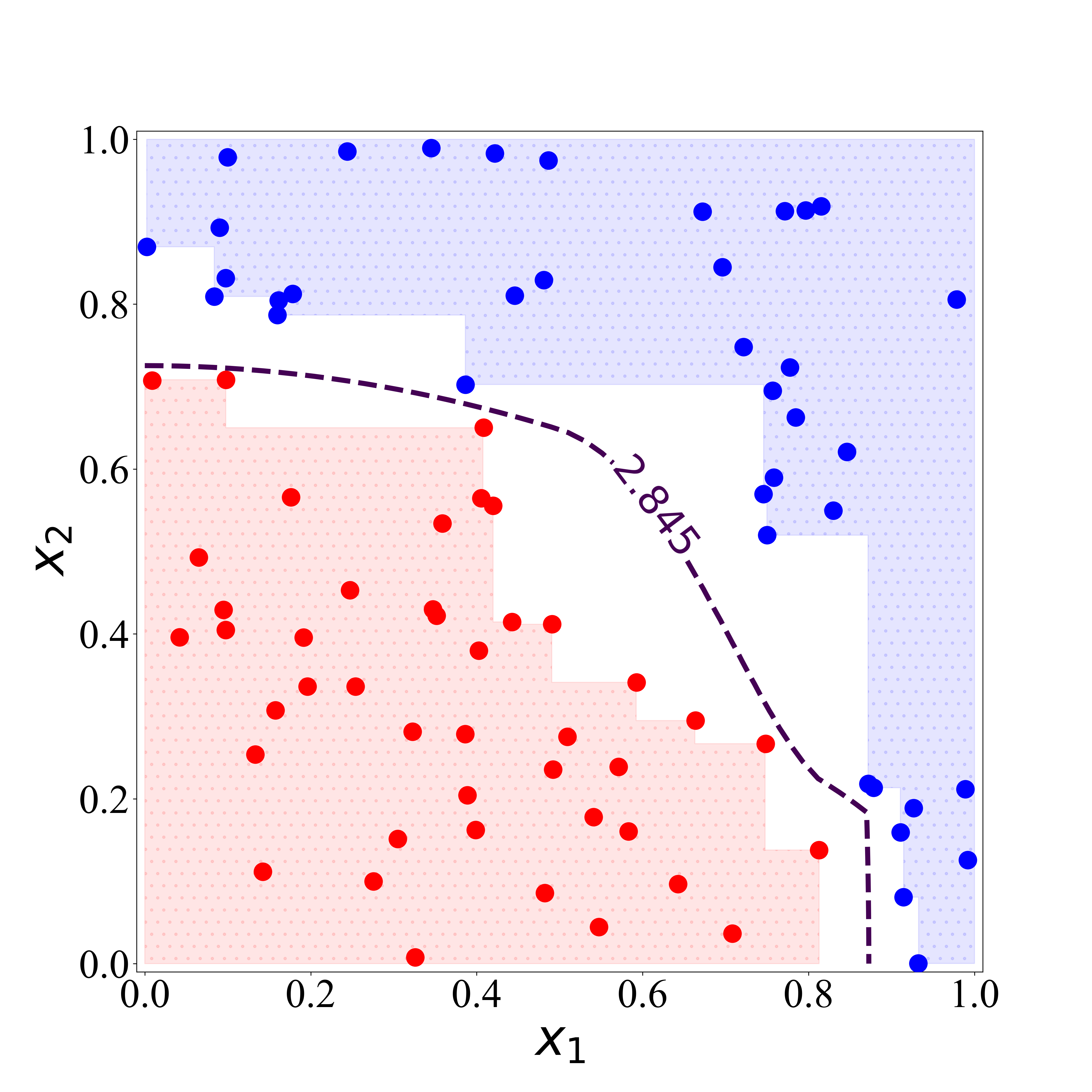}
			\label{illu_MC}
	\end{minipage}}
	\qquad
	\subfigure[$\mathbf{D}_{\text{SG},2,81}$, $v=0.188$.]{
		\begin{minipage}{0.4\linewidth}
			\centering
			\includegraphics[width=\linewidth]{illustrate/fgd.png}
			\label{illu_SG}
		\end{minipage}
	}
	\subfigure[$\mathbf{D}_{\text{SI},2,81}$, $v=0.190$.]{
		\begin{minipage}{0.4\linewidth}
			\centering
			\includegraphics[width=\linewidth]{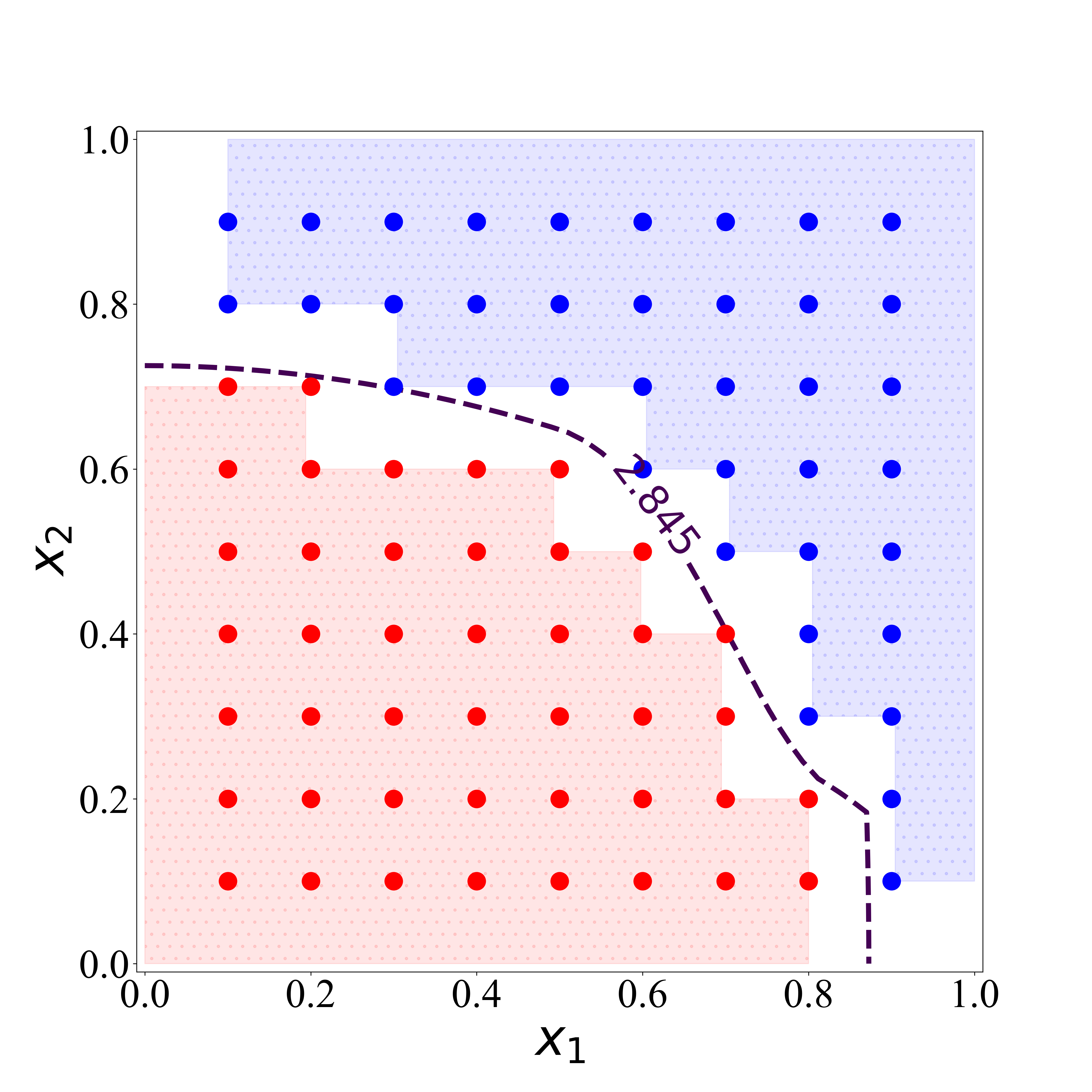}
			\label{illu_SI}
		\end{minipage}
	}
	\caption{Four static designs, showing the design points with negative response (red dots, or dots to the bottom-left of the dotted line) and positive response (blue dots, or dots to the top-right of the dotted line), the curve separating the two regions $\mathbf{A}$ and $\mathbf{B}$ (purple dotted line), the certainly negative area 
		(red shaded area, or the bottom-left shaded area), the certainly positive area 
		(blue shaded area, or the top-right shaded area), and the uncertain area (white area) after evaluating the runs corresponding to the designs. }
	\label{fig:illu:static}
\end{figure}

For better understanding, we illustrate four static designs, $\mathbf{D}_{\text{MC},2,81}$, $\mathbf{D}_{\text{LHD},2,81}$,
$\mathbf{D}_{\text{SG},2,81}$, and $\mathbf{D}_{\text{SI},2,81}$ in Fig.~\ref{fig:illu:static}, where $\mathbf{D}_{\text{LHD},2,81}$ denotes the two-dimensional Latin hypercube design (LHD)~\cite{Mckay1979} with 81 points. 
In the figure, alongside the design points, we delineate the certainly negative area $\cup_{\mathbf{x} \in \mathbf{D}, f(\mathbf{x})=-1} \prod_{k=1}^p [0,x_k]$, the certainly positive area $\cup_{\mathbf{x} \in \mathbf{D}, f(\mathbf{x})=1} \prod_{k=1}^p [x_k,1]$, and the uncertain area $\mathbf{U}$ for classifying the state function~\eqref{eq:illustrate} of the paper. 
We observe that $V\{\mathbf{U}(\mathbf{D}_{\text{MC},2,81},f)\} = 0.247$, $V\{\mathbf{U}(\mathbf{D}_{\text{LHD},2,81},f)\} = 0.248$, $V\{\mathbf{U}(\mathbf{D}_{\text{SG},2,81},f)\} = 0.188$, and $V\{\mathbf{U}(\mathbf{D}_{\text{SI},2,81},f)\}= 0.190$. 
This is consistent to our theoretical results that the SG and the SI are substantially better than the MC when $p=2$. 
Notably the $\mathbf{D}_{\text{LHD},2,81}$ is no better than the $\mathbf{D}_{\text{MC},2,81}$ in this case. 
From numerical results we will not show, in general the LHD performs nearly as well as the MC.

\subsection{Adaptive Designs}\label{sec:S1.2}

Next, we show the construction and properties of adaptive designs. 
We begin with the adaptive Monte Carlo design (AMC) summarized in \cite{DEROCQUIGNY2009363}, whose construction is outlined in Algorithm~\ref{alg:AMC}.

\begin{algorithm}
	\label{alg:AMC}
	\DontPrintSemicolon
	\KwInput{Dimension $p$, simulation function $f(\cdot)$, and number of runs $n$}
	Initialize $\mathbf{D} \leftarrow \emptyset$ and $\mathbf{U}(\mathbf{D},f) \leftarrow \emptyset$ \;
	\While{$\text{card}(\mathbf{D})<n$}
	{
		Generate $\mathbf{x}$ from the uniform distribution on $[0,1]^p$ \;
		\While{$\mathbf{x} \notin \mathbf{U}(\mathbf{D},f)$} { Generate $\mathbf{x}$ from the uniform distribution on $[0,1]^p$ }
		Evaluate $f(\mathbf{x})$, let $\mathbf{D}\leftarrow \mathbf{D}\cup\{\mathbf{x}\}$, and update $\mathbf{U}(\mathbf{D},f)$ \;
	}
	Output the adaptive Monte Carlo design $\mathbf{D}_{\text{AMC},p,n} \leftarrow \mathbf{D}$
	\caption{Steps of the adaptive Monte Carlo method}
\end{algorithm}

From the AMC method, only the runs within the uncertain area are conducted, while the others are skipped since their outputs are already known. 
Let $m_{\text{AMC}}(n)$ denote the number of functional evaluations in Step~6 when the number of $\mathbf{x}$'s tried in Steps~3 and~5 is $n$.  
Clearly, from the $m_{\text{AMC}}(n)$ functional evaluations of the AMC we obtain exactly the same information as from the $n$ runs of the MC. 
On the other hand, we expect the $m_{\text{AMC}}(n)$ to be considerably lower than the $n$, especially for large $n$. This is because as the $V(\mathbf{U})$ approaches to zero, the majority of $\mathbf{x}$'s tried in Steps~3 and~5 will not fall within $\mathbf{U}(\mathbf{D},f)$, leading to their exclusion.
Let $\gamma$ denote the Euler's constant. 
To quantify the improvement on efficiency, we establish Theorem~\ref{thm:AMC} below to connect $E\{ m_{\text{AMC}}(n) \}$ and $n$. 

\begin{theorem}\label{thm:AMC}
	For the adaptive Monte Carlo design $\mathbf{D}_{\text{AMC},p,m_{\text{AMC}}(n)}$ in classifying $\tilde f$,
	\begin{align*}
		\lim_{n\to\infty}\left[\text{E} \left\{m_{\text{AMC}}\left(n\right)\right\}-2\ln n\right]= 2\left(\gamma-\ln2\right)
	\end{align*}
	when $p=1$ and
	\begin{equation*}
		\lim_{n\to \infty} \left[ \text{E}\left\{ m_{\text{AMC}}\left(n\right) \right\} / n^{\left(p-1\right)/p} \right] = 2{p!}^{1/p-1}\Gamma\left(1/p\right) g_p p \left(p-1\right)^{-1}
	\end{equation*}
	when $p\geq 2$, where $g_p$ is given in \eqref{eqn:g}.
\end{theorem}

Combining Theorems~\ref{thm:MC} and~\ref{thm:AMC}, when $p=1$, for AMC, the $\text{E} [ V \{\mathbf{U} (\allowbreak\mathbf{D}_{\text{AMC},p,n},\tilde f ) \} ]$ is approximately $\exp(\gamma-n/2)$ for large $n$ and thus it requires approximately $2(\gamma-\ln v)$ runs to ensure that $V(\mathbf{U}) \leq v$. 
For $p\geq 2$, $\text{E} [ V \{\mathbf{U} (\mathbf{D}_{\text{AMC},p,n},\tilde f ) \} ]$ is approximately $\{ 2{p!}^{1/p-1}\Gamma(1/p)  g_p \}^{p/(p-1)}\allowbreak \{p/(p-1)\}^{1/(p-1)} n^{-1/(p-1)}$ for large $n$ and thus it requires approximately $\{2{p!}^{1/p-1} \Gamma(1/p) g_p\}^p \allowbreak\{p/(p-1)\}  v^{-(p-1)}$ runs to ensure that $V\{\mathbf{U}(\mathbf{D}_{\text{AMC},p,n},\allowbreak\tilde f)\} \leq v$. 
In both cases, the AMC requires an order of magnitude fewer functional evaluations than the MC, demonstrating its superiority.
Again, we conjecture that the $\tilde f$ is one of the functions that require the most functional evaluations for the AMC. 
If true, $\text{sup}_{f \in \mathbf{\Omega}} \text{E}[ V\{\mathbf{U}(\mathbf{D}_{\text{AMC},p,n},f)\} ] = \text{E}[ V\{\mathbf{U}(\mathbf{D}_{\text{AMC},p,n},\tilde f)\} ]$.

Active learning methods are alternatives to the AMC. 
After trying several approaches, we find that a good strategy is to sequentially supplement an initial design using the point $\mathbf{x}^\star$ that maximizes the entropy~\cite{Lewis1994}, 
\[\mathbf{x}^{\star}=\operatorname{argmax}_{\mathbf{x}\in [0,1]^p }  \left[ -p(\mathbf{x}) \ln p(\mathbf{x}) -\left\{1-p(\mathbf{x}) \right\}\ln \left\{1-p(\mathbf{x}) \right\} \right],\] 
where the probability of $f(\mathbf{x})$ being positive, $p(\mathbf{x})$, can be estimated using the Platt scaling~\cite{platt1999} with the SVC classifier introduced in Section~\ref{sec:SVC}. 
We term the ${D}_{\text{ALE},p,n}$ as the active learning design with entropy (ALE).
However, the lack of monotonicity information utilization makes the ALE potentially suboptimal for our problem. To the best of our knowledge, we have found no active learning method that exploits the monotonic property.

By substituting $\mathbf{D}_{\text{SI},p,(2^{l}-1)^p}$ for $\mathbf{D}_{\text{SG},p,(2^l+1)^p}$ and initializing $l=1$ instead of $l=0$ in Algorithm~\ref{alg:GG} of the paper, we obtain the grouped-adaptive inner grid design (GI), which has similar properties as that for GG. 
Theorem~\ref{thm:GI} below gives the exact value of $m_{\text{GI}}(g)$, the number of functional evaluations when the outcomes corresponding to the $\mathbf{D}_{\text{SI},p,(2^g-1)^p}$ are all known.

\begin{theorem}\label{thm:GI}
	\begin{equation*}
		m_{\text{GI}}\left(g\right) = \sum_{l=1}^g \left\{ (2^l-1)^p - (2^l-2)^p \right\} .
	\end{equation*}
\end{theorem}

As indicated by Theorem~\ref{thm:GI}, when $p=1$, $m_{\text{GI}}(g)=g$ and thus $V\{\mathbf{U}(\mathbf{D}_{\text{GI},p,n},\allowbreak f)\} = 2^{-n}$ and it requires 
$-\log_2 v$ functional evaluations to guarantee that $V\{\mathbf{U}(\allowbreak\mathbf{D}_{\text{GI},p,n},f)\} \leq v$. 
For $p\geq 2$ with large $g$, the $m_{\text{GI}}(g)$ is approximately $p 2^{p-1}(2^{p-1}-1)^{-1} 2^{g(p-1)} $, the same to the upper bound for $m_{\text{GG}}\left(g\right)$. 
Consequently, $V\{\mathbf{U}(\mathbf{D}_{\text{GI},p,n},\allowbreak f)\}$ is approximately $2p^{p/(p-1)}(2^{p-1}-1)^{-1/(p-1)} n^{-1/(p-1)}$ and thus it requires approximately $p^p 2^{p-1}(2^{p-1}-1)^{-1}v^{-(p-1)}$ functional evaluations to guarantee that $V\{\mathbf{U}(\allowbreak\mathbf{D}_{\text{GI},p,n},f)\} \leq v$. 
Conjecturing that the $m_{\text{GG}}(g)$ is considerably smaller than the upper bound for some $f$, we infer that the GI may be slightly inferior to the GG for certain $f$ when $n$ is large. 
Nevertheless, for small $n$, the $m_{\text{GI}}(g)$ is substantially smaller than the upper bound for the $m_{\text{GG}}\left(g\right)$, implying that the conclusion may be reversed.

All of the GG, AMC, and GI require the order of $\log(1/v)$ evaluations when $p=1$ and the order of $v^{-(p-1)}$ evaluations when $p\geq 2$. To compare their efficiencies, we summarize results in Theorems~\ref{thm:GG}, \ref{thm:AMC}, and~\ref{thm:GI} for $1\leq p\leq 6$ in Table \ref{tab:a:coeff}. 
Based on the findings, both the proposed GG and GI outperform the AMC when $p\leq 2$. 
On the other hand, although Theorem~\ref{thm:AMC} only gives a lower bound on the required number of functional evaluations, we conjecture that the AMC is better for $p\geq 3$.

\begin{table}
	\centering
	
	\caption{Volume of uncertainty area for adaptive designs in 1 to 6 dimensions. }
	\vspace{10pt}
	\resizebox{\textwidth}{!}{
		\begin{tabular}{c|ccc}
			\hline
			$p$   & 1    & 2    & 3      \\ \hline
			$ \text{E} [ V\{\mathbf{U}(\mathbf{D}_{\text{AMC},p,n},\tilde f)\} ] = $        & $1.78\exp(-n/2)+o(\exp(-n/2))$ & $12.57n^{-1}+o(n^{-1})$ & $4.65n^{-1/2}+o(n^{-1/2})$ \\ 
			$\text{sup}_{f \in \mathbf{\Omega}} V\{\mathbf{U}(\mathbf{D}_{\text{GG},p,n},f)\} \leq $  & $4\cdot2^{-n}+o(2^{-n})$ & $8n^{-1}+o(n^{-1})$ & $6n^{-1/2}+o(n^{-1/2})$ \\ 
			$V\{\mathbf{U}(\mathbf{D}_{\text{GI},p,n},f)\} = $  & $2^{-n}+o(2^{-n})$ & $8n^{-1}+o(n^{-1})$ & $6n^{-1/2}+o(n^{-1/2})$ \\ \hline$p$  & 4    & 5    & 6    \\ \hline
			$\text{E} [ V\{\mathbf{U}(\mathbf{D}_{\text{AMC},p,n},\tilde f)\} ] = $        & $4.09n^{-1/3}+o(n^{-1/3})$& $3.94n^{-1/4}+o(n^{-1/4})$ & $3.96n^{-1/5}+o(n^{-1/5})$ \\ 
			$\text{sup}_{f \in \mathbf{\Omega}} V\{\mathbf{U}(\mathbf{D}_{\text{GG},p,n},f)\} \leq$   & $6.64n^{-1/3}+o(n^{-1/3})$ & $7.60n^{-1/4}+o(n^{-1/4})$ & $8.64n^{-1/5}+o(n^{-1/5})$ \\ 
			$V\{\mathbf{U}(\mathbf{D}_{\text{GI},p,n},f)\} = $   & $6.64n^{-1/3}+o(n^{-1/3})$ & $7.60n^{-1/4}+o(n^{-1/4})$ & $8.64n^{-1/5}+o(n^{-1/5})$ \\ \hline
		\end{tabular}
	}
	\label{tab:a:coeff}
\end{table}

Similarly, by substituting the $\mathbf{D}_{\text{SI},p,(2^{l}-1)^p}$ for the $\mathbf{D}_{\text{SG},p,(2^l+1)^p}$ and initializing $l=1$ instead of $l=0$ in Algorithm~\ref{alg:AG} of the paper, we obtain the fully adaptive inner grid design (AI). 
We expect the AI to be nearly as good as the AG. 
Finally, it's worth noting that all of the AMC, ALE, AG, and AI require the runs to be carried out one-by-one. 
In contrast, with the GG and GI, runs from the same group can be carried out simultaneously. 
Hence, the GG and GI are more suitable for parallel computing. 

\begin{figure}
	\centering
	\subfigure[$\mathbf{D}_{\text{AMC},2,4}$, $v= 0.790$.]{
		\begin{minipage}[t]{0.25\linewidth}
			\centering
			\includegraphics[width=\linewidth]{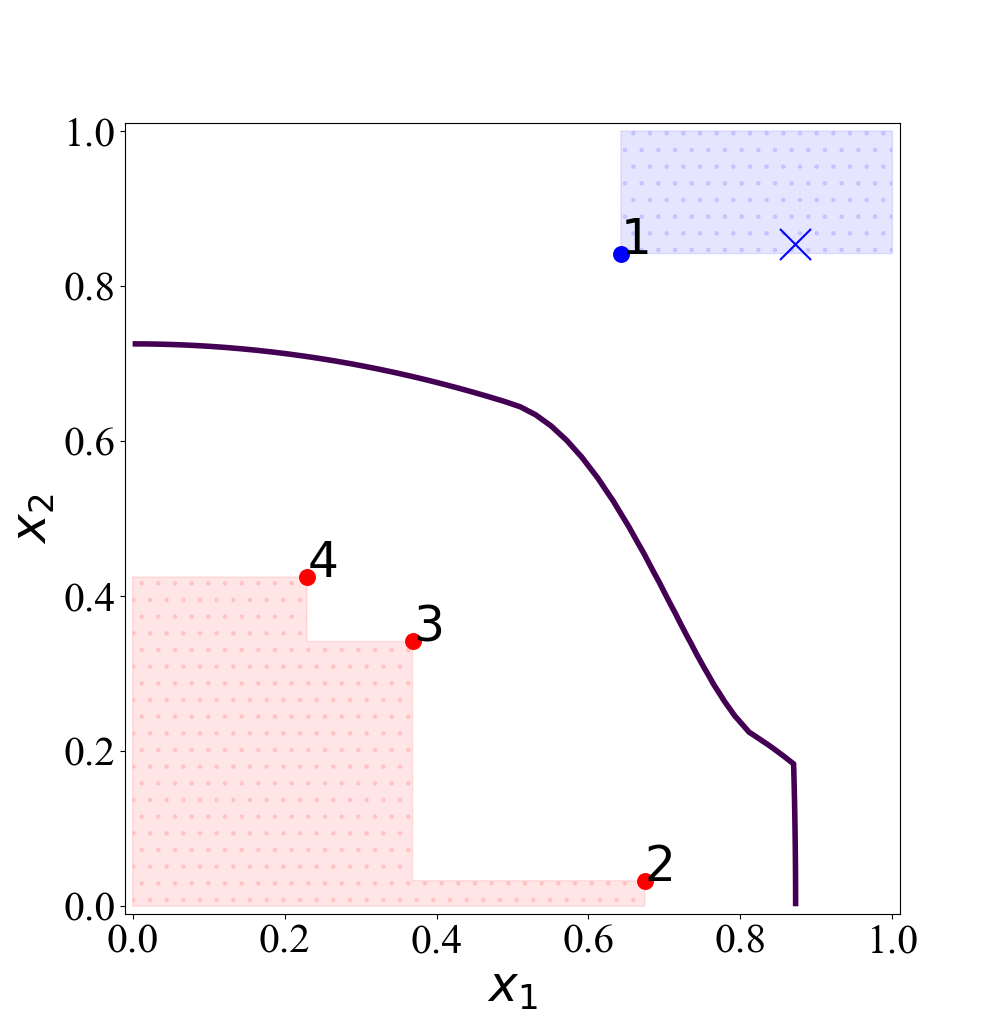}
			
			\label{illu_AMC5}
		\end{minipage}
	}
	\subfigure[$\mathbf{D}_{\text{AMC},2,8}$, $v= 0.657$.]{
		\begin{minipage}[t]{0.25\linewidth}
			\centering
			\includegraphics[width=\linewidth]{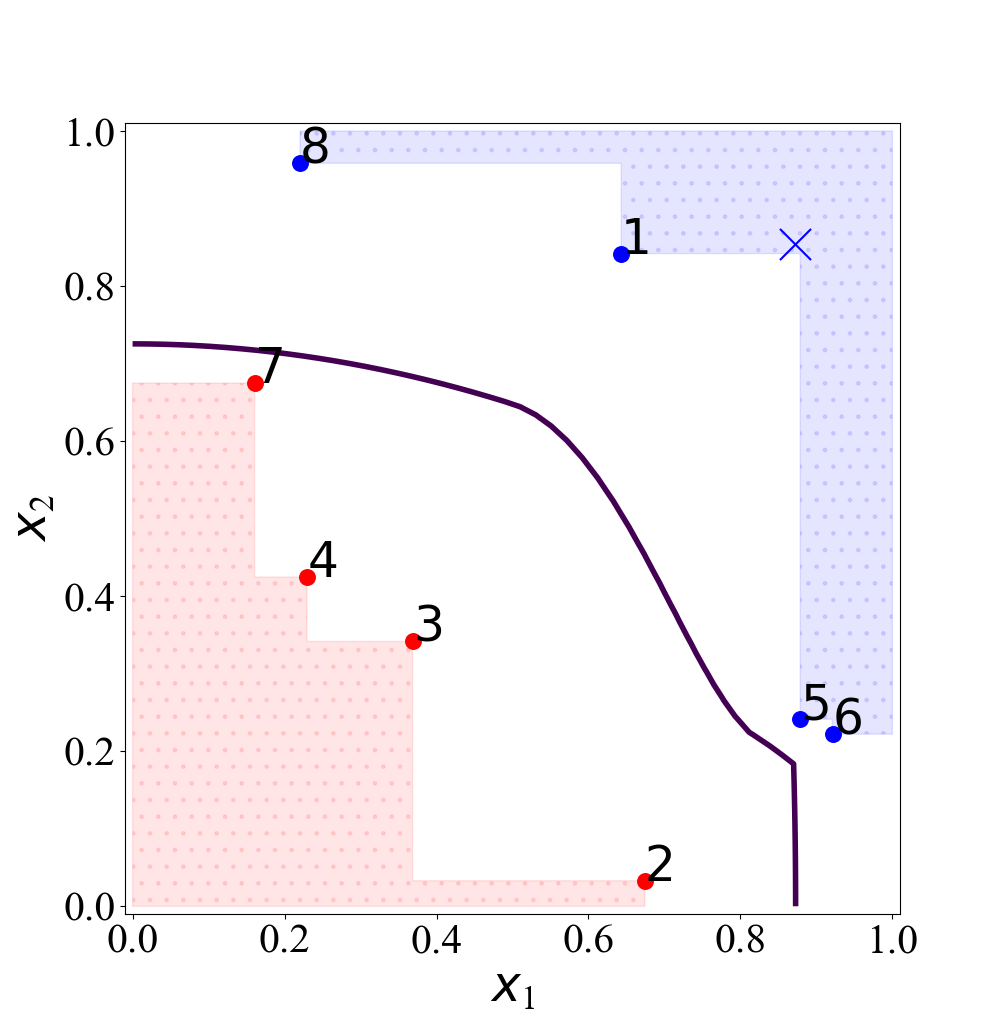}
			
			\label{illu_AMC10}
		\end{minipage}
	}
	\subfigure[$\mathbf{D}_{\text{AMC},2,16}$, $v= 0.363$.]{
		\begin{minipage}[t]{0.25\linewidth}
			\centering
			\includegraphics[width=\linewidth]{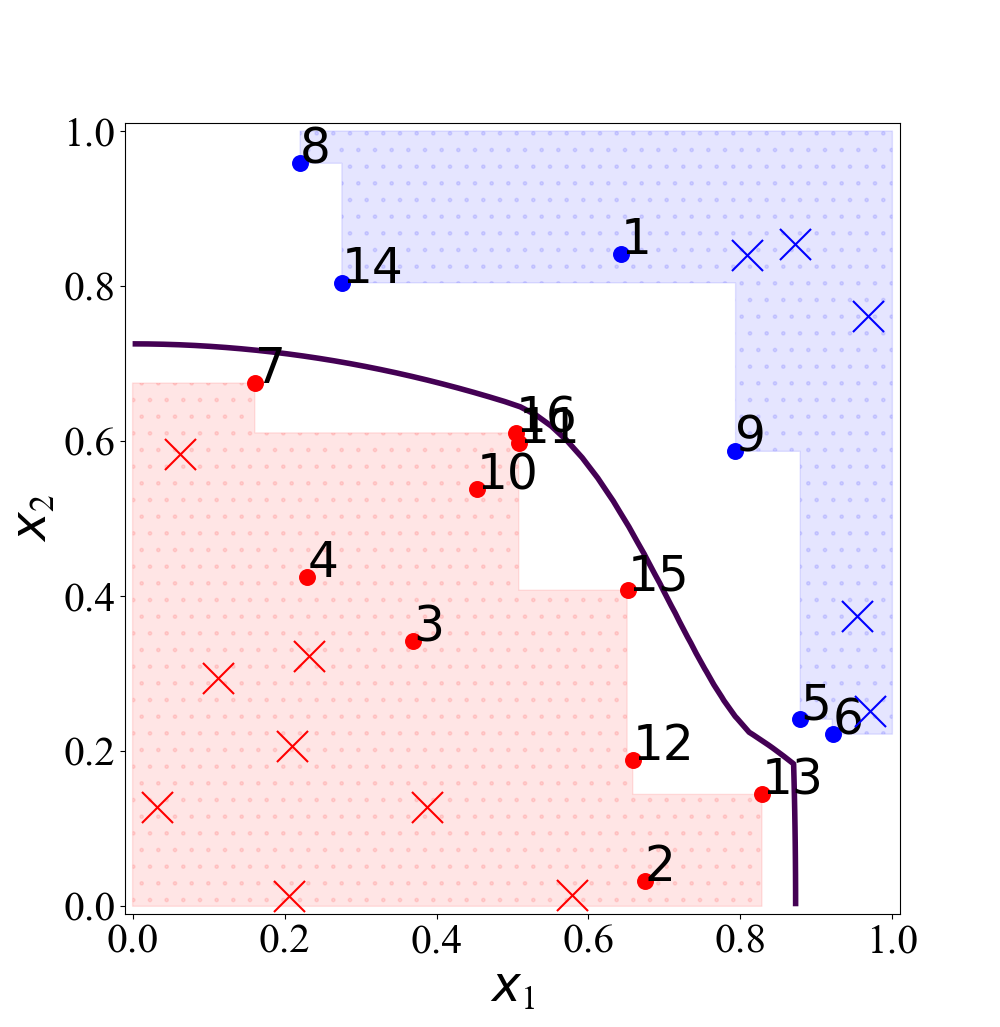}
			
			\label{illu_AMC18}
		\end{minipage}
	}
	
	\subfigure[$\mathbf{D}_{\text{ALE},2,4}$, $v= 0.689$.]{
		\begin{minipage}[t]{0.25\linewidth}
			\centering
			\includegraphics[width=\linewidth]{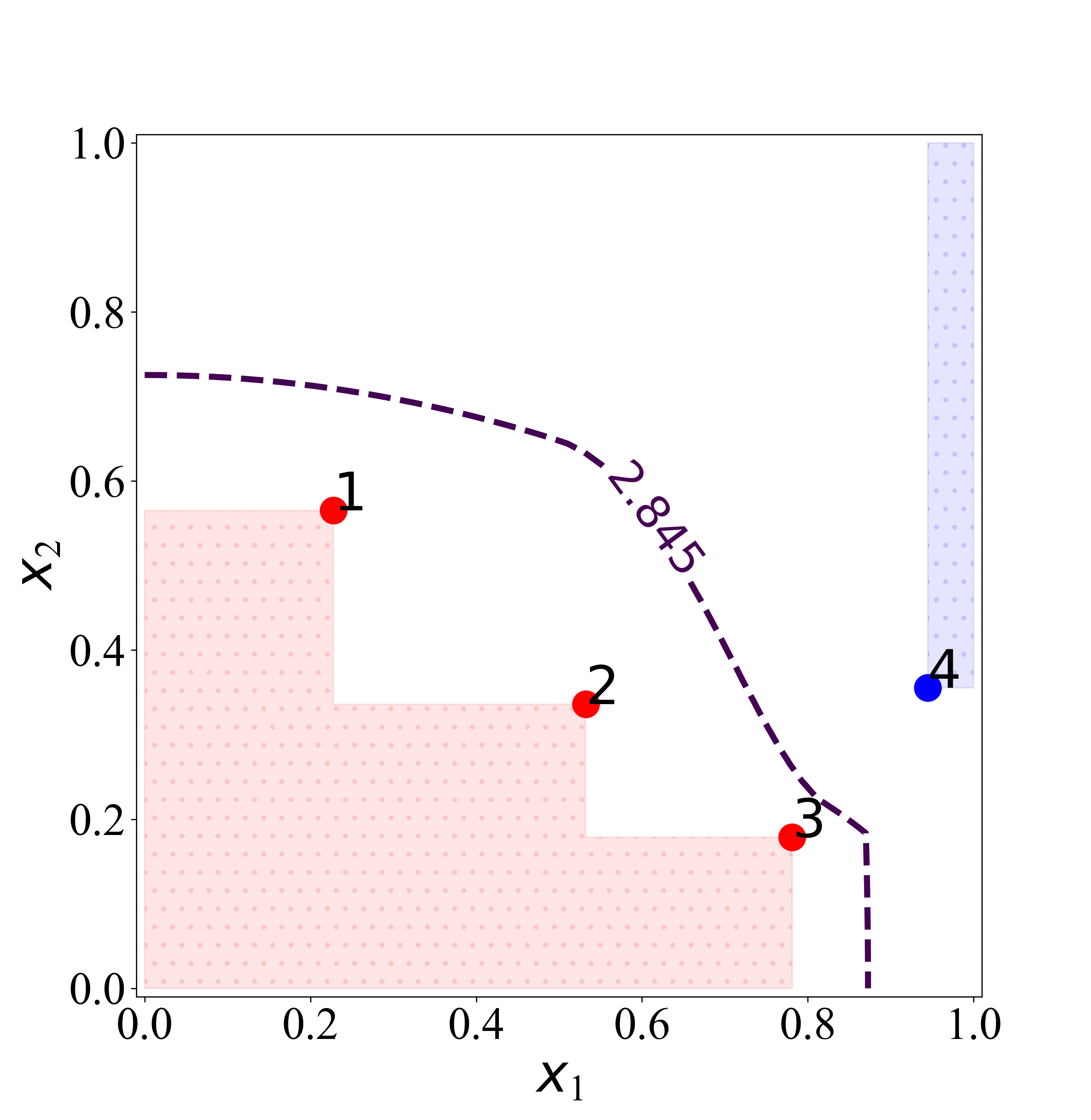}
			
			\label{illu_ALE4}
		\end{minipage}
	}
	\subfigure[$\mathbf{D}_{\text{ALE},2,8}$, $v= 0.648$.]{
		\begin{minipage}[t]{0.25\linewidth}
			\centering
			\includegraphics[width=\linewidth]{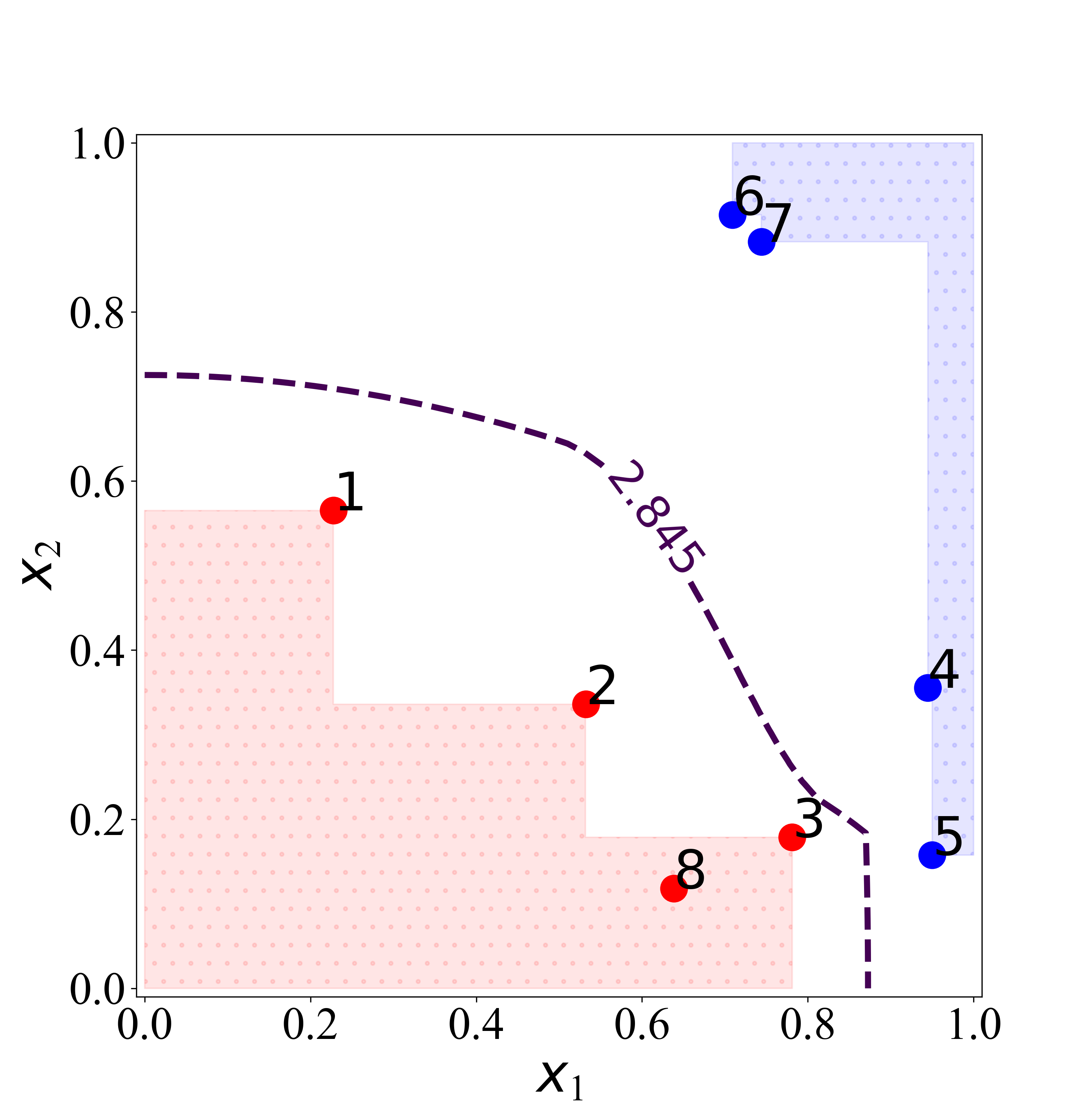}
			
			\label{illu_ALE8}
		\end{minipage}
	}
	\subfigure[$\mathbf{D}_{\text{ALE},2,16}$, $v= 0.397$.]{
		\begin{minipage}[t]{0.25\linewidth}
			\centering
			\includegraphics[width=\linewidth]{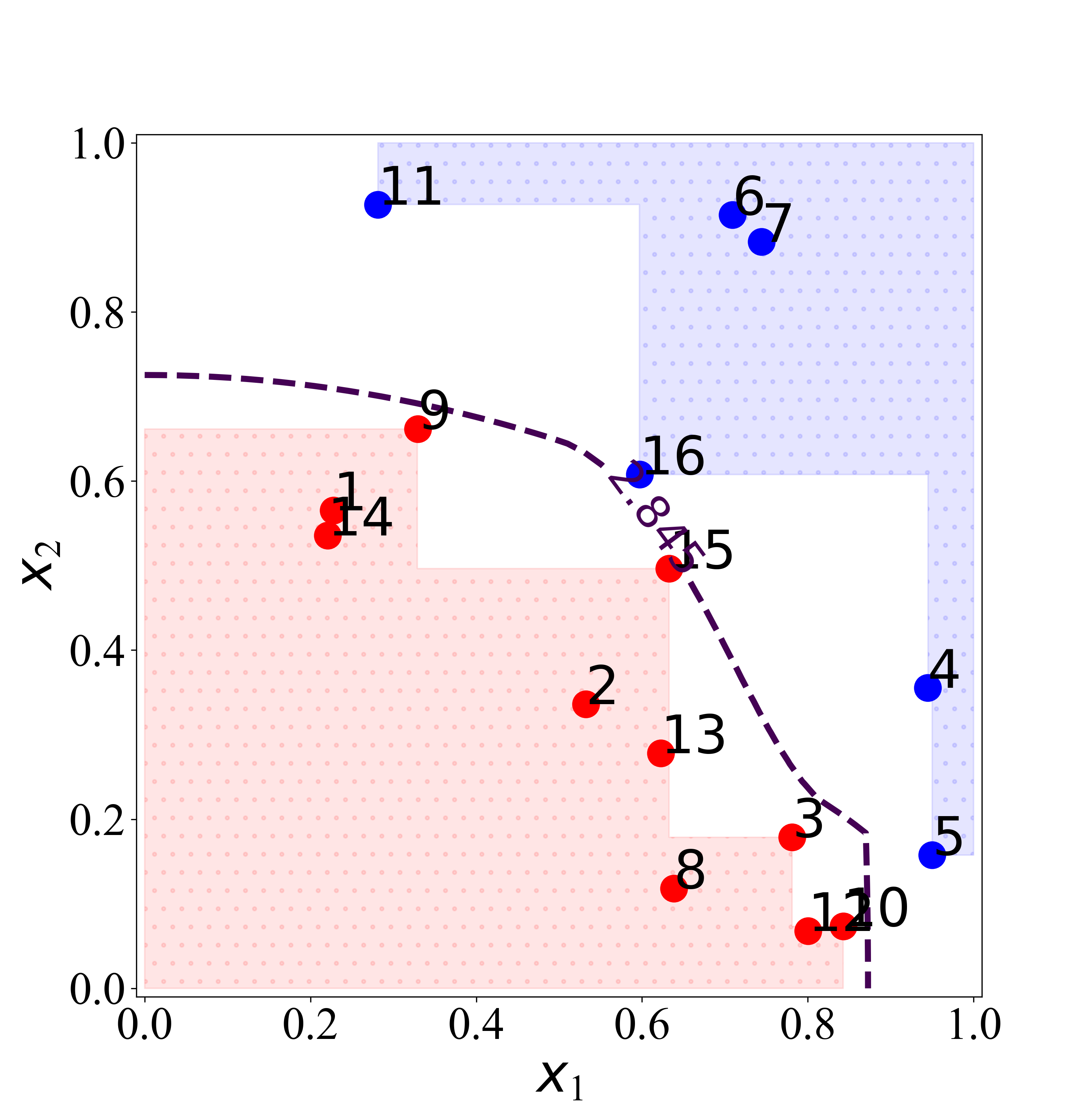}
			
			\label{illu_ALE16}
		\end{minipage}
	}
	
	\subfigure[$\mathbf{D}_{\text{AG},2,4}$, $v= 0.75$.]{
		\begin{minipage}[t]{0.25\linewidth}
			\centering
			\includegraphics[width=\linewidth]{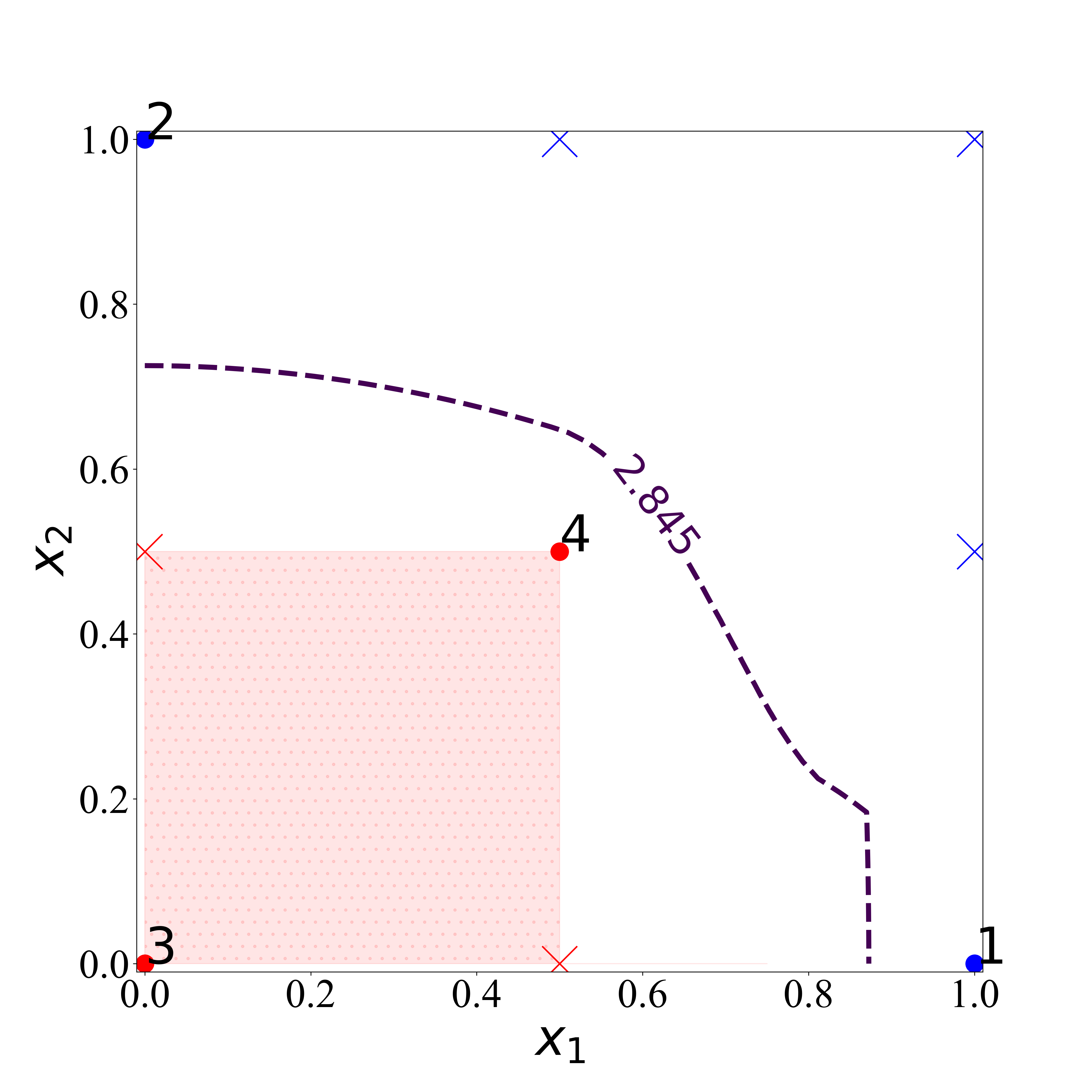}
			
			\label{illu_AG_1}
		\end{minipage}
	}
	\subfigure[$\mathbf{D}_{\text{AG},2,8}$, $v= 0.375$.]{
		\begin{minipage}[t]{0.25\linewidth}
			\centering
			\includegraphics[width=\linewidth]{illustrate/AG_2.png}
			
			\label{illu_AG_2}
		\end{minipage}
	}
	\subfigure[$\mathbf{D}_{\text{AG},2,16}$, $v= 0.188$.]{
		\begin{minipage}[t]{0.25\linewidth}
			\centering
			\includegraphics[width=\linewidth]{illustrate/AG_3.png}
			
			\label{illu_AG_3}
	\end{minipage}}
	
	\subfigure[$\mathbf{D}_{\text{AI},2,1}$, $v= 0.75$.]{
		\begin{minipage}[t]{0.25\linewidth}
			\centering
			\includegraphics[width=\linewidth]{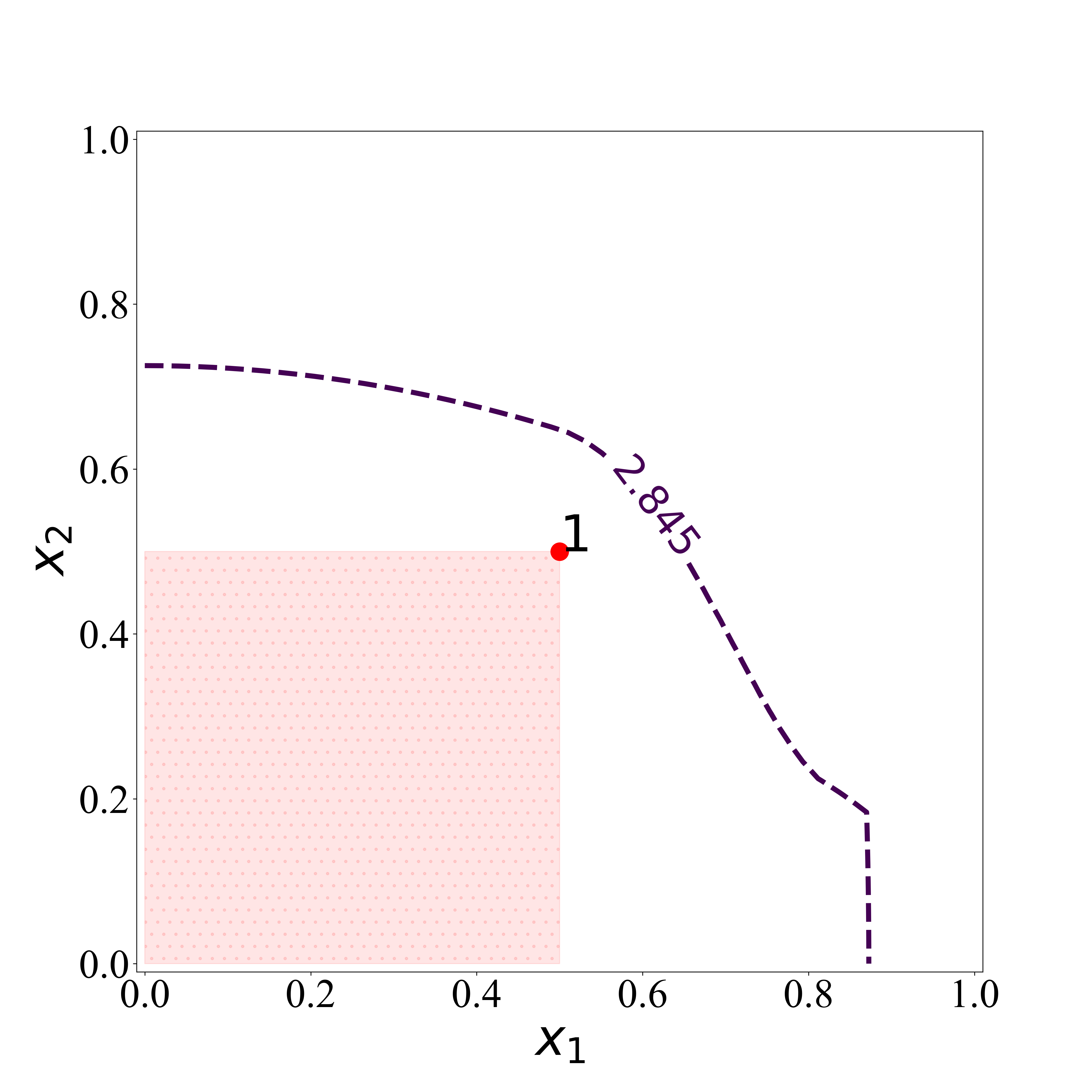}
			
			\label{illu_AI_1}
		\end{minipage}
	}
	\subfigure[$\mathbf{D}_{\text{AI},2,5}$, $v= 0.438$.]{
		\begin{minipage}[t]{0.25\linewidth}
			\centering
			\includegraphics[width=\linewidth]{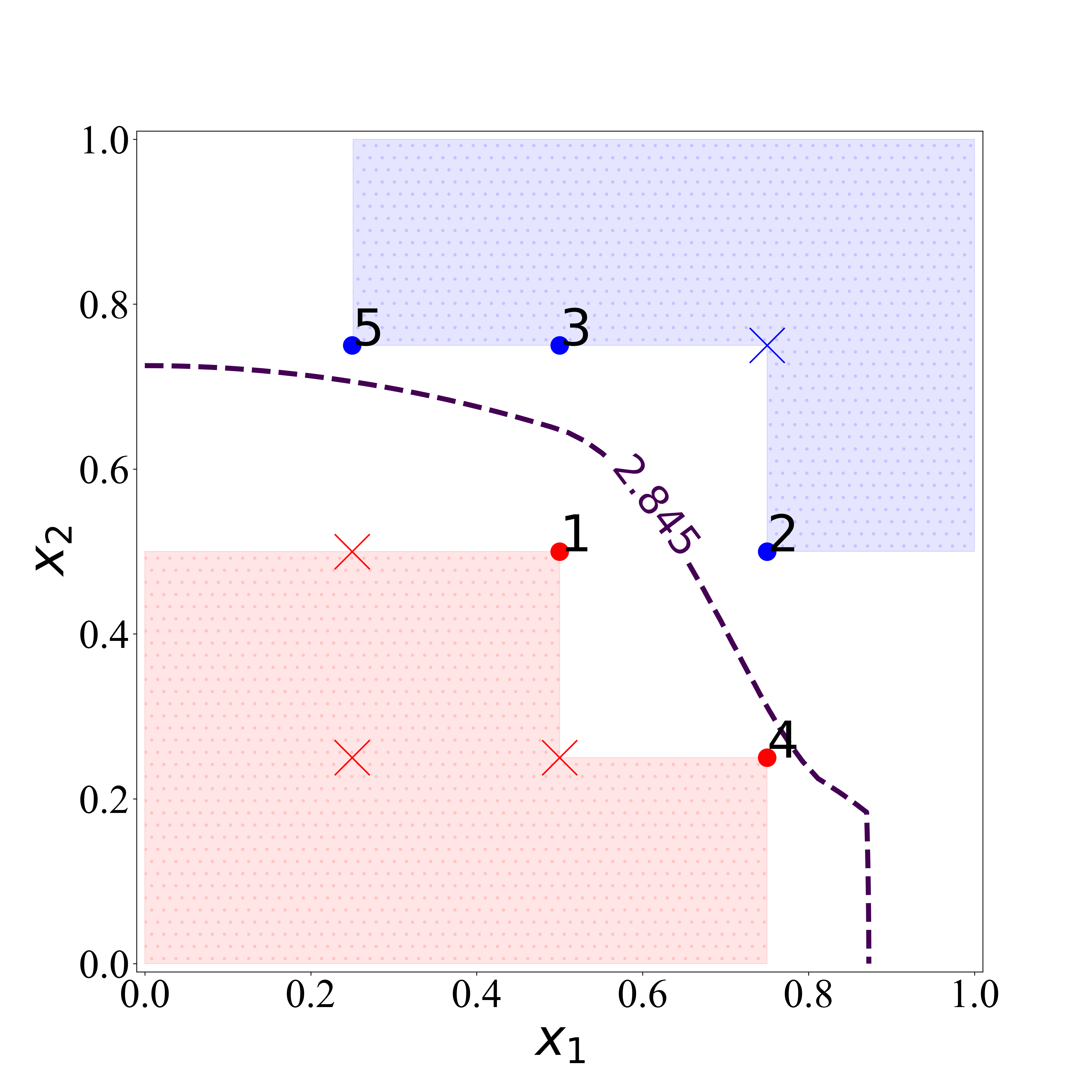}
			
			\label{illu_AI_2}
		\end{minipage}
	}
	\subfigure[$\mathbf{D}_{\text{AI},2,14}$, $v= 0.234$.]{
		\begin{minipage}[t]{0.25\linewidth}
			\centering
			\includegraphics[width=\linewidth]{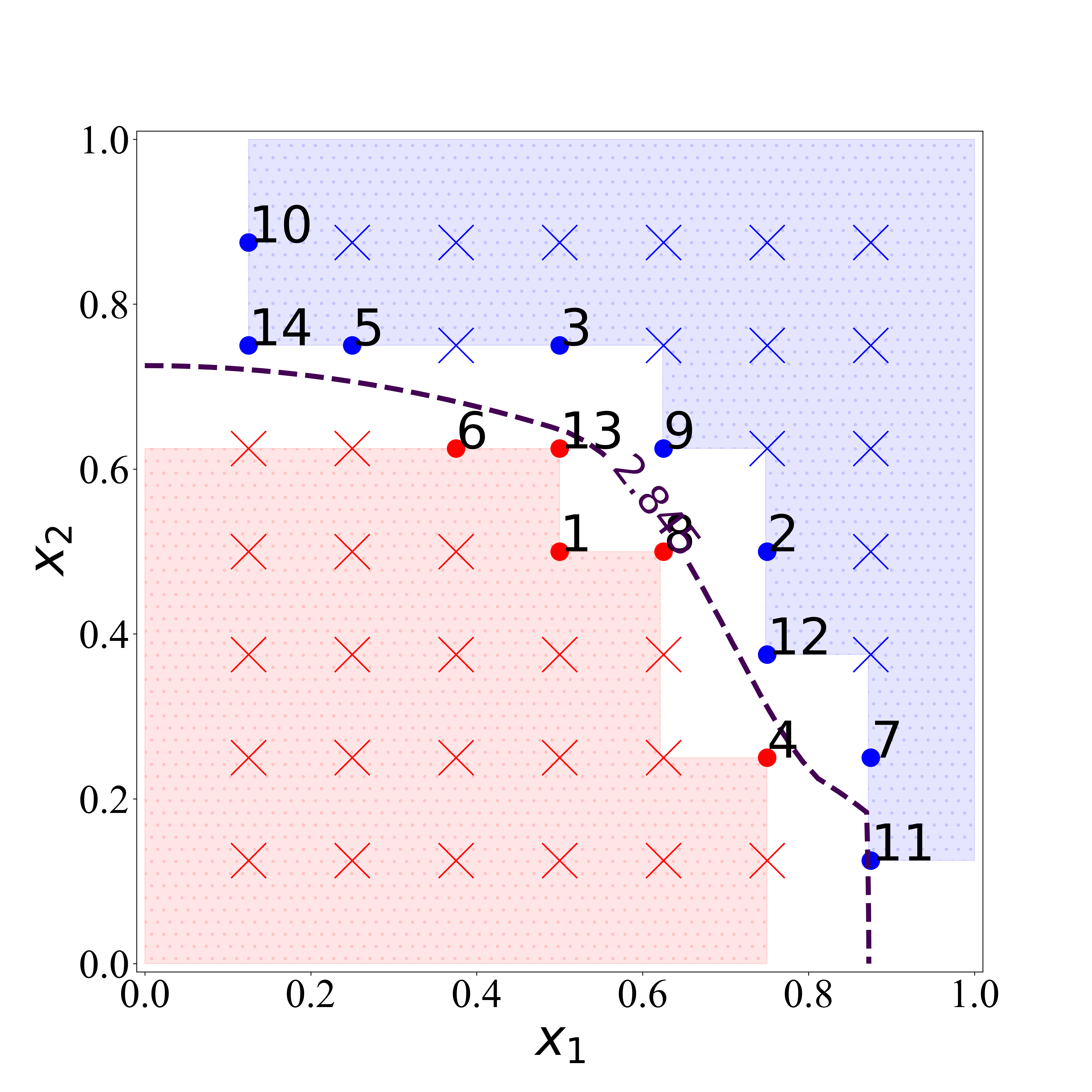}
			
			\label{illu_AI_3}
		\end{minipage}
	}
	\caption{Four adaptive designs, 
		showing the design points with negative response (red dots, or dots to the bottom-left of the dotted line) and positive response (blue dots, or dots to the top-right of the dotted line), overlaid with their order in evaluation (numbers), the skipped design points (crosses), the curve separating the two regions $\mathbf{A}$ and $\mathbf{B}$ (purple dotted line), the certainly negative area 
		(red shaded area, or the bottom-left shaded area), the certainly positive area 
		(blue shaded area, or the top-right shaded area), and the uncertain area (white area) after evaluating the runs corresponding to the designs. 
	}
	\label{fig:illu:adaptive}
\end{figure}

For better understanding, in Fig.~\ref{fig:illu:adaptive} we illustrate the AMC, ALE, AG, and AI methods for classifying the state function~\eqref{eq:illustrate} of the paper.
In the figure, alongside the design points, we delineate the certainly negative area $\cup_{\mathbf{x} \in \mathbf{D}, f(\mathbf{x})=-1} \prod_{k=1}^p [0,x_k]$, the certainly positive area $\cup_{\mathbf{x} \in \mathbf{D}, f(\mathbf{x})=1} \prod_{k=1}^p [x_k,1]$, and the uncertain area $\mathbf{U}$. 
It is apparent that for medium-to-large $n$, the AG and AI generally exhibit smaller uncertain areas compared to the AMC and ALE. 
The AMC is suboptimal in that some functional evaluations are not located around the middle line of the uncertainty area, while the ALE is suboptimal because some points fall within the certainly negative or positive area. 
On the other hand, the sequence of GG or GI ensures that all runs of the new group locate on the middle line of the uncertain area obtained from the preceding groups of runs. 

\section{Proofs}

In this section, we provide proofs of the theorems. 
Let $\lceil z \rceil$ denote the lowest integer that is no less than $z$ and $\lfloor z \rfloor$ denote the largest integer that is no more than $z$. 
Let $\mathbf{J} = \mathbf{A} \setminus \cup_{\mathbf{x}\in \mathbf{D}, f(\mathbf{x})=-1} \left\{\mathbf{z} : z_k \leq x_k, k=1,\ldots,p \right\}$ and $\mathbf{K} = \mathbf{B} \setminus \cup_{\mathbf{x} \in \mathbf{D}, f(\mathbf{x})=1} \left\{\mathbf{z} : z_k \geq x_k, k=1,\ldots,p \right\}$.

\subsection{Proof of Theorem~\ref{thm:SG}}
\begin{proof}
	When $p=1$, $\mathbf{D}_{\text{SG},1,n}=\{0/(n-1),1/(n-1),2/(n-1),\ldots,(n-1)/(n-1)\}$. 
	Consider three cases. 
	Firstly, when $0/(n-1) \in \mathbf{B}$. Then $\mathbf{U} = \emptyset$. 
	Secondly, when $(n-1)/(n-1) \in \mathbf{A}$. Then $\mathbf{U} = \emptyset$. 
	Thirdly, when there is one integer $0\leq z\leq (n-2)$ such that $z/(n-1) \in \mathbf{A}$ and $(z+1)/(n-1) \in \mathbf{B}$. 
	Then $\mathbf{U}=(z/(n-1),(z+1)/(n-1))$. 
	Combining all three cases, 
	\[\text{sup}_{f \in \mathbf{\Omega}} V\{\mathbf{U}(\mathbf{D}_{\text{SG},1,n},f)\} = 1/(n-1).\] 
	
	For $p\geq 2$, let $m=n^{1/p}-1$. For any $i_j\in\mathbb{N}$ and $0 \leq i_j \leq m$, $ 1\leq j \leq p-1$ and $j\in \mathbb{Z}$, we define $k({i_1,\cdots,i_{p-1}})$ by considering three cases:
	if $f(i_1/m,\cdots,i_{p-1}/m,1)=-1$, let $k({i_1,\cdots,i_{p-1}})=m$; if $f(i_1/m,\cdots,i_{p-1}/m,0)=1$, let $k({i_1,\cdots,i_{p-1}})=-1$; otherwise, let $k({i_1,\cdots,i_{p-1}})$ be the maximum integer that satisfies
	\[f\left(i_1/m,\cdots,i_{p-1}/m,k({i_1,\cdots,i_{p-1}})/m\right)=-1.\]
	It is worth noting that $k({i_1,\cdots,i_{p-1}})$ is non-increasing in any dimension.
	
	Because
	\begin{align*}
		V&\left(\cup_{\mathbf{x} \in \mathbf{D}_{\text{SG},p,n}, f(\mathbf{x})=-1} \left\{y : y_k \leq x_k, k=1,2,
		\cdots,p \right\}\right)\\
		=V&\left(\cup_{0\le i_1\leq m}\cdots\cup_{0\le i_{p-1}\leq m}\left\{x_1\le i_1/m,\cdots,x_{p-1}\le i_{p-1}/m ,\right.\right.\\
		&\quad\left.\left. x_p\le \max\left\{k({i_1,\cdots,i_{p-1}}),0\right\}/m\right\}\right)\\
		=\quad&\sum_{i_1,\cdots,i_{p-1}=1}^m\max\left\{k\left({i_1,\cdots,i_{p-1}}\right),0\right\}m^{-p}
	\end{align*}
	and
	\begin{align*}
		V& \left(\cup_{\mathbf{x} \in \mathbf{D}_{\text{SG},p,n}, f(\mathbf{x})=1} \{y : y_k \geq x_k, k=1,2\cdots,p \}\right)\\
		=V&\left(\cup_{0\le i_1\leq m}\cdots\cup_{0\le i_{p-1}\leq m}\left\{x_1\ge i_1/m,\cdots,x_{p-1}\ge i_{p-1}/m ,\right.\right.\\
		&\quad \left.\left.x_p \ge \min\{k({i_1,\cdots,i_{p-1}})+1,m\}/m\right\}\right)\\
		=\quad&\sum_{i_1,\cdots,i_{p-1}=0}^{m-1}\left[m-\min\left\{k({i_1,\cdots,i_{p-1}})+1,m\right\}\right]m^{-p},
	\end{align*}
	\begin{align*}
		V(\mathbf{U})=&1-V\left(\cup_{\mathbf{x} \in \mathbf{D}_{\text{SG},p,n}, f(\mathbf{x})=-1} \{y : y_k \leq x_k, k=1,2,
		\cdots,p \}\right)\\
		&-V \left(\cup_{\mathbf{x} \in \mathbf{D}_{\text{SG},p,n}, f(\mathbf{x})=1} \{y : y_k \geq x_k, k=1,2\cdots,p \}\right)\\
		=&1-\left\{\sum_{i_1,\cdots,i_{p-1}=1}^m \max\left\{k\left({i_1,\cdots,i_{p-1}}\right),0\right\}\right.\\
		&\quad\left.+\sum_{i_1,\cdots,i_{p-1}=0}^{m-1}\left[m-\min\{k({i_1,\cdots,i_{p-1}})+1,m\}\right]\right\}m^{-p}\\
		\leq& 1-\left\{\sum_{i_1,\cdots,i_{p-1}=1}^{m-1}\left[\max\left\{k({i_1,\cdots,i_{p-1}}),0\right\}+\right.\right.\\
		&\quad\left.\left. m-\min\{k({i_1,\cdots,i_{p-1}})+1,m\}\right]\right\}m^{-p}.
	\end{align*}
	Because
	\begin{align*}
		\max\left\{k({i_1,\cdots,i_{p-1}}),0\right\}+m-\min\{k({i_1,\cdots,i_{p-1}})+1,m\}\\
		=\left\{
		\begin{array}{*{2}{ll}}
			m, & k({i_1,\cdots,i_{p-1}})=-1\ \text{or}\ k({i_1,\cdots,i_{p-1}})=m,\\
			m-1,& \text{otherwise} ,
		\end{array}
		\right.
	\end{align*}
	
	\begin{align*}
		V(\mathbf{U})&\leq 1-\left\{\sum_{i_1,\cdots,i_{p-1}=1}^{m-1}\left[\max\left\{k({i_1,\cdots,i_{p-1}}),0\right\}\right.\right.\\
		&\left.\left.+m-\min\{k({i_1,\cdots,i_{p-1}})+1,m\}\right]\right\}m^{-p}\\
		&\leq 1-{(m-1)^p}m^{-p}.
	\end{align*}
	
	For arbitrary $p$ and $m$, let $f$ be the monotonic state function that 
	\begin{align*}
		f(\mathbf{x})=\left\{
		\begin{array}{*{2}{cl}}
			-1, & x_i<1 \text{ for any } i\leq p, \\
			1,& \text{otherwise} .
		\end{array}
		\right.
	\end{align*}
	By definition, $k({i_1,\cdots,i_{p-1}})$ satisfies
	\begin{align*}
		k({i_1,\cdots,i_{p-1}})=\left\{
		\begin{array}{*{2}{ll}}
			m-1, & \text{ for any } j\leq p-1, i_j\leq m-1, \\
			-1,& \text{otherwise} .
		\end{array}
		\right.
	\end{align*}
	Clearly, $k({i_1,\cdots,i_{p-1}})$ is non-increasing in any dimension. For such $k({i_1,\cdots,i_{p-1}})$,
	\begin{align*}
		V(\mathbf{U})=&1-\left\{\sum_{i_1,\cdots,i_{p-1}=1}^m \max\left\{k\left({i_1,\cdots,i_{p-1}}\right),0\right\}\right.\\
		&\left.+\sum_{i_1,\cdots,i_{p-1}=0}^{m-1}\left[m-\min\{k({i_1,\cdots,i_{p-1}})+1,m\}\right]\right\}m^{-p}\\
		=& 1-\left\{\sum_{i_1,\cdots,i_{p-1}=1}^{m-1}(m-1+m-m)\right\}m^{-p}.\\
		=&1-{(m-1)^p}m^{-p}.
	\end{align*}
	Therefore, there exists a monotonic $f$ such that the equation holds for arbitrary $m$. This completes the proof.
\end{proof}

\subsection{Proof of Theorem~\ref{thm:static}}
\begin{proof}
	When $p=1$, given $n$ sample size design $\mathbf{D}=\{x_1,x_2,\ldots,x_n\}$, let
	\[\{x_{(1)},x_{(2)},\ldots,x_{(n)}\}\] 
	denote the order statistics of design $\mathbf{D}$. Clearly, because $\mathbf{D}$ is a static design, when the outcomes of $\mathbf{D}$ is available, the maximal volume of uncertain area is
	\[\text{sup}_{f \in \mathbf{\Omega}} V\{\mathbf{U}(\mathbf{D},f)\}=\max \{x_{(1)},x_{(2)}-x_{(1)},x_{(3)}-x_{(2)},\ldots,x_{(n)}-x_{(n-1)},1-x_{(n)}\}.\]
	Let $\mathbf{D}_{n}$ with sample size $n$ be:
	\[\mathbf{D}_n=\{(n+1)^{-1},2(n+1)^{-1},\ldots,n(n+1)^{-1}\}.\]
	Therefore,
	\begin{align*}
		\text{sup}_{f \in \mathbf{\Omega}} V\{\mathbf{U}(\mathbf{D}_n,f)\}&=\max \{x_{(1)},x_{(2)}-x_{(1)},x_{(3)}-x_{(2)},\ldots,x_{(n)}-x_{(n-1)},1-x_{(n)}\}\\
		&=(n+1)^{-1}.
	\end{align*}
	Suppose there exists a better $n$-point design $\mathbf{Y}=\{y_1,y_2,\ldots,y_n\}$ such that
	\[\text{sup}_{f \in \mathbf{\Omega}} V\{\mathbf{U}(\mathbf{Y},f)\}<(n+1)^{-1},\]
	because
	\[\max \{y_{(1)},y_{(2)}-y_{(1)},y_{(3)}-y_{(2)},\ldots,y_{(n)}-y_{(n-1)},1-y_{(n)}\}<(n+1)^{-1},\]
	\[y_{(n)}>n(n+1)^{-1}.\]
	Because $y_{(1)}<(n+1)^{-1}$ and for $k=2,3,\ldots,n$
	\[y_{(k)}-y_{(k-1)}<(n+1)^{-1},\]
	\[y_{(n)}<y_{(1)}+(n-1)(n+1)^{-1}<n(n+1)^{-1}.\]
	This contradicts with $y_{(n)}>n(n+1)^{-1}$. Therefore,
	\[\text{sup}_{f \in \mathbf{\Omega}} V\{\mathbf{U}(\mathbf{D},f)\}\geq (n+1)^{-1}.\]
	This completes the case with $p=1$.

	Let $v = \sup_{f\in\Omega}V\{\mathbf{U}(\mathbf{D},f)\}$. When $p\geq 2$.
	If $v>2^{-p/(p-1)}5^{-1}{p!}^{-1}$, because $n\geq 10^{p-1}p^p$,
	\begin{align*}
		2^{-1/(p-1)}10^{-1/p}{(p-1)!}^{-1}n^{-1/p}\leq2^{-p/(p-1)}5^{-1}{p!}^{-1}<v.
	\end{align*}
	Therefore, the conclusion holds for $v>2^{-p/(p-1)}5^{-1}{p!}^{-1}$.
	
	If $v\leq 2^{-p/(p-1)}5^{-1}{p!}^{-1}$,
	for any $d$ such that $0<d<1-(p-1)!v$, 
	let $f \in \mathbf{\Omega}$ be the function that outputs $-1$ if and only if $\sum_{k=1}^p x_k \leq d$. 
	Clearly, 
	\begin{align*}
		&\cup_{\mathbf{x} \in \mathbf{D}, f(\mathbf{x})=-1} \{\mathbf{z} : 0 \leq z_k \leq x_k, k=1,\ldots,p \} \\
		&\subset \left\{\mathbf{z} : \sum_{k=1}^p z_k \leq d - (p-1)!v \right\} \\ 
		&\quad \cup \left\{ \cup_{\mathbf{x} \in \mathbf{D}, d - (p-1)!v < \sum_{k=1}^p x_k \leq d } \left(\{\mathbf{z} : 0 \leq z_k \leq x_k, k=1,\ldots,p \} \right.\right.\\
		&\quad \quad\left.\left.\setminus \left\{\mathbf{z} : 0 \leq z_k \leq x_k, k=1,\ldots,p , \sum_{k=1}^p z_k \leq d - (p-1)!v \right\} \right) \right\}. 
	\end{align*}
	
	Consequently, 
	\begin{align*}
		V(\mathbf{J}) =& V\left( \left\{\mathbf{z}: \sum_{k=1}^p z_k \leq d, 0 \leq z_k \leq 1, k=1,\ldots,p \right\} \right) \\
		&- V\left( \cup_{\mathbf{x} \in \mathbf{D}, f(\mathbf{x})=-1} \left\{\mathbf{z} : 0 \leq z_k \leq x_k, k=1,\ldots,p \right\} \right) \\
		\geq& V\left( \left\{\mathbf{z}: d-(p-1)!v < \sum_{k=1}^p z_k \leq d, 0 \leq z_k \leq 1, k=1,\ldots,p \right\} \right) \\
		&-\sum_{\mathbf{x} \in \mathbf{D}, d - (p-1)!v < \sum_{k=1}^p x_k \leq d} V\left( \left\{\mathbf{z} : 0 \leq z_k \leq x_k, k=1,\ldots,p \right\} \setminus\right.\\
		&\quad \quad\left. \left\{\mathbf{z} : 0 \leq z_k \leq x_k, k=1,\ldots,p , \sum_{k=1}^p z_k \leq d - (p-1)!v \right\} \right). 
	\end{align*}
	Because
	\begin{align*}
		V&\left( \left\{\mathbf{z}: d-(p-1)!v < \sum_{k=1}^p z_k \leq d, 0 \leq z_k \leq 1, k=1,\ldots,p \right\} \right)\\ &=\left[d^p-\left\{d-(p-1)!v\right\}^p\right]{p!}^{-1}
	\end{align*}
	and for $\mathbf{x} \in \mathbf{D}$ such that $d - (p-1)!v < \sum_{k=1}^p x_k \leq d$,
	\begin{align*}
		V&\left( \left\{\mathbf{z} : 0 \leq z_k \leq x_k, k=1,\ldots,p \right\}\right. \\
		&\left.\setminus \left\{\mathbf{z} : 0 \leq z_k \leq x_k, k=1,\ldots,p , \sum_{k=1}^p z_k \leq d - (p-1)!v \right\} \right)\\
		&\leq \left\{(p-1)!v\right\}^p{p!}^{-1},
	\end{align*}
	\begin{align*}
		V(\mathbf{J}) &\geq \left[d^p-\left\{d-(p-1)!v\right\}^p\right]{p!}^{-1}\\
		&- \text{card}\left( \left\{ \mathbf{x} \in \mathbf{D}: d - (p-1)!v < \sum_{k=1}^p x_k \leq d \right\} \right) \left\{(p-1)!v\right\}^p{p!}^{-1}.
	\end{align*}
	
	Similarly, 
	\begin{align*}
		V(\mathbf{K}) &\geq \left[\left\{d+(p-1)!v\right\}^p-d^p\right]{p!}^{-1}\\ 
		&- \text{card}\left( \left\{ \mathbf{x} \in \mathbf{D}: d < \sum_{k=1}^p x_k < d +(p-1)! v \right\} \right) \left\{(p-1)!v\right\}^p{p!}^{-1}. 
	\end{align*}
	Because $\text{sup}_{f \in \mathbf{\Omega}} V\{\mathbf{U}(\mathbf{D},f)\} \leq v$, $V(\mathbf{J})+V(\mathbf{K})=V(\mathbf{U}) \leq v$. 
	Therefore, 
	\begin{align}\label{eqn:plugin}
		\text{card}\left( \left\{ \mathbf{x} \in \mathbf{D}: d - (p-1)!v < \sum_{k=1}^p x_k < d + (p-1)!v \right\} \right)\nonumber\\
		\geq \left\lceil \left[\left\{d+(p-1)!v\right\}^p-\left\{d-(p-1)!v\right\}^p-vp!\right]\left\{(p-1)!v\right\}^{-p}\right\rceil . 
	\end{align}

	Let $z^{*}=[1-(1/2)^{1/(p-1)}]/[2(p-1)!v]$. Plugging $d=1-2z(p-1)!v-(p-1)!v$ in~\eqref{eqn:plugin}, we have that for any integer $z$ with $0 \leq z \leq z^{*}$, 
	\begin{align*}
		\text{card}\left( \left\{ \mathbf{x} \in \mathbf{D}: 1-(z+1)(p-1)!v < \sum_{k=1}^p x_k < 1-z(p-1)!v \right\} \right)\\
		\geq \left\lceil \left[\left\{1-z(p-1)!v\right\}^p-\left\{1-(z+1)(p-1)!v\right\}^p-vp!\right]\left\{(p-1)!v\right\}^{-p}\right\rceil .
	\end{align*}
	Summing the inequalities for every integer $z$ such that $0 \leq z \leq z^{*}$, we have 
	\begin{align*}
		&\text{card}\left( \left\{ \mathbf{x} \in \mathbf{D}: 0 < \sum_{k=1}^p x_k < 1 \right\} \right)\\
		&\geq \left\lceil\sum_{z \in \mathbb{N}, z\le \lfloor z^*\rfloor }\text{card}\left( \left\{ \mathbf{x} \in \mathbf{D}: 1-2z(p-1)!v-2(p-1)!v < \sum_{k=1}^p x_k\right.\right.\right.\\
		&\quad \left.\left.\left.< 1-2z(p-1)!v \right\} \right) \right\rceil\\
		&\geq \left\lceil\left[1^p-\left\{1-2\lfloor z^*\rfloor(p-1)!v-2(p-1)!v\right\}^p-(\lfloor z^*\rfloor+1)p!v\right]{\left\{(p-1)!v\right\}^{-p}}\right\rceil\\
		&\geq \left\lceil\left[1^p-\left\{1-2z^*(p-1)!v\right\}^p-z^*p!v\right]{\left\{(p-1)!v\right\}^{-p}}-{p!}{(p-1)!^{-p}v^{-(p-1)}}\right\rceil\\
		&\geq \left\lceil\left\{1^p-({1}/{2})^{{p}/({p-1})}-{p}/{2}+p({1}/{2})^{{p}/({p-1})}\right\}{\left\{(p-1)!v\right\}^{-p}}-{p!}{(p-1)!^{-p}v^{-(p-1)}}\right\rceil.
	\end{align*}
	
	Because $v\leq 2^{-p/(p-1)}5^{-1}{p!}^{-1}$,
	\begin{align*}
		&\text{card}\left( \left\{ \mathbf{x} \in \mathbf{D}: 0 < \sum_{k=1}^p x_k < 1 \right\} \right)\\
		&\quad\geq\left\lceil\left\{1^p-({1}/{2})^{{p}/({p-1})}-{p}/{2}+p({1}/{2})^{{p}/({p-1})}\right\}\left\{(p-1)!v\right\}^{-p}\right.\\
		&\quad-\left.{p!}{(p-1)!^{-p}v^{-(p-1)}}\right\rceil\\
		&\quad\geq \left(2^{{p}/{(p-1)}}-1-2^{{1}/{(p-1)}}p+p-1/5\right){2^{-{p}/{(p-1)}}(p-1)!^{-p}v^{-p}}.
	\end{align*}
	Let $t=1/(p-1)$, where $p\ge 2$, $0<t\le 1$,
	\begin{align*}
		2^{{p}/{(p-1)}}-1-2^{{1}/{(p-1)}}p+p&=2^{{1}/{(p-1)}}+(1-p)2^{{1}/{(p-1)}}+p-1\\
		&=2^t-{2^t}/{t}+{1}/{t}.
	\end{align*}
	Let $g_1(t)=2^t-2^t/t+1/t$. The first derivative of $g_1(t)$ is
	\begin{align*}
		{\mathrm{d}g_1}/{\mathrm{d}t}=2^t \ln 2-t^{-2}-{t^{-1}}2^t\ln2 +{2^t}{t^{-2}}=\left(t^22^t\ln2-t2^t\ln2+2^t-1\right)/{t^2}.
	\end{align*}
	Let $g_2(t)=t^22^t\ln2-t2^t\ln2+2^t-1$. Then
	\begin{align*}
		{\mathrm{d}g_2}/{\mathrm{d}t}&=2t2^t \ln 2+t^22^t\ln^22-t2^t\ln^22 =2^t\ln 2\left\{t(2-\ln2)+t^2\ln2\right\}>0.
	\end{align*}
	Therefore $g_2(t)> g_2(0)=0$ and thus
	\begin{align*}
		{\mathrm{d}g_1}/{\mathrm{d}t}=g_2(t)/t^2>0.
	\end{align*}
	Put it in another way, $g_1(t)$ is strictly increasing for $0<t\le 1$. Therefore, 
	\begin{align*}
		g_1(t)>g_1(0)=\lim_{t\rightarrow0}2^t-2^t/t+1/t=1-\ln2.
	\end{align*}
	
	Consequently,
	\begin{align*}
		&\text{card}\left( \left\{ \mathbf{x} \in \mathbf{D}: 0 < \sum_{k=1}^p x_k < 1 \right\} \right) \\
		&\geq ({2^{{p}/{(p-1)}}-1-2^{{1}/{(p-1)}}p+p-1/5}){2^{-{p}/{(p-1)}}(p-1)!^{-p}v^{-p}}\\
		&\geq{ 2^{-{p}/({p-1})}10^{-1}(p-1)!^{-p}v^{-p}}.
	\end{align*}
	
	Therefore, when $v\leq 2^{-p/(p-1)}5^{-1}{p!}^{-1}$,
	\[n\ge{2^{-{p}/({p-1})}10^{-1}(p-1)!^{-p}v^{-p}}.\]
	This completes the proof.
\end{proof}

\subsection{Proof of Theorem~\ref{thm:GG}}

\begin{proof}
	{Let $m=2^l$.} First, we prove that 
	\begin{align}\label{eqn:GG}
		m_{\text{GG}}(l+1)-m_{\text{GG}}(l)\leq p\left(2m+1\right)^{p-1} 
	\end{align}
	for every $l$. 
	We prove \eqref{eqn:GG} by inducing on $p$. 
	
	When $p=1$, consider three cases. 
	Firstly, when $0/m \in \mathbf{B}$. Then $\mathbf{D}_{\text{GG},1,m_{\text{GG}}(l+1)}\setminus \mathbf{D}_{\text{GG},1,m_{\text{GG}}(l)} = \emptyset$. 
	Secondly, when $m/m \in \mathbf{A}$. Then $\mathbf{D}_{\text{GG},1,m_{\text{GG}}(l+1)}\setminus \mathbf{D}_{\text{GG},1,m_{\text{GG}}(l)} = \emptyset$. 
	Thirdly, when there is one integer $0\leq z\leq m-1$ such that $z/m \in \mathbf{A}$ and $(z+1)/m \in \mathbf{B}$. 
	Then $\mathbf{D}_{\text{GG},1,m_{\text{GG}}(l+1)}\setminus \mathbf{D}_{\text{GG},1,m_{\text{GG}}(l)} = \left\{ (2z+1)/(2m) \right\}$. 
	In all three cases, $m_{\text{GG}}\left(l+1\right)-m_{\text{GG}}(l)\leq 1\allowbreak={1(2m+1)^0}$.
	
	Suppose \eqref{eqn:GG} holds for $1,\ldots,p-1$. 
	Clearly, $\mathbf{D}_{\text{GG},p,m_{\text{GG}}(l+1)}\setminus \mathbf{D}_{\text{GG},p,m_{\text{GG}}(l)}$ can be uniquely partitioned by 
	\[ \mathbf{D}_{\text{GG},p,m_{\text{GG}}(l+1)}\setminus \mathbf{D}_{\text{GG},p,m_{\text{GG}}(l)} = \cup_{z=0}^{2 m} \left\{ \mathbf{W}_{z} \times\left \{ z/(2 m) \right\} \right\}. \]
	From induction, for any even integer $z$ such that $0\leq z\leq 2m$, 
	$ \text{card}(\mathbf{W}_z) \leq (p-1) \left( 2 m +1 \right)^{p-2}$. 
	Suppose $z$ and $\tilde z$ are odd integers and $0<z<\tilde z<2m$. 
	Because $( \mathbf{W}_{\tilde z} \setminus \mathbf{W}_{\tilde z-1} ) \times \{(\tilde z-1)/(2m)\} \subset {\mathbf{A}}$ and ${(\mathbf{W}_z\setminus \mathbf{W}_{\tilde z-1}) }\times \{(\tilde z-1)/(2m)\} \subset {\mathbf{B}}$, 
	$(\mathbf{W}_z \setminus \mathbf{W}_{z-1}) \cap (\mathbf{W}_{\tilde z} \setminus \mathbf{W}_{\tilde z-1}) = \emptyset$. 
	Consequently, $\sum_{0\leq z\leq 2m, z \text{ is odd}} \text{card}(\mathbf{W}_z \setminus \mathbf{W}_{z-1}) \leq \left( 2 m +1 \right)^{p-1}$. 
	Therefore, 
	\begin{align*}
		m_{\text{GG}}(l+1)-m_{\text{GG}}(l) &= \sum_{z=0}^{2m} \text{card}(\mathbf{W}_z) 
		= \sum_{0\leq z\leq 2m, z \text{ is even}} \text{card}(\mathbf{W}_z) \\
		&+ \sum_{0\leq z\leq 2m, z \text{ is odd}} \text{card}(\mathbf{W}_z) \\
		&\leq\sum_{0\leq z\leq 2m, z \text{ is even}} \text{card}(\mathbf{W}_z) + \sum_{0\leq z\leq 2m, z \text{ is odd}} \text{card}(\mathbf{W}_{z-1})\\
		& + \sum_{0\leq z\leq 2m, z \text{ is odd}} \text{card}(\mathbf{W}_z \setminus \mathbf{W}_{z-1})\\
		&\leq (2m+1) (p-1) \left( 2 m +1 \right)^{p-2} + \left( 2 m +1 \right)^{p-1}\\ 
		&= p \left( 2 m +1 \right)^{p-1}.
	\end{align*}
	That is, \eqref{eqn:GG} holds for $p$. 
	
	Consequently,
	\[m_{\text{GG}}(g)= 2^p+\sum_{i=1}^g\left\{m_{\text{GG}}\left(i\right)-m_{\text{GG}}\left(i-1\right)\right\}\leq2^p+\sum_{l=1}^g \left\{ p \left(2^{l}+1\right)^{p-1}\right\}.\]
	This completes the proof. 
\end{proof}

\subsection{Proof of Theorem~\ref{thm:adaptive}}

\begin{proof}
	When $p=1$, we prove the theorem by inducing on $n$. Let $f_{\alpha}\in \mathbf{\Omega}$ be the function that outputs -1 if and only if $x<\alpha$. 
	When $n=1$, let $\mathbf{D}_1=\{x\}$ be an adaptive design with 1 sample, where $0\leq x\leq 1$. If $x<1/2$, 
	\[\text{sup}_{f\in \mathbf{\Omega}}V\{\mathbf{U}(\mathbf{D}_1,f)\}\geq V\{\mathbf{U}(\mathbf{D}_1,f_{3/4})\}\geq 1/2,\]
	if $x\geq 1/2$,
	\[\text{sup}_{f\in \mathbf{\Omega}}V\{\mathbf{U}(\mathbf{D}_1,f)\}\geq V\{\mathbf{U}(\mathbf{D}_1,f_{1/4})\}\geq 1/2.\]
	Combining the two cases,
	\[\text{sup}_{f\in \mathbf{\Omega}}V\{\mathbf{U}(\mathbf{D}_1,f)\}\geq 1/2.\]

	Suppose the theorem holds for $1,\ldots, n-1$. Let 
	\[\mathbf{D}_{n-1}=\{x_1,x_2,\ldots,x_{n-1}\}\]
	be an adaptive design with $n-1$ samples. Let $x_n$ be the newly added sample and $\mathbf{D}_n=\mathbf{D}_{n-1}\cup \{x_n\}$ be the adaptive design with $n$ samples.
	Define $x^-$ and $x^+$ by considering three cases:
	if $f(x_i)=-1$ for any $1\leq i\leq n-1$, let
	\[x^-=\max_{x_j\in \mathbf{D}_{n-1}}x_j\]
	and $x^+=1$; if $f(x_i)=1$ for any $1\leq i\leq n-1$, let
	\[x^+=\min_{x_j\in \mathbf{D}_{n-1}}x_j\]
	and $x^-=0$; otherwise, let
	\[x^-=\max_{x_j\in \mathbf{D}_{n-1},f(x_j)=-1}x_j\]
	and
	\[x^+=\min_{x_j\in \mathbf{D}_{n-1},f(x_j)=1}x_j.\]
	Clearly, $\mathbf{U}(\mathbf{D}_{n-1},f)=(x^-,x^+)$. From induction, $\text{sup}_{f\in\mathbf{\Omega}}(x^+-x^-)\geq 2^{-(n-1)}$.
	If $x_n\in [0,x^-]\cup [x^+,1]$, $V\{\mathbf{U}(\mathbf{D}_n,f)\}=V\{\mathbf{U}(\mathbf{D}_{n-1},f)\}\geq2^{-(n-1)}\geq 2^{-n}$.
	If $x^-<x_n\leq (x^++x^-)/2$,
	\[\text{sup}_{f\in\mathbf{\Omega}}V\{\mathbf{U}(\mathbf{D}_n,f)\}\geq V\{ \mathbf{U}(\mathbf{D}_n,f_{(x^++x^-)/2+(x^+-x^-)/10})\}=(x^+-x_n)\geq 2^{-n}.\]
	If $x^+>x_n> (x^++x^-)/2$, 
	\[\text{sup}_{f\in\mathbf{\Omega}}V\{\mathbf{U}(\mathbf{D}_n,f)\}\geq V\{\mathbf{U}(\mathbf{D}_n,f_{(x^++x^-)/2-(x^+-x^-)/10})\}=(x_n-x^-)\geq 2^{-n}.\]
	Combining the three cases,
	\[\text{sup}_{f\in\mathbf{\Omega}}V
	\{\mathbf{U}(\mathbf{D}_n,f)\}\geq 2^{-n}. \]
	This completes the case with $p=1$.

	Let $v = \sup_{f\in\Omega}V\{\mathbf{U}(\mathbf{D},f)\}$. When $p\geq 2$, if $v> (8p!)^{-1}$, because $n\geq 4^{p-2}p^p$,
	\begin{align*}
		p^{1/(p-1)}2^{-(p+1)/(p-1)}(p-1)!^{-1}n^{-1/(p-1)}\leq (8p!)^{-1}<v
	\end{align*}
	Therefore, the conclusion holds when $v> (8p!)^{-1}$.
	
	If $v\leq(8p!)^{-1}$. Let $f\in \mathbf{\Omega}$ be the function that outputs -1 if and only if $\sum_{k=1}^p x_k\leq 1$. Then
	\begin{align*}
		&\cup_{\mathbf{x} \in \mathbf{D}, f(\mathbf{x})=-1} \{\mathbf{z} : 0 \leq z_k \leq x_k, k=1,\ldots,p \} \\
		&\subset \left\{\mathbf{z}  : \sum_{k=1}^p z_k \leq 1 - 2(p-1)!v \right\} \\ 
		&\quad \cup \left\{ \cup_{\mathbf{x} \in \mathbf{D}, 1 - 2(p-1)!v < \sum_{k=1}^p x_k \leq 1 } \left[ \{\mathbf{z}  : 0 \leq z_k \leq x_k, k=1,\ldots,p \}\right.\right.\\
		&\quad \quad\left.\left. \setminus \left\{\mathbf{z} : 0 \leq z_k \leq x_k, k=1,\ldots,p , \sum_{k=1}^p z_k \leq 1 - 2(p-1)!v \right\} \right] \right\}. 
	\end{align*}
	
	Consequently, 
	\begin{align*}
		V(\mathbf{J}) =& V\left( \left\{\mathbf{z} : \sum_{k=1}^p z_k \leq 1, 0 \leq z_k \leq 1, k=1,\ldots,p \right\} \right) \\
		&\quad - V\left( \cup_{\mathbf{x} \in \mathbf{D}, f(\mathbf{x})=-1} \left\{\mathbf{z}  : 0 \leq z_k \leq x_k, k=1,\ldots,p \right\} \right) \\
		\geq& V\left( \left\{\mathbf{z} : 1-2(p-1)!v < \sum_{k=1}^p z_k \leq d, 0 \leq z_k \leq 1, k=1,\ldots,p \right\} \right) \\
		&-\sum_{\mathbf{x} \in \mathbf{D}, 1 - 2(p-1)!v < \sum_{k=1}^p x_k \leq 1} V\left( \{\mathbf{z}  : 0 \leq z_k \leq x_k, k=1,\ldots,p \} \right.\\
		&\quad\left.\setminus \left\{\mathbf{z}  :0 \leq z_k \leq x_k, k=1,\ldots,p , \sum_{k=1}^p z_k \leq 1 - 2(p-1)!v \right\} \right). 
	\end{align*}

	Because
	\begin{align*}
		&V\left( \left\{\mathbf{z} : 1-2(p-1)!v < \sum_{k=1}^p z_k \leq 1, 0 \leq z_k \leq 1, k=1,\ldots,p \right\} \right)\\ &=\left[1-\left\{1-2(p-1)!v\right\}^p\right]{p!}^{-1}
	\end{align*}
	and
	\begin{align*}
		&V\left( \{\mathbf{z}  : 0 \leq z_k \leq x_k, k=1,\ldots,p \} \right.\\
		&\quad\left.\setminus \left\{\mathbf{z}  :0 \leq z_k \leq x_k, k=1,\ldots,p , \sum_{k=1}^p z_k \leq 1 - 2(p-1)!v \right\} \right)\\
		&\leq \left\{2(p-1)!v\right\}^p{p!}^{-1},
	\end{align*}
	\begin{align*}
		V(\mathbf{J}) &\geq \left[1-\left\{1-2(p-1)!v\right\}^p\right]{p!}^{-1}\\
		&\quad- \text{card}\left( \left\{ \mathbf{x} \in \mathbf{D}: 1 - 2(p-1)!v < \sum_{k=1}^p x_k \leq 1 \right\} \right) \left\{2(p-1)!v\right\}^p{p!}^{-1}. 
	\end{align*}
	
	Because $v\leq{(8p!)}^{-1}$, $0<1-2(p-1)!v<1$. Also because $\text{sup}_{f \in \mathbf{\Omega}} V\{\mathbf{U}(\mathbf{D},f)\}=v\geq V(\mathbf{J})+V(\mathbf{K})\geq V(\mathbf{J})$,
	
	\begin{align*}
		&\text{card}\left( \left\{ \mathbf{x} \in \mathbf{D}: 1 - 2(p-1)!v < \sum_{k=1}^p x_k \leq 1 \right\} \right)\\
		&\geq\left[1^p-\left\{1-2(p-1)!v\right\}^p-p!v\right]\left\{2(p-1)!v\right\}^{-p}\\
		&=\left[2(p-1)!v\sum_{i=0}^{p-1}\left\{1-2(p-1)!v\right\}^{i}-p!v\right]\left\{2(p-1)!v\right\}^{-p}\\
		&\geq \left[2p!v\left\{1-2(p-1)!v\right\}^{p-1}-p!v\right]\left\{2(p-1)!v\right\}^{-p}.\\
	\end{align*}
	Because 
	\begin{align*}
		\left\{1-2(p-1)!v\right\}^{p-1}-1+2p!v&=2p!v-2(p-1)!v\sum_{i=0}^{p-2}\left\{1-2(p-1)!v\right\}^{i}\\
		&\geq 2p!v-2(p-1)!v (p-1)\\
		&=2(p-1)!v\geq 0,
	\end{align*}
	\begin{align*}
		&\text{card}\left( \left\{ \mathbf{x} \in \mathbf{D}: 1 - 2(p-1)!v < \sum_{k=1}^p x_k \leq 1 \right\} \right)\\
		&\geq\left[2p!v\left\{1-2(p-1)!v\right\}^{p-1}-p!v\right]\left\{2^{-p}(p-1)!^{-p}v^{-p}\right\}\\
		&\geq \{2p!v(1-2p!v)-p!v\}\{2^{-p}(p-1)!^{-p}v^{-p}\}\\
		&\geq{\{p!v(1-4p!v)\}\{2^{-p}(p-1)!^{-p}v^{-p}\}.}
	\end{align*}
	Also because $v\leq {(8p!)}^{-1}$,
	\begin{align*}
		\text{card}\left( \left\{ \mathbf{x} \in \mathbf{D}: 1 - 2(p-1)!v < \sum_{k=1}^p x_k \leq 1 \right\} \right)&\geq {p}{2^{-(p+1)}(p-1)!^{-(p-1)}v^{-(p-1)}}.
	\end{align*}
	Therefore, when $v\leq {(8p!)}^{-1}$,
	\[n\geq{p}{2^{-(p+1)}(p-1)!^{-(p-1)}v^{-(p-1)}}.\]
	This completes the proof.
\end{proof}

\subsection{Proof of Theorem~\ref{thm:MC}}

\begin{proof} 
	When $p=1$, consider three cases. 
	Firstly, when $f(0)=1$. Then for any $x \in [0,1]$, the probability of $x \in \mathbf{U}$ is $(1-x)^n$. 
	Therefore, the expectation of uncertain volume is 
	\[\int_0^1 (1-x)^n dx =1/(n+1). \]
	Secondly, when $f(1)=-1$. Then for any $x \in [0,1]$, the probability of $x \in \mathbf{U}$ is $x^n$. 
	Therefore, the expectation of uncertain volume is 
	\[\int_0^1 x^n dx =1/(n+1). \]
	Thirdly, when $f(0)=-1$ and $f(1)=1$. 
	Let $z = \sup \{x: f(x)=-1\}$. Clearly, $0< z< 1$.
	Then for any $x \in [0,z)$, the probability of $x \in \mathbf{\mathbf{U}}$ is $\{1-(z-x)\}^n$. 
	Similarly, for any $x \in (z,1]$, the probability of $ x \in \mathbf{U}$ is $\{1-(x-z)\}^n$. 
	Therefore, the expectation of uncertain volume is 
	\[
	\int_0^z \left\{1-\left(z-x\right)\right\}^n dx + \int_z^1 \left\{1-\left(x-z\right)\right\}^n dx 
	= \left\{ 2 - \left(1-z\right)^{n+1} - z^{n+1} \right\}/\left(n+1\right). \]
	Because $\sup_{z\in(0,1)} \{ 2 - (1-z)^{n+1} - z^{n+1} \}/(n+1) = \{ 2 - (1-1/2)^{n+1} - (1/2)^{n+1} \}/(n+1) 
	= (2 - 2^{-n}) /(n+1)>1/(n+1)$ for $n\geq 1$,
	\[ \text{sup}_{f \in \mathbf{\Omega}} \text{E} \left[ V\left\{\mathbf{U}\left(\mathbf{D}_{\text{MC},1,n},f\right)\right\}\right ] = \left(2 - 2^{-n}\right) /\left(n+1\right) . \]
	
	When $p\geq 2$, 
	for any $\mathbf{x}$ with $\sum_{k=1}^p x_k < p/2$, let 
	\[ W(\mathbf{x}) = V\left( \left\{ \mathbf{y}: x_k \leq y_k \leq 1 \text{ for any } k = 1,\ldots,p, \sum_{k=1}^p y_k < p/2 \right\} \right). \] 
	Then the probability of $\mathbf{x} \in \mathbf{U}$ is $ \left\{ 1 - W(\mathbf{x}) \right\}^n $. 
	Let $\mathbf{S} = \left\{\mathbf{y}: \sum_{k=1}^p y_k < p/2 \right\}$, we have 
	\[ \text{E} \left[ V\left\{\mathbf{U}\left(\mathbf{D}_{\text{MC},p,n},\tilde f\right)\right\} \right] = 2 \text{E} \left[ V\left\{ \mathbf{U}(\mathbf{D}_{\text{MC},p,n},\tilde f) \cap \mathbf{S} \right\} \right] 
	= 2 \int_{\mathbf{x} \in \mathbf{S}} \left\{ 1 - W(\mathbf{x}) \right\}^n d\mathbf{x}. \]
	Partition $\mathbf{S}$ into six subregions, 
	\[\mathbf{S}_1 = [0,1-n^{-1/p+\delta}]^p \cap \{\mathbf{y}: \sum_{k=1}^p y_k \leq p/2-n^{-1/p+\delta} \},\]
	\begin{align*}
		\mathbf{S}_2 = \{&\mathbf{y}: \mathbf{y}\in [0,1]^p, \sum_{k=1}^p y_k \leq p/2-n^{-1/p+\delta}, \text{ there exist at least two } k\\
		& \text{ such that }y_k > 1-n^{-1/p+\delta} \},
	\end{align*}
	\begin{align*}
		\mathbf{S}_3 = \{&\mathbf{y}: \mathbf{y} \in [0,1]^p, \sum_{k=1}^p y_k \leq p/2-n^{-1/p+\delta}, \allowbreak\text{ there exists exactly one } k \text{ such that } \\
		& y_k > 1-n^{-1/p+\delta} \text{ and } y_k <1- n^{-1/p-\tilde\delta} \text{ for this } k \},
	\end{align*}
	\begin{align*}
		\mathbf{S}_4 = \{&\mathbf{y}: \mathbf{y} \in [0,1]^p, \sum_{k=1}^p y_k \leq p/2-n^{-1/p+\delta}, \text{ there exists exactly one } k \text{ such that } \\
		&y_k > 1-n^{-1/p+\delta}  \text{ and } y_k > 1-n^{-1/p-\tilde\delta} \text{ for this }  k \},		
	\end{align*}
	\begin{align*}
		\mathbf{S}_5 = \{&\mathbf{y}: \mathbf{y} \in [0,1]^p, p/2-n^{-1/p+\delta} \leq \sum_{k=1}^p y_k < p/2, \text{ there exists at least one } k \\
		&\text{ such that }  y_k > 1-n^{-1/p+\delta} \},
	\end{align*}
	and 
	\[\mathbf{S}_6 = [0,1-n^{-1/p+\delta}]^p \cap \{\mathbf{y}: p/2-n^{-1/p+\delta} \leq \sum_{k=1}^p y_k < p/2 \},\]
	where $\delta$ and $\tilde\delta$ are chosen such that $0<\delta<1/(2p)$ and $0<\tilde\delta<\delta(p-1)$. 
	
	Because $W(\mathbf{x}) \geq (n^{-1/p+\delta})^p/p! = n^{-1+\delta p}/p!$ for $\mathbf{x} \in \mathbf{S}_1$, $\left\{ 1 - W(\mathbf{x}) \right\}^n / n^{-1/p}\leq n^{1/p}/\exp({n^{\delta p}/p!})$ converges to 0 as $n \to \infty$ uniformly for all $\mathbf{x} \in \mathbf{S}_1$. 
	Therefore, \[ \int_{\mathbf{x} \in \mathbf{S}_1} \left\{ 1 - W(\mathbf{x}) \right\}^n d\mathbf{x} = o\left(n^{-1/p}\right). \]
	Because $V(\mathbf{S}_2) = o\left(n^{-1/p}\right)$ and $0 \leq \left\{ 1 - W(\mathbf{x}) \right\}^n \leq 1$, 
	\[ \int_{\mathbf{x} \in \mathbf{S}_2} \left\{ 1 - W(\mathbf{x}) \right\}^n d\mathbf{x} = o\left(n^{-1/p}\right). \]
	Suppose that $\mathbf{x} \in \mathbf{S}_3$ and $1-n^{-1/p+\delta} < x_k < 1-n^{-1/p-\tilde\delta}$. 
	Then $W(\mathbf{x}) \geq (1-x_k) (n^{-1/p+\delta})^{p-1}/{(p-1)}! \geq n^{-1+\delta (p-1)-\tilde\delta}/(p-1)!$ 
	and thus $\left\{ 1 - W(\mathbf{x}) \right\}^n / n^{-1/p}$ converges to 0 as $n \to \infty$ uniformly for all $\mathbf{x} \in \mathbf{S}_3$. 
	Therefore, \[ \int_{\mathbf{x} \in \mathbf{S}_3} \left\{ 1 - W(\mathbf{x}) \right\}^n d\mathbf{x} = o\left(n^{-1/p}\right). \]
	Because $V(\mathbf{S}_4) = o\left(n^{-1/p}\right)$ and $0 \leq \left\{ 1 - W(\mathbf{x}) \right\}^n \leq 1$, 
	\[ \int_{\mathbf{x} \in \mathbf{S}_4} \left\{ 1 - W(\mathbf{x}) \right\}^n d\mathbf{x} = o\left(n^{-1/p}\right). \]
	Because $V(\mathbf{S}_5)=o(n^{-1/p})$, $0\leq \{1-W(\mathbf{x})\}^n\leq 1$, $ 0\leq\left( p/2 - \sum_{k=1}^p x_k \right)^p / p!\leq 1 $ , 
	\[ \int_{\mathbf{x} \in \mathbf{S}_5} \left\{ 1 - W(\mathbf{x}) \right\}^n d\mathbf{x}= \int_{\mathbf{x} \in \mathbf{S}_5} \left\{ 1 - \left( p/2 - \sum_{k=1}^p x_k \right)^p / p! \right\}^n d\mathbf{x}+o(n^{-1/p}). \]
	Because $W(\mathbf{x}) = \left( p/2 - \sum_{k=1}^p x_k \right)^p / p! $ for $\mathbf{x} \in \mathbf{S}_6$, 
	\[ \int_{\mathbf{x} \in \mathbf{S}_6} \left\{ 1 - W(\mathbf{x}) \right\}^n d\mathbf{x} = \int_{\mathbf{x} \in \mathbf{S}_6} \left\{ 1 - \left( p/2 - \sum_{k=1}^p x_k \right)^p / p! \right\}^n d\mathbf{x}. \]
	Combining these results, 
	\[ \text{E} \left[ V\left\{\mathbf{U}\left(\mathbf{D}_{\text{MC},p,n},\tilde f\right)\right\} \right] = 2 \int_{\mathbf{S}_5 \cup \mathbf{S}_6} \left\{ 1 - \left( p/2 - \sum_{k=1}^p x_k \right)^p / p! \right\}^n d\mathbf{x}. \]
	
	For any set $\mathbf{T} \subset \{1,\ldots,p\}$ with $\text{card}(\mathbf{T})<p/2$, let 
	\begin{align*}
		V_\mathbf{T} &= 2 \int_{p/2 - n^{-1/p+\delta} \leq \sum_{k=1}^p x_k < p/2, x_k>0 \text{ for any } k , x_k>1 \text{ for } k \in \mathbf{T}} \left\{ 1\right.\\
		& \quad\left.- \left( p/2 - \sum_{k=1}^p x_k \right)^p / p! \right\}^n d\mathbf{x}.
	\end{align*}
	Then \[ 2\int_{\mathbf{S}_5 \cup \mathbf{S}_6} \left\{ 1 - \left( p/2 - \sum_{k=1}^p x_k \right)^p / p! \right\}^n d\mathbf{x} = \sum_{0\leq \text{card}(\mathbf{T})<p/2} \left\{\left(-1\right)^{\text{card}(\mathbf{T})} V_\mathbf{T} \right\}.\]
	
	Clearly, 
	\begin{align*}
		V_\mathbf{T} &= 2 \int_{p/2 - \text{card}(\mathbf{T}) - n^{-1/p+\delta} \leq \sum_{k=1}^p x_k < p/2 - \text{card}(\mathbf{T}), x_k>0 \text{ for any } k} \left\{ 1 \right.\\
		&\quad\left.- \left( p/2 - \text{card}(\mathbf{T}) - \sum_{k=1}^p x_k \right)^p / p! \right\}^n d\mathbf{x}\\
		&= 2 \int_{\sum_{k=1}^p x_k < p/2 - \text{card}(\mathbf{T}), x_k>0 \text{ for any } k} \left\{ 1 - \left( p/2 - \text{card}(\mathbf{T}) - \sum_{k=1}^p x_k \right)^p / p! \right\}^n d\mathbf{x} \\
		&\quad- 2 \int_{\sum_{k=1}^p x_k < p/2 - \text{card}(\mathbf{T}) - n^{-1/p+\delta} , x_k>0 \text{ for any } k} \left\{ 1 \right.\\
		&\quad\left.- \left( p/2 - \text{card}(\mathbf{T}) - \sum_{k=1}^p x_k \right)^p / p! \right\}^n d\mathbf{x}\\
		&= 2 \int_{0}^{ p/2 - \text{card}(\mathbf{T})} du 
		\int_{0}^{ u } dx_1 
		\int_{0}^{ u-x_1 } dx_2 
		\cdots
		\int_{0}^{ u-\sum_{k=1}^{p-2}x_k } dx_{p-1} \left\{ 1\right.\\
		&\quad\left. - \left( p/2 - \text{card}(\mathbf{T}) - u \right)^p / p! \right\}^n \\
		&\quad- 2 \int_{0}^{ p/2 - \text{card}(\mathbf{T}) - n^{-1/p+\delta}} du 
		\int_{0}^{ u } dx_1 
		\int_{0}^{ u-x_1 } dx_2 
		\cdots
		\int_{0}^{ u-\sum_{k=1}^{p-2}x_k } dx_{p-1}\\
		&\quad\left\{ 1 - \left( p/2 - \text{card}(\mathbf{T}) - u \right)^p / p! \right\}^n \\
		&= 2 \int_{p/2-\text{card}(\mathbf{T})-n^{-1/p+\delta}}^{p/2-\text{card}(\mathbf{T})} u^{p-1} \left \{ 1 - \left(p/2-\text{card}(\mathbf{T})-u \right)^p/p! \right\}^n du /\left(p-1\right)!\\
		& = 2\int_{0}^{n^{-1/p+\delta}} \left\{p/2-\text{card}\left(\mathbf{T}\right)-\tilde u\right\}^{p-1} \left ( 1 - \tilde u^p/p! \right)^n d \tilde u /\left(p-1\right)! 
	\end{align*}
	for $\mathbf{T}$ such that $\text{card}(\mathbf{T})<p/2$.
	
	Because $0\leq \tilde u\leq n^{-1/p+\delta}$,
	\[\left\{p/2-\text{card}\left(\mathbf{T}\right)- n^{-1/p+\delta}\right\}^{p-1}\leq\left\{p/2-\text{card}\left(\mathbf{T}\right)- \tilde u\right\}^{p-1}\leq\left\{p/2-\text{card}\left(\mathbf{T}\right)\right\}^{p-1}.\]
	Also because 
	\[\lim_{n\to\infty}\left\{p/2-\text{card}\left(\mathbf{T}\right)- n^{-1/p+\delta}\right\}^{p-1}=\left\{p/2-\text{card}\left(\mathbf{T}\right)\right\}^{p-1}\]
	and
	\[V_\mathbf{T}\leq 2\int_{0}^{n^{-1/p+\delta}} \left\{p/2-\text{card}\left(\mathbf{T}\right)\right\}^{p-1} \left ( 1 - \tilde u^p/p! \right)^n d \tilde u /\left(p-1\right)!,\]
	\[V_\mathbf{T}\geq2\int_{0}^{n^{-1/p+\delta}} \left\{p/2-\text{card}\left(\mathbf{T}\right)- n^{-1/p+\delta}\right\}^{p-1} \left ( 1 - \tilde u^p/p! \right)^n d \tilde u /\left(p-1\right)!,\]
	according to squeeze theorem, as $n\to\infty$,
	\[V_\mathbf{T}/\left[2\int_{0}^{n^{-1/p+\delta}} \left\{p/2-\text{card}\left(\mathbf{T}\right)\right\}^{p-1} \left ( 1 - \tilde u^p/p! \right)^n d \tilde u /\left(p-1\right)!\right]\rightarrow 1.\]
	Clearly,
	\[\int_0^1 \left (1 - u^p/p! \right)^n du=\int_{0}^{n^{-1/p+\delta}} \left ( 1 - u^p/p! \right)^n du +\int_{n^{-1/p+\delta}}^{1} \left ( 1 - u^p/p! \right)^n du. \]
	Because $\left ( 1 - u^p/p! \right)^n\leq \left ( 1 - n^{-1+p\delta}/p! \right)^n$ for $n^{-1/p+\delta}\leq u\leq 1$ and
	\[\lim_{n\to\infty}\left ( 1 - n^{-1+p\delta}/p! \right)^n=\exp({-n^{p\delta}/p!})=o\left(n^{-1/p}\right),\]
	when $n\to\infty$,
	\[\int_{n^{-1/p+\delta}}^1 \left( 1 - u^p/p! \right)^n du=o\left(n^{-1/p}\right).\]
	Because when $n\to\infty$
	\[\int_0^1 \left ( 1 - u^p/p! \right)^n du=\int_{0}^{n^{-1/p+\delta}} \left ( 1 - u^p/p! \right)^n du +o\left(n^{-1/p}\right)\]
	and
	\begin{align*}
		\int_0^1 \left (1 - u^p/p! \right)^n du&={p!}^{1/p}\int_0^{{1/p!}}\left(1-t\right)^n t^{1/p-1}dt/p\\
		&={p!}^{1/p}\text{Beta}_{{1/p!}}\left(1/p,n+1\right)/p,
	\end{align*} 
	where $\text{Beta}_{x}(a,b)$ is the incomplete Beta function. According to \cite{Gautschi1967}, we have
	\[\lim_{n\to\infty} V_\mathbf{T}=2\left\{p/2-\text{card}\left(\mathbf{T}\right)\right\}^{p-1}p!^{1/p-1}\Gamma\left(1/p\right)n^{-1/p},\]
	where $\Gamma(\cdot)$ is the Gamma function.
	
	Because 
	\[\text{E} \left[ V\left\{\mathbf{U}\left(\mathbf{D}_{\text{MC},p,n},\tilde f\right)\right\}\right ] = \sum_{0\leq \text{card}(\mathbf{T})<p/2} \left\{\left(-1\right)^{\text{card}(\mathbf{T})} V_\mathbf{T}\right\},\]
	\begin{align*}
		\lim_{n\to\infty}\left\{ \text{E} \left[ V\left\{\mathbf{U}\left(\mathbf{D}_{\text{MC},p,n},\tilde f\right)\right\}\right ]/n^{-1/p}\right\}=2p!^{1/p-1}\Gamma\left(1/p\right)g_p.
	\end{align*}
\end{proof}

\subsection{Proof of Theorem~\ref{thm:SI}}
\begin{proof}
	When $p=1$, $\mathbf{D}_{\text{SI},1,n}=\{1/(n+1),2/(n+1),\ldots,n/(n+1)\}$. 
	Consider three cases. 
	Firstly, when $1/(n+1) \in \mathbf{B}$. Then $\mathbf{U} = [0,1/(n+1))$. 
	Secondly, when $n/(n+1) \in \mathbf{A}$. Then $\mathbf{U} = (n/(n+1),1]$. 
	Thirdly, when there is one integer $1\leq z\leq (n-1)$ such that $z/(n+1) \in \mathbf{A}$ and $(z+1)/(n+1) \in \mathbf{B}$. 
	Then $\mathbf{U}=(z/(n+1),(z+1)/(n+1))$. 
	For all three cases,
	\[V\{\mathbf{U}(\mathbf{D}_{\text{SI},1,n},f)\} =(n+1)^{-1}.\] 
	
	For $p\geq 2$, let $m=n^{1/p}+1$. For any $i_j\in\mathbb{N}$ and $1 \leq i_j \leq m-1$, $ 1\leq j \leq p-1$ and $j\in \mathbb{Z}$, let $k({i_1,\cdots,i_{p-1}})$ denote the maximum integer that satisfies
	\[f\left(i_1/m,\cdots,i_{p-1}/m,k({i_1,\cdots,i_{p-1}})/m\right)=-1.\]
	When $f\left(i_1/m,\cdots,i_{p-1}/m,1/m\right)=1$, let $k({i_1,\cdots,i_{p-1}})$ be 0.
	
	Because
	\begin{align*}
		V&\left(\cup_{\mathbf{x} \in \mathbf{D}_{\text{SI},p,n}, f(\mathbf{x})=-1} \{y : y_k \leq x_k, k=1,2,
		\cdots,p \}\right)\\
		=V&\left(\cup_{1\le i_1\leq m-1}\cdots\cup_{1\le i_{p-1}\leq m-1}\left\{x_1\le i_1/m,\cdots,x_{p-1}\le i_{p-1}/m ,\right.\right.\\
		&\quad\left.\left. x_p\le k({i_1,\cdots,i_{p-1}})/m\right\}\right)\\
		=\quad&\sum_{i_1,\cdots,i_{p-1}=1}^{m-1}k\left({i_1,\cdots,i_{p-1}}\right)m^{-p}
	\end{align*}
	and
	\begin{align*}
		V& \left(\cup_{\mathbf{x} \in \mathbf{D}_{\text{SI},p,n}, f(\mathbf{x})=1} \{y : y_k \geq x_k, k=1,2\cdots,p \}\right)\\
		=V&\left(\cup_{1\leq i_1\leq m-1}\cdots\cup_{1\le i_{p-1}\leq m-1}\left\{x_1\ge i_1/m,\cdots,x_{p-1}\ge i_{p-1}/m ,\right.\right.\\
		&\quad \left.\left.x_p \ge \{k({i_1,\cdots,i_{p-1}})+1\}/m\right\}\right)\\
		=\quad&\sum_{i_1,\cdots,i_{p-1}=1}^{m-1}\left[m-\left\{k({i_1,\cdots,i_{p-1}})+1\right\}\right]m^{-p},
	\end{align*}
	\begin{align*}
		V(\mathbf{U})=&1-V\left(\cup_{\mathbf{x} \in \mathbf{D}_{\text{SI},p,n}, f(\mathbf{x})=-1} \{y : y_k \leq x_k, k=1,2,
		\cdots,p \}\right)\\
		&-V \left(\cup_{\mathbf{x} \in \mathbf{D}_{\text{SI},p,n}, f(\mathbf{x})=1} \{y : y_k \geq x_k, k=1,2\cdots,p \}\right)\\
		=&1-\left\{\sum_{i_1,\cdots,i_{p-1}=1}^{m-1} k\left({i_1,\cdots,i_{p-1}}\right)\right.\\
		&\quad\left.+\sum_{i_1,\cdots,i_{p-1}=1}^{m-1}\left[m-\{k({i_1,\cdots,i_{p-1}})+1\}\right]\right\}m^{-p}\\
		=& 1-(m-1)^pm^{-p}.
	\end{align*}
	Therefore, $V\{\mathbf{U}(\mathbf{D}_{\text{SI},p,n},f)\} =1-n/(n^{1/p}+1)^p $ for $p\geq 2$.
	This completes the proof.
\end{proof}

\subsection{Proof of Theorem~\ref{thm:AMC}}
\begin{proof}
	When $p=1$, let $\tilde f(x)$ denote the function that outputs -1 if and only if $x< 1/2$. From Theorem~\ref{thm:MC}, 
	\[ \text{E} [ V\{\mathbf{U}(\mathbf{D}_{\text{MC},1,n}, \tilde f)\} ] = (2 - 2^{-n}) /(n+1).\]
	Consequently, 
	\[\text{E}\left\{m_{\text{AMC}}\left(n\right)\right\}=\sum_{i=1}^n V\left\{\mathbf{U}\left(\mathbf{D}_{\text{MC},1,i-1}, \tilde f\right)\right\}= \sum_{i=1}^n\left(2 - 2^{-(i-1)}\right) /i.\]
	Because
	\[\lim_{n\to\infty}\sum_{i=1}^n 2^{-(i-1)}/i=2\lim_{n\to\infty}\sum_{i=1}^n 2^{-i}/i=2\lim_{n\to\infty}\lim_{x\to1/2}\sum_{i=1}^n x^i/i\]
	\[=2\lim_{x\to1/2}\int_0^x \sum_{i=1}^{\infty} y^{i-1}dy=2\int_0^{1/2}\sum_{i=0}^{\infty} y^idy=2\int_{0}^{1/2}1/(1-y)dy=2\ln 2\]
	and according to \cite{eulerconstant},
	\[\lim_{n\to\infty}\left\{ \left(\sum_{i=1}^n 1/i\right)- \ln n\right\}=\gamma,\]
	\[\lim_{n\to\infty}\left[\text{E}\left\{m_{\text{AMC}}\left(n\right)\right\}-2\ln n\right]= 2\left(\gamma-\ln2\right).\]
	
	When $p\geq 2$, let $c = 2{p!}^{1/p-1}\Gamma(1/p)g_p$. From Theorem~\ref{thm:MC}, 
	\[ \lim_{n\to \infty} \text{E} \left\{\left[ V\left\{\mathbf{U}(\mathbf{D}_{\text{MC},p,n},\tilde f)\right\}\right] /  n^{-1/p}  \right\} = c. \]
	Consequently, for any $\epsilon>0$, there exist an integer $N$ such that for any $n\geq N$, 
	\begin{equation}\label{eqn:V:MC:n}
		c(1-\epsilon)n^{-1/p} \leq \text{E} \left[ V\left\{\mathbf{U}(\mathbf{D}_{\text{MC},p,n},\tilde f)\right\}\right] \leq c(1+\epsilon)n^{-1/p} . 
	\end{equation} 
	Let $m_t$ denote the number of functional evaluations required for trying the $(t^pN +1)$-th to $((t+1)^pN )$-th $\mathbf{x}$'s in the algorithm. 
	Then \[ \text{E}(m_t) = \sum_{i=t^pN +1}^{(t+1)^pN} V\left\{\mathbf{U}(\mathbf{D}_{\text{MC},p,i-1},\tilde f)\right\}. \]
	From \eqref{eqn:V:MC:n}, 
	\[ c(1-\epsilon) \{(t+1)^p-t^p\} N^{(p-1)/p}/(t+1) \leq \text{E}(m_t) \leq c(1+\epsilon) \{(t+1)^p-t^p\} N^{(p-1)/p}/t. \]
	Therefore, 
	\[ \text{E}\left\{m_{\text{AMC}}(k^p N)\right\} = \text{E}\left\{m_{\text{AMC}}(N)\right\} + \sum_{t=1}^{k-1} E(m_k) \]
	\begin{align*}
		\in \left[ c(1-\epsilon) N^{(p-1)/p} \sum_{t=1}^{k-1} \left[\left\{\left(t+1\right)^p-t^p\right\}/(t+1)\right],\right.\\
		\left. c(1+\epsilon) N^{(p-1)/p} \sum_{t=1}^{k-1} \left[\left\{\left(t+1\right)^p-t^p\right\}/t\right] + N \right].
	\end{align*}
	Because 
	\[ \lim_{k \to \infty} \left\{ \sum_{t=1}^{k-1} [\{(t+1)^p-t^p\}/(t+1)] / \{p/(p-1)k^{p-1}\} \right\} \] 
	\[ = \lim_{k \to \infty} \left\{ \sum_{t=1}^{k-1} [\{(t+1)^p-t^p\}/t] / \{p/(p-1)k^{p-1}\} \right\} = 1, \]
	for any $\tilde \epsilon>0$, there exists an integer $K$ such that for any $k>K$, 
	\begin{align*}
		&c(1-\epsilon)(1-\tilde\epsilon) \{p/(p-1)k^{p-1}\} N^{(p-1)/p} \leq \text{E}\{m_{\text{AMC}}(k^p N)\}\\
		&\leq c(1+\epsilon)(1+\tilde\epsilon) \{p/(p-1)k^{p-1}\} N^{(p-1)/p}+N.
	\end{align*}
	Consequently, 
	\[ \lim_{n \to \infty} \left[\text{E}\left\{m_{\text{AMC}}(n)\right\}/n^{(p-1)/p}\right] =  2{p!}^{1/p-1}\Gamma(1/p)g_pp(p-1)^{-1}. \]
\end{proof}

\subsection{Proof of Theorem~\ref{thm:GI}}
\begin{proof}
	{Let $m=2^l$.}
	First, we prove that
	\begin{align}\label{eqn:GI}
		m_{\text{GI}}\left(l+1\right) - m_{\text{GI}}\left(l\right) = (2m-1)^p - (2m-2)^p
	\end{align}
	for every $l$. We prove \eqref{eqn:GI} by inducing on $p$. 
	
	When $p=1$, consider three cases. 
	Firstly, when $1/m \in \mathbf{B}$. Then $\mathbf{D}_{\text{GI},1,m_{\text{GI}}(l+1)}\setminus \mathbf{D}_{\text{GI},1,m_{\text{GI}}(l)} = \left\{ 1/(2m) \right\}$. 
	Secondly, when $(m-1)/m \in \mathbf{A}$. Then $\mathbf{D}_{\text{GI},1,m_{\text{GI}}(l+1)}\setminus \mathbf{D}_{\text{GI},1,m_{\text{GI}}(l)} = \left\{ (2m-1)/(2m) \right\}$. 
	Thirdly, when there is one integer $1\leq z\leq m-2$ such that $z/m \in \mathbf{A}$ and $(z+1)/m \in \mathbf{B}$. 
	Then $\mathbf{D}_{\text{GI},1,m_{\text{GI}}(l+1)}\setminus \mathbf{D}_{\text{GI},1,m_{\text{GI}}(l)} = \left\{ (2z+1)/(2m) \right\}$. 
	In all three cases, $m_{\text{GI}}\left(l+1\right)-m_{\text{GI}}(l) = 1 = (2m-1)^1 - (2m-2)^1$. 
	
	Suppose \eqref{eqn:GI} holds for $1,\ldots,p-1$. 
	Clearly, $\mathbf{D}_{\text{GI},p,m_{\text{GI}}(l+1)}\setminus \mathbf{D}_{\text{GI},p,m_{\text{GI}}(l)}$ can be uniquely partitioned by 
	\[ \mathbf{D}_{\text{GI},p,m_{\text{GI}}(l+1)}\setminus \mathbf{D}_{\text{GI},p,m_{\text{GI}}(l)} = \cup_{z=1}^{2 m-1} \left\{\mathbf{W}_{z} \times \left\{ z/\left(2 m\right) \right\} \right\}. \]
	From induction, for any even integer $z$ such that $1< z< 2m-1$, 
	$ \text{card}(\mathbf{W}_z) = (2m-1)^{p-1} - (2m-2)^{p-1}$. 
	Suppose $z$ and $\tilde z$ are odd integers and $1\leq z<\tilde z\leq 2m-1$. 
	Because $( \mathbf{W}_{\tilde z} \setminus \mathbf{W}_{\tilde z-1} ) \times \{(\tilde z-1)/(2m)\} \subset {\mathbf{A}}$ and ${(\mathbf{W}_z\setminus \mathbf{W}_{\tilde z-1}) }\times \{(\tilde z-1)/(2m)\} \subset {\mathbf{B}}$, 
	$\mathbf{W}_z \cap (\mathbf{W}_{\tilde z} \setminus \mathbf{W}_{\tilde z-1}) = \emptyset$.
	Consequently, $\sum_{3\leq z\leq 2m-1, z \text{ is odd}} \text{card}(\mathbf{W}_z \setminus \mathbf{W}_{z-1}) +\text{card}(\mathbf{W}_1) = \left( 2 m -1 \right)^{p-1}$. 
	Therefore, 
	\begin{align*}
		m_{\text{GI}}\left(l+1\right)-m_{\text{GI}}\left(l\right) &= \sum_{z=1}^{2m-1} \text{card}(\mathbf{W}_z) 
		= \sum_{1\leq z\leq 2m-1, z \text{ is even}} \text{card}(\mathbf{W}_z) \\
		&\quad+ \sum_{1\leq z\leq 2m-1, z \text{ is odd}} \text{card}(\mathbf{W}_z) \\
		&= \sum_{1\leq z\leq 2m-1, z \text{ is even}} \text{card}(\mathbf{W}_z) + \sum_{3\leq z\leq 2m-1, z \text{ is odd}} \text{card}(\mathbf{W}_{z-1}) \\
		&\quad+ \sum_{3\leq z\leq 2m-1, z \text{ is odd}} \text{card}(\mathbf{W}_z \setminus \mathbf{W}_{z-1})+\text{card}(\mathbf{W}_1)\\
		&= (2m-2) \{ (2m-1)^{p-1} - (2m-2)^{p-1} \} + \left( 2 m -1 \right)^{p-1} \\
		&= (2m-1)^p - (2m-2)^p.
	\end{align*}
	From the induction, \eqref{eqn:GI} holds for any $p$. 
	Consequently, 
	\[m_{\text{GI}}(g)= 1+\sum_{i=2}^g\left\{m_{\text{GI}}\left(i\right)-m_{\text{GI}}\left(i-1\right)\right\}
	= \sum_{l=1}^g \left\{ (2^l-1)^p - (2^l-2)^p \right\},\]
	which completes the proof. 
	
\end{proof}

\bibliographystyle{unsrt}
\bibliography{wpref}

\end{document}